\documentclass[twocolumn]{aa}
\usepackage{graphicx}
\usepackage{txfonts}
\usepackage[english]{babel}
\usepackage{color}
\usepackage{natbib}
\usepackage{lscape}
\usepackage{longtable}
\usepackage{setspace}
\usepackage{subfigure}

\bibpunct{(}{)}{;}{a}{}{,} 

\begin{document}

\title{Open clusters: III. Fundamental parameters of B stars in NGC\,6087, NGC\,6250, NGC\,6383 and NGC\,6530.}
\subtitle{B type stars with circumstellar envelopes.\thanks{Observations taken at CASLEO, operating under agreement of CONICET and the Universities of La Plata, C\'ordoba and San Juan, Argentina}}
\titlerunning{Open clusters: Fundamental parameters of B and Be stars}

\author{Y. Aidelman\inst{1,2}\fnmsep\thanks{Fellow of CONICET, Argentina}\and 
L. S. Cidale\inst{1,2}\fnmsep\thanks{Member of the Carrera del Investigador Cient\'{\i}fico, CONICET, Argentina}
\and
J. Zorec\inst{3,4}
\and
J. A. Panei\inst{1,2}$^{,\star\star\star}$
}

\authorrunning{Y. Aidelman et al.}

\institute{
Instituto de Astrof\'{\i}sica La Plata, CCT La Plata, CONICET-UNLP, Paseo del Bosque S/N, B1900FWA, La Plata, Argentina.
\and
Departamento de Espectroscop\'{\i}a, Facultad de Ciencias Astron\'omicas y Geof\'{\i}sicas, Universidad Nacional de La Plata (UNLP), Paseo del Bosque S/N, B1900FWA, La Plata, Argentina.
\and
Sorbonne Universit\'e, UPMC  Univ. Paris 06, 934 UMR7095-IAP, 75014 Paris, France.
\and
CNRS, 934 UMR7095-IAP, Institut d'Astrophysique de Paris, 98bis Bd. Arago, 75014 Paris, France.
}
\offprints{Y. Aidelman \email{aidelman@fcaglp.unlp.edu.ar}}

\date{Received / Accepted }

\abstract
{Stellar physical properties of star clusters are poorly known and the cluster parameters are often quite uncertain.}
{Our goals  are  to perform a spectrophotometric study of the B star population in open clusters to derive accurate stellar parameters, search for the presence of circumstellar envelopes and discuss the characteristics of these stars.}
{The BCD spectrophotometric system is a powerful method to obtain stellar fundamental parameters from direct measurements of the Balmer discontinuity.
To this end, we wrote the interactive code MIDE3700.
The BCD parameters can also be used to infer the main properties of open clusters: distance modulus, color excess and age.
Furthermore, we inspect the aspect of the Balmer discontinuity to put in evidence the presence of circumstellar disks and identify Be-star candidates.
An additional set of high resolution spectra, in the H$\alpha$ region, is used to confirm the Be nature of these stars.}
{In this work we provide $T_{\rm {eff}}$, $\log\,g$, $M_{\rm v}$, $M_{\rm {bol}}$ and spectral types for a sample of $68$ stars in the field of the open clusters \object{NGC\,6087}, \object{NGC\,6250}, \object{NGC\,6383}, and \object{NGC\,6530}, as well as the cluster distances, ages and reddening.
Then, based on a sample of $230$ B stars in the direction of the eleven open clusters studied along these three series of papers, we report new six Be stars, four blue straggler candidates, and fifteen B type stars (called Bdd) with a double Balmer discontinuity which indicates the presence of circumstellar envelopes.
We discuss the distribution of the fraction of B, Be and Bdd star cluster members per spectral subtype.
The majority of the Be stars are dwarfs and present a maximum at the spectral type B2-B4 in young and intermediate-age open clusters ($< 40$~Myr).
Another maximum of Be stars is observed at the spectral type B6-B8 in open clusters older than $40$~Myr, where the population of Bdd stars also becomes relevant.
The Bdd stars seem to be in  a ``passive'' emission phase.}
{Our results support previous statements that the Be phenomenon is present along the whole main sequence band and occurs in very different evolutionary states.
We find clear evidence of an enhance of stars with circumstellar envelopes with cluster age.
The Be phenomenon reaches it maximum in clusters of intermediate-age ($10-40$~Myr) and the number of B stars with circumstellar envelopes (Be plus Bdd stars) keeps also high for the older clusters ($40-100$~Myr).}


\maketitle


\section{Introduction}
\label{Int}

This is the third of a series of papers devoted to the spectroscopic study of a total of $230$ B-type stars in the field of view of eleven open clusters.
This study is based on the spectrophotometric BCD (Barbier-Chalonge-Divan) system, developed by \citet{Barbier1939} and \citet{Chalonge1973}.

Our main goals were  to derive accurate stellar parameters ($T_{\rm {eff}}$, $\log\,g$, $M_{\rm v}$, $M_{\rm {bol}}$ and spectral types) of the B star population in open clusters and search for the presence of B stars with circumstellar envelopes.
A secondary objective was to provide an  improved set of cluster parameters and identify their star memberships.

Particularly, this work presents a spectrophotometric investigation of $68$ B-type stars in the region of the galactic clusters \object{NGC\,6087}, \object{NGC\,6250}, \object{NGC\,6383}, and \object{NGC\,6530}.
The results related with other seven clusters were published by \citet[][hereafter Paper I and Paper II, respectively]{Aidelman2012, Aidelman2015}.
In addition, all of the  information gathered along this series of papers  allows us to analyze here the relative frequency of cluster star members per spectral subtypes and age, the distribution of stars with circumstellar envelopes in a $T_{\rm eff}-\log\,g$ diagram and discuss some characteristics of the  peculiar group of Be stars.
In Paper I, we introduced the BCD method as a tool to evaluate precise physical parameters of the cluster's stars and the cluster properties, in a fast and direct way, by measuring the Balmer discontinuity (BD).
On the other hand, as the BCD parameters are  not affected by interstellar absorption and/or circumstellar emission/absorption \citep{Zorec1991}, the location of stars in the Hertzsprung-Russell (HR) diagram is more accurate than classical methods based on plain photometry, thus allowing for better determinations of cluster ages.
Moreover, the BCD method also enables us to characterize each star in very crowded stellar regions with fundamental parameters that cannot be strongly perturbed by the light of the nearest objects, so that it is possible to discuss properly their cluster membership by taking color excesses and distance estimates, as done in Paper II.

Another advantage of the study of the BD is that it allows us to recognize B stars with circumstellar envelopes since some B-type stars display a second component of the BD \citep[called hereafter SBD. For more details see][and Paper I]{Divan1979, Zorec1991, Cidale2001, Zorec2005, Cidale2007}.
The main BD, in absorption, occurs as in normal dwarf stars and is attributed to the central star.
The SBD is situated at shorter wavelengths and is originated in an envelope having low pressure \citep{Divan1979}.
A SBD in absorption is often related with spectral line ``shell'' signatures, while a SBD in emission is generally accompanied by line emission features.
We can then use the SBD, when is present, as a criterion to detect stars having circumstellar envelopes or disks. This is possible even though they do not exhibit clear signs of emission or shell-absorption features among the first attainable members of the Balmer lines, as a consequence of a late B spectral type or the low resolution of the available spectrophotometric spectra.
Therefore, as a complement of this work, we present spectra of B stars in open clusters that exhibit a SBD and discuss the properties of this star population (see details in sections \S \ref{Dis_SBD} and \ref{CE}).

The paper is organized as follow: data acquisition and reduction procedure are given in section \S~\ref{Obs}.
In section \S~\ref{Met} we briefly describes the BCD method and its application to our star sample.
Spectrophotometric results on the cluster parameters  (\object{NGC\,6087}, \object{NGC\,6250}, \object{NGC\,6383}, and \object{NGC\,6530}) and their memberships are given in section \S~\ref{Res}.
In section \S~\ref{Dis} we analyze the relative frequency of B and Be stars per spectral subtypes and age, using data of  the eleven open clusters studied.
We discuss the derived stellar parameters and the presence of circumstellar envelopes in the late B-type stars.
Our conclusions are summarized in section \S~\ref{Con}.


\section{Observations}
\label{Obs}

Low-resolution spectra of additional $68$ B-type stars in four open clusters were obtained in long slit mode during multiple observing runs between 2003 and 2013 with a Boller \& Chivens spectrograph\footnote{The optical layout of this instrument is similar to the Boller \& Chivens spectrograph offered for the du Pont 100-inch telescope of Las Campanas Observatory (Chile), http://www.lco.cl/telescopes-information/irenee-du-pont/instruments/website/boller-chivens-spectrograph-manuals/user-manual/the-boller-and-chivens-spectrograph} attached to the J. Sahade $2.15$~m telescope at the Complejo Astron\'omico El Leoncito (CASLEO), San Juan, Argentina. The log of observations is listed in Table~\ref{Log-Obs-1}. The spectra were taken with a Bausch and Lomb replica diffraction grating of $600$~l\,mm$^{-1}$ (\#$80$, blazed at 4000\,\AA) and slit widths of $250~\mu$m  and $350~\mu$m oriented along east-west. The slit widths were set to match an aperture on  the sky of about $2\farcs3$ and $3\farcs3$, respectively, according to the average seeing at CASLEO.

Before 2011, we used a PM $512$ CCD detector co\-ve\-ring the spectral wavelength range of $3500-4700$~\AA.
The rest of the observations were carried out with the CCD detector TEK $1024$ that covered the wavelength interval $3500-5000$~\AA.
The effective spectral resolutions are $4.53$~\AA\, and $2.93$~\AA\, every two pixels ($R \sim 900$, $R \sim 1\,400$), respectively.
A comparison lamp of He-Ne-Ar was taken after each science stellar spectrum.

The observations were performed at the lowest possible zenith distance to minimize refraction effects due to the Earth's atmosphere and, thus, to avoid light losses as a function of wavelength. For an air mass of 1.5 the expected differential refraction effect between 4000\,\AA\, and 6000\,\AA\, is $1\farcs08$ (http://www.eso.org/sci/observing/tools/ca\-len\-dar/ParAng.html). This assures that the light-loss is minimized.
Each night we observed at least two flux standard stars (\object{HR\,3454}, \object{HR\,5501}, \object{HR\,7596}, \object{HR\,4468} and \object{HR\,5694}) to perform the spectral flux calibrations.

The reduction procedure (overscan, bias, trimming, and flat-field corrections) was carried out with the {\sc iraf}\footnote{{\sc iraf} is distributed by the National Optical Astronomy Observatory, which is operated by the Association of Universities for Research in Astronomy (AURA), Inc., under cooperative agreement with the National Science Foundation} software package (version 2.14.1) and all spectra were wavelength calibrated and corrected for atmospheric extinction.
We used the atmospheric extinction coefficients published in CASLEO web-page (http://www.casleo.gov.ar/info-obs.php).

An addition set of high-resolution spectra ($R = 12\,600$) were taken with the echelle spectrograph REOSC in the spectral range $4225-6700$~\AA\, to seek the H$\alpha$ emission in the B stars with a SBD. These observations were done only for some of the star members identified in the eleven studied Galactic clusters (see Table~\ref{Log-Obs-2}).
The instrumental configuration was a $400$~l\,mm$^{-1}$ grating (\#$580$), a $250~\mu$m slit width and a TK $1$K CCD detector.
A Th-Ar comparison lamp was used to apply the wavelength calibration.

\onllongtab{
\begin{longtable}{lllllll}
\caption{\label{Log-Obs-1} Log of observations of low-resolution spectra. The coordinates $\alpha$ and $\delta$ correspond to the ICRS system (ep = J2000 and eq = 2000). $AM$ is the air mass and $N$ is the  number of spectra taken for each star with the Boller \& Chivens spectrograph. Nomenclatures for \object{NGC\,6087} corresponds to \citet{Fernie1961} and \citet{Breger1966}, for \object{NGC\,6250} to \citet{Moffat1975} and \citet{Herbst1977}, for \object{NGC\,6383} to \citet{Eggen1961, The1965} and \citet{LloydEvans1978}, and for \object{NGC\,6530} to \citet{Walker1957} and \citet{Kilambi1977}.}\\
\hline
\hline
ID & Other       & $\alpha$ & $\delta$ & $AM$ & dates    & N\\
   & designation &              &               &          &              &\\

\hline
\endfirsthead

\multicolumn{7}{l}%
{\tablename\ \thetable\ -- \textit{Continued from previous page.}} \\

\hline
\hline
ID & Other       & $\alpha$ & $\delta$ & $AM$ & dates   & N\\
   & designation &          &          &     &  &\\

\hline
\endhead
\hline \multicolumn{7}{r}{\textit{Continued on next page.}} \\
\endfoot
\hline
\endlastfoot

\multicolumn{7}{l}{\bf \object{NGC\,6087}}\\
\hline
001      & \object{HD\,146\,271}       & $16:18:37.60$ & $-57:41:46.06$ & $1.18$ & 2013-07-20 & 1\\
007      & \object{HD\,146\,483}       & $16:19:38.46$ & $-57:54:22.57$ & $1.1 $, $1.1 $, $1.14$ & 2003-04-26, 2003-06-04, 2013-07-22 & 3\\
008      & \object{HD\,146\,448}       & $16:19:25.14$ & $-57:53:08.59$ & $1.3 $ & 2003-04-25 & 1\\
009      & \object{HD\,146\,484}       & $16:19:36.67$ & $-57:56:05.46$ & $1.1 $, $1.1 $, $1.13$ & 2003-04-26, 2003-06-04, 2013-07-22 & 3\\
010      & \object{HD\,146\,324}       & $16:18:49.01$ & $-57:55:51.52$ & $1.1 $, $1.17$ & 2003-04-25,  2013-07-20 & 2\\
011      & \object{HD\,146\,294}       & $16:18:37.89$ & $-57:56:20.30$ & $1.1 $ & 2003-04-25 & 2\\
013      & \object{HD\,146\,261}       & $16:18:24.89$ & $-57:49:32.97$ & $1.14$ & 2013-07-20 & 1\\
014      & \object{CPD$-$57\,7791}     & $16:18:22.81$ & $-57:51:28.67$ & $1.2 $ & 2003-04-26 & 1\\
015      & \object{HD\,146\,204}       & $16:18:06.40$ & $-57:51:00.39$ & $1.2 $ & 2003-04-26 & 1\\
022      & \object{HD\,146\,531}       & $16:19:55.47$ & $-58:10:01.87$ & $1.16$ & 2013-07-20 & 1\\
025      & \object{CPD$-$57\,7817}     & $16:18:49.69$ & $-57:54:30.31$ & $1.1 $ & 2003-04-25 & 1\\
033      & \object{GEN\#\,+2.60870033} & $16:19:47.00$ & $-57:56:12.00$ & $1.1 $ & 2003-06-04 & 1\\
035      & \object{HD\,146\,428}       & $16:19:21.24$ & $-57:54:23.72$ & $1.2 $ & 2003-04-25 & 1\\
036      & \object{GEN\#\,+2.60870036} & $16:19:20.00$ & $-57:52:18.00$ & $1.2 $ & 2003-04-25 & 1\\
101      & \object{GEN\#\,+2.60870101} & $16:18:16.62$ & $-57:48:42.76$ & $1.2 $ & 2003-04-26 & 1\\
128      & \object{CD$-$57\,6341}      & $16:18:50.58$ & $-57:56:39.71$ & $1.1 $ & 2003-04-25 & 1\\
129      & \object{GEN\#\,+2.60870129} & $16:18:47.06$ & $-57:56:44.89$ & $1.1 $ & 2003-04-25 & 1\\
156      & \object{CD$-$57\,6346}      & $16:19:18.33$ & $-57:53:36.54$ & $1.1 $ & 2003-04-25 & 1\\
\hline
\multicolumn{7}{l}{\bf \object{NGC\,6250}}\\
\hline
001      & \object{HD\,152\,853}       & $16:58:07.93$ & $-45:58:56.47$ & $1.25$ & 2011-05-12 & 1\\
002      & \object{HD\,152\,799}       & $16:57:42.84$ & $-45:57:41.26$ & $1.2 $ & 2011-05-12 & 1\\
003      & \object{HD\,152\,822}       & $16:57:55.89$ & $-45:56:09.27$ & $1.14$ & 2011-05-12 & 1\\
004      & \object{HD\,152\,917}       & $16:58:33.03$ & $-45:55:13.28$ & $1.25$, $1.08$ & 2011-05-13, 2013-07-21 & 2\\
017      & \object{HD\,329\,271}       & $16:57:41.27$ & $-45:58:49.91$ & $1.04$ & 2013-07-21 & 1\\
018      & \object{CD$-$45\,11088}     & $16:57:43.04$ & $-46:00:27.09$ & $1.03$ & 2011-05-12 & 1\\
021      & \object{HD\,152\,743}       & $16:57:20.45$ & $-45:51:57.87$ & $1.09$ & 2011-05-13 & 1\\
022      & \object{HD\,152\,706}       & $16:57:12.10$ & $-45:52:35.73$ & $1.14$ & 2011-05-13 & 1\\
032      & \object{HD\,152\,687}       & $16:57:04.46$ & $-46:08:37.49$ & $1.2 $ & 2011-05-13 & 1\\
033      & \object{HD\,152\,561}       & $16:56:15.54$ & $-46:02:57.73$ & $1.03$ & 2011-05-13 & 1\\
034      & \object{HD\,153\,073}       & $16:59:26.70$ & $-45:53:04.29$ & $1.06$ & 2011-05-13 & 1\\
035      & \object{HD\,152\,979}       & $16:58:56.81$ & $-46:07:44.38$ & $1.04$ & 2011-05-12 & 1\\
037      & \object{HD\,329\,379}       & $17:01:05.88$ & $-45:42:04.32$ & $1.04$, $1.09$ & 2011-05-13, 2013-07-21 & 2\\
$\cdots$ & \object{CD$-$49\,11096}     & $16:59:36.86$ & $-49:45:25.45$ & $1.2 $ & 2013-07-23 & 1\\
$\cdots$ & \object{HD\,329\,211}       & $16:59:32.55$ & $-44:52:37.52$ & $1.08$ & 2011-05-12 & 1\\
\hline
\multicolumn{7}{l}{\bf \object{NGC\,6383}}\\
\hline
001      & \object{CD$-$32\,12935}     & $17:34:42.49$ & $-32:34:53.99$ & $1.03$, $1.0$, $1.12$ & 2012-06-06, 2012-06-08, 2013-07-21 & 3\\
002      & \object{CD$-$32\,12931}     & $17:34:34.95$ & $-32:33:47.13$ & $1.46$ & 2012-06-08 & 1\\
003      & \object{CD$-$32\,12927}     & $17:34:34.20$ & $-32:36:08.90$ & $1.  $ & 2012-06-08 & 1\\
006      & \object{CD$-$32\,12921}     & $17:34:22.02$ & $-32:33:38.85$ & $1.04$, $1.02$ & 2012-06-06, 2012-06-08 & 2\\
010      & \object{CD$-$32\,12943}     & $17:34:57.22$ & $-32:40:20.92$ & $1.11$ & 2012-06-08 & 1\\
014      & \object{CD$-$32\,12929}     & $17:34:38.56$ & $-32:34:59.30$ & $1.27$ & 2012-06-07 & 1\\
047      & \object{CD$-$32\,12954}     & $17:35:10.24$ & $-32:29:04.08$ & $1.06$ & 2012-06-08 & 1\\
057      & \object{CD$-$32\,12946}     & $17:35:02.16$ & $-32:37:36.30$ & $1.25$ & 2013-07-23 & 1\\
076      & \object{CD$-$32\,12908}     & $17:34:02.23$ & $-32:40:39.68$ & $1.  $ & 2012-06-07 & 1\\
083      & \object{CD$-$32\,12919}     & $17:34:16.86$ & $-32:36:32.06$ & $1.02$ & 2012-06-07 & 1\\
085      & \object{CD$-$32\,12910}     & $17:34:02.69$ & $-32:37:49.14$ & $1.06$ & 2012-06-07 & 1\\
100      & \object{CD$-$32\,12924}     & $17:34:24.51$ & $-32:30:15.98$ & $1.09$, $1.03$ & 2012-06-06, 2012-06-08 & 2\\
\hline
\multicolumn{7}{l}{\bf \object{NGC\,6530}}\\
\hline
007      & \object{HD\,164\,794}       & $18:03:52.44$ & $-24:21:38.63$ & $1.22$, $1.18$ & 2012-06-06, 2013-07-21 & 2\\
009      & \object{HD\,164\,816}       & $18:03:56.87$ & $-24:18:45.22$ & $1.3 $ & 2011-05-12 & 1\\
032      & \object{GEN\#\,+2.65300032} & $18:04:11.16$ & $-24:21:45.20$ & $1.01$ & 2011-05-10 & 1\\
042      & \object{HD\,315\,032}       & $18:04:15.03$ & $-24:23:27.71$ & $1.01$ & 2011-05-10 & 1\\
043      & \object{HD\,315\,026}       & $18:04:14.51$ & $-24:14:37.02$ & $1.26$ & 2011-05-13 & 1\\
045      & \object{HD\,164\,865}       & $18:04:15.22$ & $-24:11:00.09$ & $1.14$, $1.2 $ & 2011-05-11, 2013-07-21 & 2\\
055      & \object{HD\,315\,023}       & $18:04:20.56$ & $-24:13:54.90$ & $1.01$ & 2011-05-12 & 1\\
056      & \object{CD$-$24\,13829}     & $18:04:21.30$ & $-24:21:18.18$ & $1.02$ & 2011-05-10 & 1\\
059      & \object{HD\,315\,033}       & $18:04:23.23$ & $-24:26:16.71$ & $1.22$, $1.18$ & 2011-05-11, 2013-07-20 & 2\\
060      & \object{LSS\,4615}          & $18:04:24.00$ & $-24:21:26.70$ & $1.04$ & 2011-05-10 & 1\\
061      & \object{ALS\,16976}         & $18:04:24.29$ & $-24:20:59.50$ & $1.08$ & 2011-05-10 & 1\\
065      & \object{HD\,164\,906}       & $18:04:25.84$ & $-24:23:08.35$ & $1.09$ & 2011-05-09 & 1\\
066      & \object{CD$-$24\,13831}     & $18:04:25.55$ & $-24:20:45.00$ & $1.03$ & 2011-05-11 & 1\\
068      & \object{CD$-$24\,13830}     & $18:04:22.74$ & $-24:22:09.80$ & $1.01$, $1.24$ & 2011-05-11, 2013-07-21 & 2\\
070      & \object{ALS\,18783}         & $18:04:27.21$ & $-24:22:49.56$ & $1.05$ & 2011-05-12 & 1\\
073      & \object{HD\,315\,031}       & $18:04:28.05$ & $-24:21:42.98$ & $1.14$ & 2011-05-09 & 1\\
076      & \object{HD\,315\,024}       & $18:04:29.28$ & $-24:19:25.37$ & $1.03$ & 2011-05-13 & 1\\
080      & \object{CD$-$24\,13837}     & $18:04:32.99$ & $-24:23:12.52$ & $1.15$ & 2011-05-13 & 1\\
085      & \object{HD\,164\,933}       & $18:04:33.03$ & $-24:09:38.27$ & $1.1 $ & 2011-05-11 & 1\\
086      & \object{CD$-$24\,13840}     & $18:04:34.20$ & $-24:22:00.60$ & $1.17$ & 2011-05-12 & 1\\
093      & \object{HD\,315\,021}       & $18:04:36.00$ & $-24:19:52.19$ & $1.27$ & 2011-05-09 & 1\\
100      & \object{HD\,164\,947}       & $18:04:41.66$ & $-24:20:56.95$ & $1.12$, $1.33$ & 2012-06-06, 2012-06-06 & 2\\
$\cdots$ & \object{LSS\,4627}          & $18:04:41.03$ & $-21:04:52.33$ & $1.06$ & 2011-05-13 & 1\\
\hline
\end{longtable}
}

\onllongtab{
\begin{longtable}{llllll}
\caption{\label{Log-Obs-2} Log of high-resolution spectra observations. The coordinates $\alpha$ and $\delta$ correspond to the ICRS system (ep = J2000 and eq = 2000). $X$ indicates the values of the air mass of star. The last column, N, lists the number of spectra taken for each star with the REOSC DC spectrograph. Clusters \object{NGC\,3766} and \object{NGC\,4755} were published in Paper I, \object{Collinder\,223}, \object{Hogg\,16}, \object{NGC\,3114}, and \object{NGC\,6025} were published in Paper II.}\\
\hline
\hline
ID & $\alpha$ & $\delta$ & $AM$ & dates  & N\\
   &          &          &     &  &\\

\hline
\endfirsthead

\multicolumn{6}{l}%
{\tablename\ \thetable\ -- \textit{Continued from previous page.}} \\

\hline
\hline
ID & $\alpha$ & $\delta$ & $AM$ & dates    & N\\
   &          &          &     & &\\

\hline
\endhead
\hline \multicolumn{6}{r}{\textit{Continued on next page.}} \\
\endfoot
\hline
\endlastfoot

\multicolumn{6}{l}{\bf \object{Collinder\,223}}\\
\hline
002           & $10:29:33.699 $ & $-60:12:37.42 $ & $1.15$ & 2013-04-22 & 1\\
080           & $10:30:23.8684$ & $-60:04:40.429$ & $1.21$ & 2013-04-22 & 1\\
106           & $10:30:20.3124$ & $-60:01:48.515$ & $1.17$ & 2013-04-22 & 1\\
HD\,305\,296  & $10:31:39.8575$ & $-60:17:10.938$ & $1.14$ & 2013-04-23 & 1\\

\hline
\multicolumn{6}{l}{\bf \object{Hogg\,16}}\\
\hline
003 & $13:29:34.28722$ & $-61:11:36.1245$ & $1.20$ & 2013-04-22 & 1\\
009 & $13:29:17.565  $ & $-61:11:51.84  $ & $1.20$ & 2013-04-24 & 1\\
014 & $13:29:09.1238 $ & $-61:05:41.931 $ & $1.15$ & 2013-06-22 & 1\\
052 & $13:28:43.9231 $ & $-61:01:25.820 $ & $1.15$ & 2013-04-22 & 1\\
068 & $13:28:01.49589$ & $-61:03:45.0368$ & $1.15$ & 2013-04-22 & 1\\

\hline
\multicolumn{6}{l}{\bf \object{NGC\,2645}}\\
\hline
01 & $08:39:04.174$ & $-46:13:36.51$ & $1.03$ & 2013-04-21 & 1\\
04 & $08:39:05.711$ & $-46:14:45.44$ & $1.04$ & 2013-04-21 & 1\\
05 & $08:39:04.341$ & $-46:14:52.09$ & $1.06$ & 2013-04-22 & 1\\

\hline
\multicolumn{6}{l}{\bf \object{NGC\,3114}}\\
\hline
003         & $10:03:08.21073$ & $-59:50:27.8712$ & $1.17$ & 2013-04-23 & 1\\
004         & $10:01:41.76408$ & $-60:06:46.5882$ & $1.14$ & 2013-04-23 & 1\\
011         & $10:00:21.091  $ & $-59:59:52.59  $ & $1.14$ & 2013-04-24 & 1\\
015         & $10:04:08.9394 $ & $-59:52:54.735 $ & $1.14$ & 2013-04-24 & 1\\
028         & $10:02:48.9721 $ & $-59:50:18.995 $ & $1.20$ & 2013-04-23 & 1\\
033         & $10:05:39.7720 $ & $-60:00:37.716 $ & $1.18$ & 2013-04-24 & 1\\
047         & $10:03:50.7764 $ & $-59:51:23.325 $ & $1.15$ & 2013-04-24 & 1\\
091         & $10:00:57.517  $ & $-59:58:23.81  $ & $1.24$ & 2013-04-24 & 1\\
129         & $10:02:35.580  $ & $-60:08:51.49  $ & $1.16$ & 2013-04-23 & 1\\
HD\,87\,801 & $10:05:33.4110 $ & $-59:54:23.665 $ & $1.17$ & 2013-04-23 & 1\\

\hline
\multicolumn{6}{l}{\bf \object{NGC\,3766}}\\
\hline
026           & $11:36:09.565  $ & $-61:35:38.20  $ & $1.20$ & 2013-04-23 & 1\\
027           & $11:36:11.903  $ & $-61:35:50.05  $ & $1.18$ & 2013-04-23 & 1\\
151           & $11:36:12.35   $ & $-61:32:44.9   $ & $1.16$ & 2013-04-23 & 1\\
232           & $11:36:28.37117$ & $-61:39:54.5183$ & $1.24$ & 2013-04-24 & 1\\
239           & $11:36:09.36848$ & $-61:41:41.5868$ & $1.17$ & 2013-06-23 & 1\\
240           & $11:36:05.48263$ & $-61:42:06.0590$ & $1.20$ & 2013-06-21 & 1\\
264           & $11:35:15.1721 $ & $-61:41:59.607 $ & $1.16$ & 2013-06-21 & 1\\
291           & $11:35:22.0682 $ & $-61:32:11.048 $ & $1.21$ & 2013-06-23 & 1\\
301           & $11:37:48.854  $ & $-61:45:05.15  $ & $1.17$ & 2013-06-24 & 1\\
HD\,306\,787  & $11:36:48.37   $ & $-61:26:41.7   $ & $1.24$ & 2013-06-25 & 1\\
HD\,308\,852  & $11:37:41.9600 $ & $-61:45:42.733 $ & $1.19$ & 2013-04-24 & 1\\

\hline
\multicolumn{6}{l}{\bf \object{NGC\,4755}}\\
\hline
001  & $12:53:21.89463$ & $-60:19:42.5630$ & $1.18$ & 2013-04-23 & 1\\
106  & $12:53:37.62181$ & $-60:21:25.3912$ & $1.17$ & 2013-04-23 & 1\\
306  & $12:53:51.57   $ & $-60:23:16.9   $ & $1.18$ & 2013-04-24 & 1\\

\hline
\multicolumn{6}{l}{\bf \object{NGC\,6025}}\\
\hline
01  & $16:03:44.46671$ & $-60:29:54.4704$ & $1.18$ & 2013-04-22 & 1\\
03  & $16:03:28.89312$ & $-60:29:12.7768$ & $1.16$ & 2013-04-22 & 1\\
25  & $16:03:31.9865 $ & $-60:20:11.228 $ & $1.15$ & 2013-04-22 & 1\\

\hline
\multicolumn{6}{l}{\bf \object{NGC\,6087}}\\
\hline
007 & $16:19:38.4600 $ & $-57:54:22.571 $ & $1.12$ & 2013-04-22 & 1\\
009 & $16:19:36.6658 $ & $-57:56:05.457 $ & $1.11$ & 2013-04-22 & 1\\
010 & $16:18:49.0134 $ & $-57:55:51.524 $ & $1.17$, $1.11$ & 2013-04-22, 2013-04-24 & 2\\
011 & $16:18:37.8903 $ & $-57:56:20.299 $ & $1.16$ & 2013-04-23 & 1\\
014 & $16:18:22.8167 $ & $-57:51:28.668 $ & $1.13$ & 2013-04-23 & 1\\
156 & $16:19:18.329  $ & $-57:53:36.54  $ & $1.11$ & 2013-04-23 & 1\\

\hline
\multicolumn{6}{l}{\bf \object{NGC\,6250}}\\
\hline
01 & $16:58:07.92884$ & $-45:58:56.4699$ & $1.05$ & 2013-06-22 & 1\\
02 & $16:57:42.84338$ & $-45:57:41.2602$ & $1.04$ & 2013-06-22 & 1\\
03 & $16:57:55.891  $ & $-45:56:09.27  $ & $1.03$ & 2013-06-22 & 1\\
17 & $16:57:41.2747 $ & $-45:58:49.914 $ & $1.07$ & 2013-06-24 & 1\\
32 & $16:57:04.45812$ & $-46:08:37.4944$ & $1.04$ & 2013-06-22 & 1\\
33 & $16:56:15.5437 $ & $-46:02:57.727 $ & $1.05$ & 2013-06-22 & 1\\
35 & $16:58:56.80575$ & $-46:07:44.3797$ & $1.07$ & 2013-06-25 & 1\\

\hline
\multicolumn{6}{l}{\bf \object{NGC\,6383}}\\
\hline
001 & $17:34:42.49348$ & $-32:34:53.9926$ & $1.00$ & 2013-04-23 & 1\\
047 & $17:35:10.244  $ & $-32:29:04.08  $ & $1.01$ & 2013-04-23 & 1\\
076 & $17:34:02.227  $ & $-32:40:39.68  $ & $1.00$ & 2013-04-24 & 1\\
100 & $17:34:24.51266$ & $-32:30:15.9812$ & $1.02$, $1.07$ & 2013-04-24, 2013-06-24 & 2\\

\hline
\multicolumn{6}{l}{\bf \object{NGC\,6530}}\\
\hline
009 & $18:03:56.866  $ & $-24:18:45.22  $ & $1.21$ & 2013-06-24 & 1\\
032 & $18:04:11.16   $ & $-24:21:45.2   $ & $1.01$ & 2013-06-25 & 1\\
042 & $18:04:15.02620$ & $-24:23:27.7087$ & $1.03$ & 2013-06-25 & 1\\
045 & $18:04:15.2176 $ & $-24:11:00.090 $ & $1.12$ & 2013-06-25 & 1\\
055 & $18:04:20.56   $ & $-24:13:54.9   $ & $1.09$ & 2013-06-24 & 1\\
086 & $18:04:34.20   $ & $-24:22:00.6   $ & $1.11$ & 2013-06-25 & 1\\
100 & $18:04:41.658  $ & $-24:20:56.95  $ & $1.32$ & 2013-06-25 & 1\\
\hline
\end{longtable}
}

\section{Methodology}
\label{Met}

As was previously described in Papers I and II, we derived from low-resolution spectroscopic data the fundamental parameters of open clusters and individual stars using the BCD spectrophotometric method.

For each star of our sample we measured the BCD parameters: $\lambda_{1}$, $D$, and $\Phi_{\rm b}$ (or $\Phi_{\rm bb}$).
The first two parameters describe the average spectral position and the height of the BD, respectively.
The third parameter is the color gradient that measures the slope of the Paschen continuum.
Then using the BCD calibrations (see Paper I) we obtain the stellar fundamental parameters: spectral type, luminosity class, effective temperature ($T_{\rm eff}$), logarithm of surface gravity ($\log g$), absolute visual magnitude ($M_{\rm v}$), absolute bolometric magnitude ($M_{\rm bol}$) and intrinsic color gradient ($\Phi_{\rm b}^{0}$).

To perform all these measurements, one of us (YA) wrote the interactive code MIDE3700\footnote{The authors gently offer the code MIDE3700 to whom wish to use it.}, in {\it Python} language.
Its name corresponds to the Spanish acronym for \textit{Medici\'on Interactiva de la Discontinuidad En $3700$~\AA} (Interactive Discontinuity Measurement at $3700$~\AA).
With this tool the user can work over the stellar spectrum and choose the best fits to Balmer and Paschen continua.
The code then automatically delivers the values of $D$, $\lambda_{1}$ and $\Phi_{\rm b}$ and the corresponding fundamental parameters of a star.

From observed and intrinsic color gradients we estimated the color excess, $E(B-V)= 2.1\,(\Phi_{\rm b} - \Phi_{\rm b}^{0})$ or $E(B-V)= 2.3\,(\Phi_{\rm bb} - \Phi_{\rm bb}^{0})$, of each star and its distance modulus, using the apparent visual magnitude available in the literature.
In addition, we calculated the stellar luminosity (${\cal L}_{\star} / {\cal L}_{\odot}$), stellar mass ($M_{\star} / M_{\odot}$) and age of each star interpolating in the stellar evolutionary models, for a non-rotating case and $Z = 0.014$, provided by \citet{Ekstrom2012}.
For more details see Paper II.

From these data we derived for each open cluster, the true distance modulus, $(m_{\rm v} - M_{\rm v})_{0}$, the color excess $E(B-V)$ and the age, with their respective error bars.
The distance modulus and memberships are calculated using an iterative procedure as explained in Paper II,  based on the statistical distribution of individual distance modulus for the observed full sample of stars in the direction of each cluster. The first step is the calculation of a mean distance modulus ($DM^0$) and its standard deviation ($\sigma^0$). Then, we select a smaller sample that consist of stars with distance moduli at $1 \sigma^0$ of $DM^0$. With this reduced sample we recalculate new values, which are the adopt mean true distance modulus of the cluster $DM_c$= $(m_{\rm v} - M_{\rm v})_{0}$ and its standard deviation $\sigma$. The color excesses $E(B-V)$ are calculated using the same small sample.
  
The membership criterion is then based on the distance modulus of each star and its error bars. When this distance modulus is outside the region defined by $3 \sigma$ size around the mean cluster distance modulus, the star is considered a non-member ($nm$). If the star is inside a region between $2 \sigma$ and  $3 \sigma$ size, the star is considered a probable non-member ($pnm$). If  the star is between $1 \sigma$ and $2 \sigma$ size it is a probable  member ($pm$) and if its distance modulus is less than $ 1\sigma$ size the star is considered as member ($m$).

On the other hand, cluster membership probabilities, $p$, of individual stars can be estimated using the statistical P-method approach. Assuming the null hypothesis is true, if the P-value is less than (or equal to) a given significance level, $\alpha$, then the null hypothesis is rejected instead if the P-value is large it is accepted. As we are dealing with a rather modest number of objects in each cluster, to define a well distribution of distances we adopt the standard statistical level of significance of $5 \%$ to classify a star as a cluster member.

The cluster age was estimated from the HR diagram by setting the isochrones  \citep{Ekstrom2012} to the locations of the stars classified as members and probable members of the cluster. These stars are identified as $m$ and $pm$ in column 5 of  Tables \ref{Dist-6087}, \ref{Dist-6250}, \ref{Dist-6383}, and  \ref{Dist-6530}.

\begin{figure*}[th!]
\centering
\includegraphics[width=16cm,angle=0]{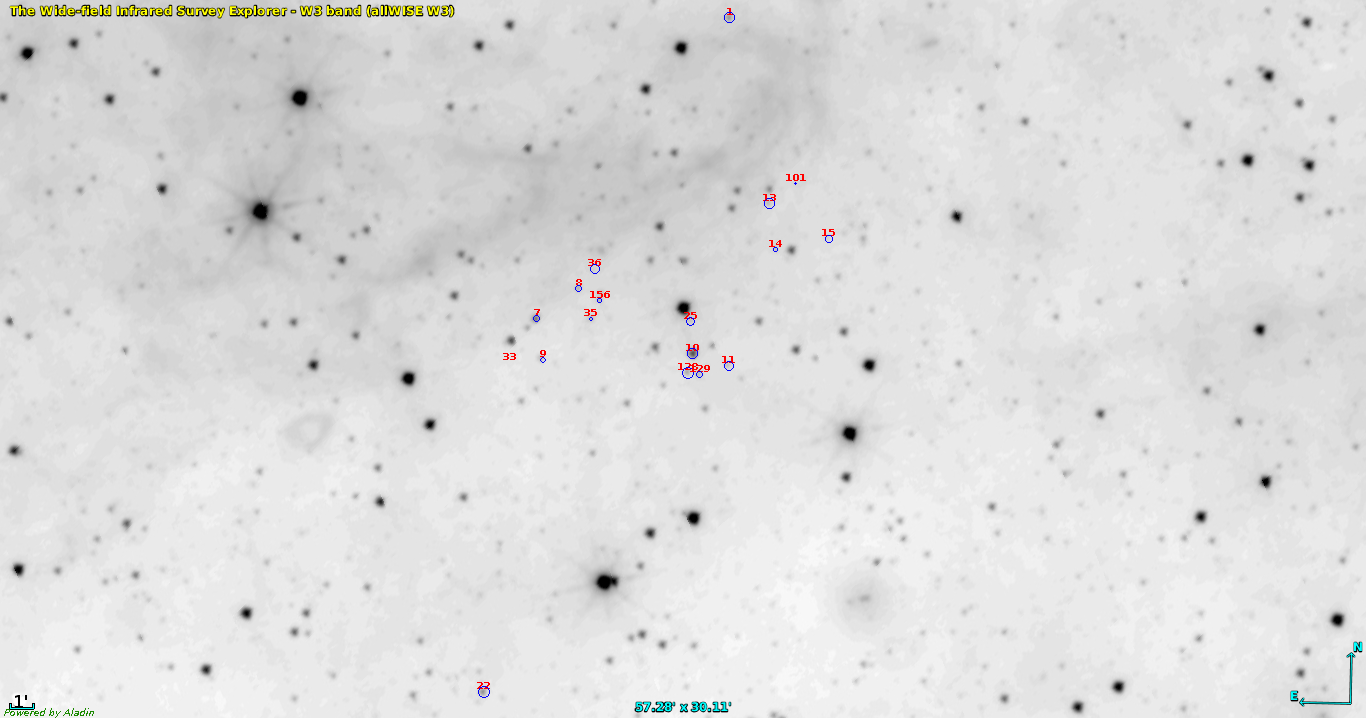}
\caption{\object{NGC\,6087}: WISE band 3 image showing a non uniform dust distribution. The circles indicate the positions of the stars which are proportional to the measured $E(B-V)$.} \label{NGC6087_wise3}
\end{figure*}
\section{Results}
\label{Res}
 
For each cluster analyzed here: \object{NGC\,6087}, \object{NGC\,6250}, \object{NGC\,6383}, and \object{NGC\,6530}, we present a brief summary of previous stu\-dies and the results obtained with the BCD system of each individual star.
This information is listed in two tables.
The first table shows the star identification, the observed BCD parameters $D$, $\lambda_1$, $\Phi_{\rm {b}}$ and the stellar physical properties (spectral type and luminosity class, $T_{\rm{eff}}$,  $\log\,g$,  $M_{\rm{v}}$,  $M_{\rm{bol}}$, and the intrinsic color gradient $\Phi_{\rm b}^0$).
The BCD spectral types have been tested, when it was possible, with those of the standard MK system.
In some occasions we were not able to interpolate or extrapolate the BCD parameters because they were out of the range of the BCD calibrations.
In those cases we used the MK-classification system to derive the spectral types and the corresponding fundamental parameters \citep{Cox2000}.
In addition, the intrinsic color gradients were estimated using the relation between $\Phi_{\rm {b}}$ and $(B-V)$ given by \citet{Moujtahid1998}.

The second table includes for each star: the visual apparent magnitude $m_{\rm{v}}$, the obtained color excess $E(B-V)$, the true distance modulus $(m_{\rm v} - M_{\rm v})_{0}$, the angular distance to the cluster center $AD$, the membership probability $p$, the luminosity $\cal L_{\star}$ relative to the solar one, the stellar mass $M_{\star}$ (in unit of solar mass), and the age.
The coordinates of the cluster and their angular diameters were taken from \citet{Dias2002}.
  
The cluster parameters are presented and discussed in the following subsections.

\subsection{\object{NGC\,6087}}


\object{NGC\,6087} ($\alpha\,=\,16^{\mathrm{h}}\,18^{\mathrm{m}}\,50$ and $\delta\,=\,-57^{\circ}\,56\,\fm\,1;\;$ J$2\,000$) is located near the southern edge of the Normae constellation in a region that appears to be uniformly reddened near the Galactic plane \citep{Sagar1997}.
It is one of the brightest open clusters in this region and it is very rich in Be stars.
Several authors studied this cluster and its parameters were estimated with different techniques.
In Table~\ref{Datos-6087} we can notice that the $E(B-V)$ values range from $0.13$~mag to $0.22$~mag and the cluster distance modulus is between $9.375$~mag and $10.233$~mag ($750$~pc $<\,d\,<\,1000$~pc).
The cluster age presents big uncertainties, ranging from $10-20$~Myr to $146$~Myr.

The spectral classification of the stars in the region of \object{NGC\,6087} was performed only by \citet{Feast1957} who reported the H$\beta$ line in emission in the star No.\,10.
Additional Be stars (Nos.\,9, 14, and 22) were reported by \citet{Mermilliod1982a}, while the star No.\,25 was classified as a CP2 star by \citet{Maitzen1985} using the photometric $\Delta a$-system.

\begin{table*}[th!]
\begin{center}
\caption{\object{NGC\,6087}: Previous and current determinations of color excess, distance, and age.\label{Datos-6087}}
\begin{tabular}{lcccc}
\hline
\hline
Reference & $E(B-V)$ & $(m_{\rm v} - M_{\rm v})_{0}$ & $d$ & $age$\\
         & [mag] & [mag] & [pc] & [Myr]\\
\hline
\citet{Trumpler1930}\tablefootmark{sp}  & $\cdots$                        & $\cdots$        & $870$        & $\cdots$\\
\citet{Fernie1961}\tablefootmark{ph}    & $0.22$                          & $9.4$           & $750$        & $10-20$\\
\citet{Landolt1963}\tablefootmark{ph}   & $0.20$                          & $9.8\pm0.1$     & $\cdots$     & $\cdots$\\
\citet{Breger1966}\tablefootmark{ph}    & $0.20\pm0.01$                   & $9.7$           & $\cdots$     & $\cdots$\\
\citet{Graham1967}\tablefootmark{ph}    & $\cdots$                        & $9.9$           & $950$        & $\cdots$\\
\citet{Lindoff1968}\tablefootmark{sc}   & $\cdots$                        & $\cdots$        & $930$        & $\sim 41$\\
\citet{Becker1971}\tablefootmark{ph}    & $0.13$                          & $9.96$          & $820$        & $\cdots$\\
\citet{Tosi1979}\tablefootmark{*}       & $\cdots$                        & $\cdots$        & $\cdots$     &  $55\pm5$; $55\pm9.6$;$146\pm11$\\
\citet{Schmidt1980}\tablefootmark{ph}   & $0.17\pm0.01$\tablefootmark{**} & $9.60\pm0.09$   & $\cdots$     & $\cdots$\\
\citet{Janes1982}\tablefootmark{ph}     & $0.21$                          & $10.23$         & $\cdots$     & $\sim 40$\\
\citet{Turner1986}\tablefootmark{ph}    & $\cdots$                        & $9.78\pm0.13$   & $902\pm10$   & $\cdots$\\
\citet{Sagar1987}\tablefootmark{ph,sp}  & $0.18\pm0.05$                   & $9.8$           & $\cdots$     & $20$\\
\citet{Meynet1993}\tablefootmark{i}     & $\cdots$                        & $\cdots$        & $\cdots$     & $\sim 71$\\
\citet{Sagar1997}\tablefootmark{ph}     & $0.22$                          & $\cdots$        & $1000\pm100$ & $65$\\
\citet{Robichon1999}\tablefootmark{h}   & $\cdots$                        & $\cdots$        & $769$        & $\cdots$\\
\citet{Rastorguev1999}\tablefootmark{h} & $\cdots$                        & $\cdots$        & $820$        & $\cdots$\\
\citet{Baumgardt2000}\tablefootmark{h}  & $\cdots$                        & $\cdots$        & $819.67$     & $\cdots$\\
\citet{Loktin2001}\tablefootmark{h}     & $\cdots$                        & $9.375$         & $\cdots$     & $\sim 95$\\
\citet{Kharchenko2005}\tablefootmark{h} & $\cdots$                        & $\cdots$        & $901$        & $\sim 85$\\
\citet{Piskunov2007}\tablefootmark{h}   & $0.18$                          & $\cdots$        & $\cdots$     & $\cdots$\\
\citet{An2007}\tablefootmark{ph}        & $0.132\pm0.007$                 & $9.708\pm0.048$ & $\cdots$     & $\cdots$\\
\hline
This work                               & $0.35\pm0.03$                   & $9.00\pm0.19$   & $629\pm54$   & $\sim 55$\\
\hline
\end{tabular}
\tablefoot{
\tablefoottext{h}{Results based on {\sc Hipparcos} data.}
\tablefoottext{i}{Results based on the isochrone fitting method.}
\tablefoottext{ph}{Results based on photometric data.}
\tablefoottext{sc}{Results based on synthetic clusters.}
\tablefoottext{sp}{Results based on spectroscopic data.}
\tablefoottext{*}{The age of the cluster was derived by three different methods: isochrone fitting, position of the turn-off point, and period of 
Cepheid stars ($55$ Myr, $146$ Myr and $55$ Myr, respectively). The author argued that the most accurate method is the isochrone fitting followed by the 
Cepheid periods.}
\tablefoottext{**}{The color excess $E(B-V)$ was calculated in this work, using the $E(b-y)$ value given by \citet{Schmidt1980} and the relation $E(b-y) = 0.74\, E(B-V)$.}
}
\end{center}
\end{table*}

Although some preliminary stellar parameters for this cluster were previously reported by \citet{Aidelman2010}, here we provide an improved set of BCD measurements (see Tables~\ref{BCD-6087} and \ref{Dist-6087}) that were obtained using the MIDE3700 code and the new BCD $M_{\rm v}$-calibration given by \citet{Zorec1991}, instead of that reported by \citet{Zorec1986}.
From these data we obtained the following cluster parameters: $(m_{\rm v} - M_{\rm v})_{0} = 9.00 \pm 0.19$~mag, $d = 629 \pm 54$~pc and $E(B-V) = 0.35 \pm 0.03$~mag.
As we derived a large value of $E(B-V)$ our distance modulus is slightly lower than those previously reported in the literature.
We attribute this large $E(B-V)$ to particularities of the selected sample.
Apart of the fact that many of the stars of the sample present evidence of  circumstellar envelopes, the large $E(B-V)$ excess could be attained to the non uniform dust distribution (see Fig. \ref{NGC6087_wise3}), revealed in the WISE \citep[Wide-field Infrared Survey Explorer,][]{Wright2010} band-3 image, where we identified our sample.

Considering our criterion for selecting star members (see section \S~\ref{Met}), based on $(m_{\rm v} - M_{\rm v})_{0} = 9.00$~mag and $\sigma = 0.5$~mag, we find $12$ star members ($m$, Nos.\,1, 8, 9, 10, 11, 13, 14, 15, 25, 36, 128, 156).
Furthermore, taking into account the $p$ values, we can confirm that the stars Nos.\,8, 9, 13, 14, 25 and 128 are cluster members (see Table \ref{Dist-6087}), and stars Nos.\,33 and 35 are non-member ($nm$).
 
On the other hand, according to the calculated angular distances and considering that the mean cluster diameter is $14\arcmin$ \citep{Dias2002}, the most distant stars are Nos. 1 and 22.

In the literature we found that stars reported as $pnm$ and/or $nm$ are Nos.\,7, 10, and 129 \citep{Landolt1964, Schmidt1980, Turner1986}.
However, as both Nos.\,7 and 10 are Be stars (see section \S~\ref{Dis}), these authors  could have been underestimated the  absolute magnitudes due to the presence of the circumstellar envelope.
From proper motion measurements \citet{Dias2014} classified stars Nos. 7, 8, 9, 10, 11, 13, 14, 15, 25, 128, 129, 156, as members and assigned to the star No. 35 a probability of $11~\%$.

\begin{figure}[]
\centering
\includegraphics[width=9cm,angle=0]{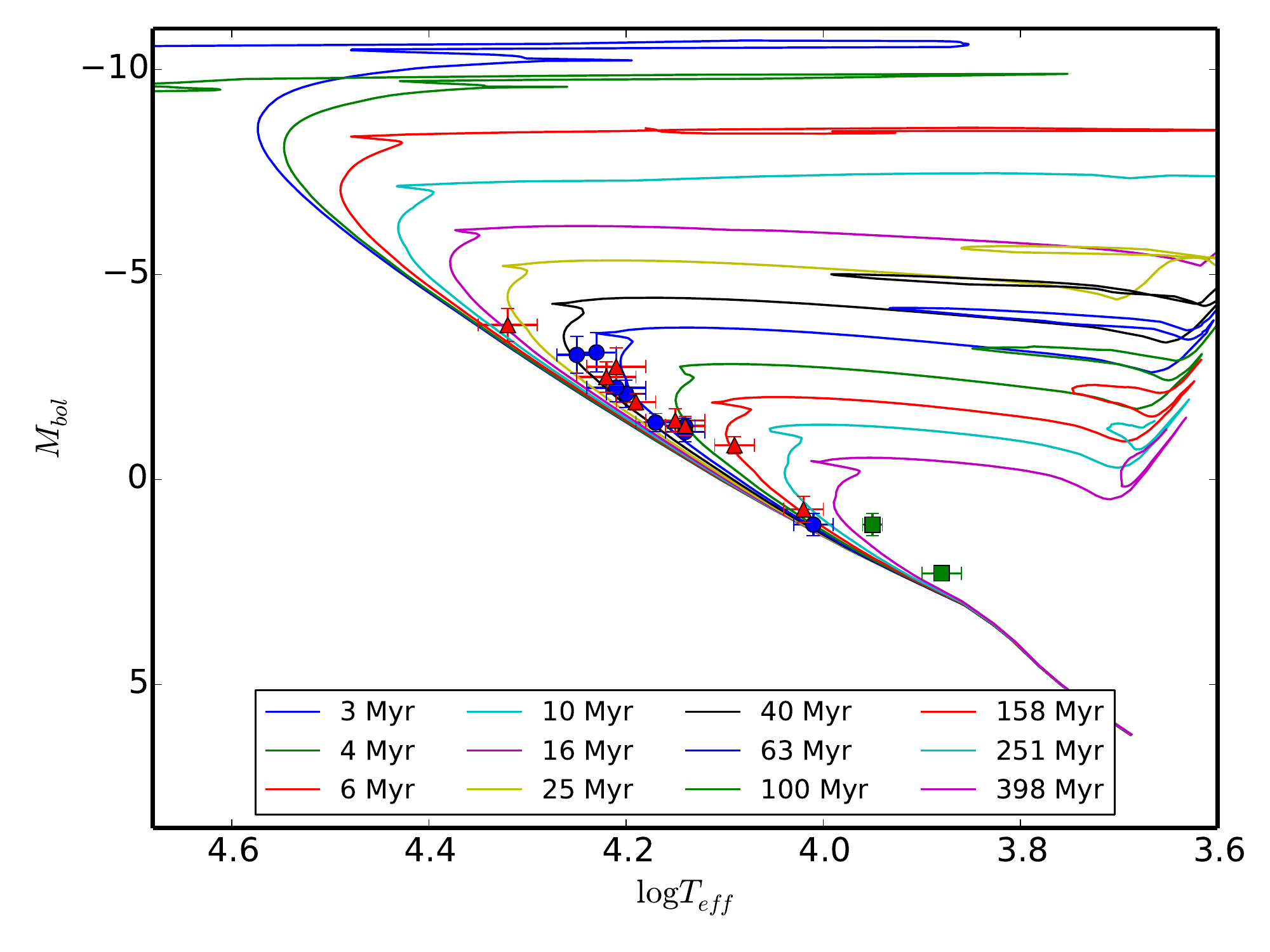}
\caption{\object{NGC\,6087}: HR diagram. The estimated age for this cluster is $\sim 55$~Myr. The isochrone curves are given by \citet{Ekstrom2012}. Members and probable members of the clusters are denoted in $\bullet$ (blue) symbols and probable non-members and non-members in $\blacksquare$ (green). Star members with circumstellar envelope are indicated in $\blacktriangle$ (red).}\label{HR-6087}
\end{figure}

Our studied sample of stars defined well the cluster HR diagram over which we had plotted the isochrones for non-rotating star models given by \citet[][see Fig.~\ref{HR-6087}]{Ekstrom2012}.
The Be star No.\,22, classified as $pm$, is the most luminous object in the HR diagram and is located in the most external region of the cluster (see Fig.~\ref{NGC6087_wise3}).
This star was also classified as a blue straggler.
Therefore, if we do not consider this object in the sample, the inferred cluster age is $\sim 55$~Myr.
This value agrees very well with those obtained by \citet{Tosi1979}.

The BCD spectral types agree with the spectral classification reported by \citet{Feast1957}, considering the sample of stars that are in common.
In addition, we have determined the spectral type of four stars (Nos.\,25, 33, 101, and 128) for the first time.
In our sample we distinguish seven stars showing a SBD: Nos.\,7, 9, 10, 11, 14, 101, and 156 (see Fig. \ref{Bdd_Espectros} and  \ref{Bdd_Espectros2}).

\begin{figure*}[]
\centering
\includegraphics[width=16cm,angle=0]{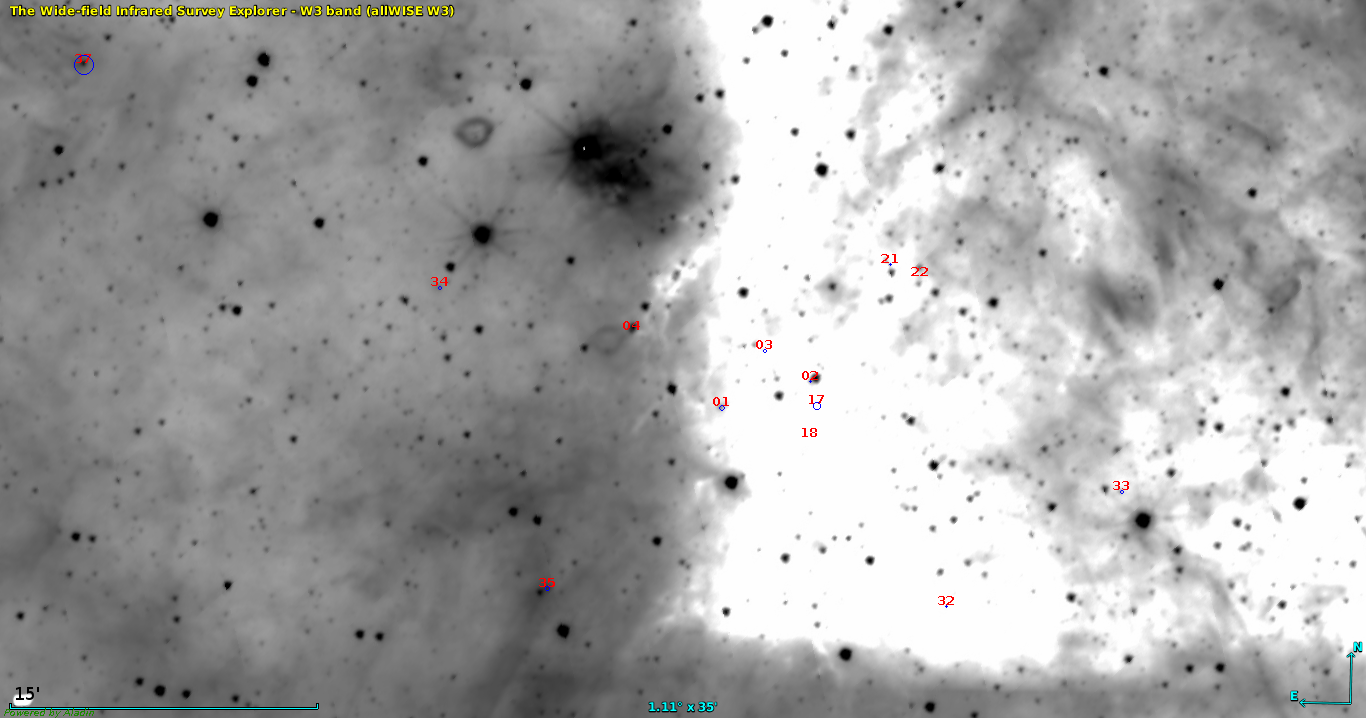}
\caption{Idem Fig. \ref{NGC6087_wise3} but for \object{NGC\,6250}. Stars \object{HD\,329\,211} and \object{CD$-$49\,11096} are not included due to its large angular distance from the center of the cluster.} \label{NGC6250_wise3}
\end{figure*}
 
\subsection{\object{NGC\,6250}}


\object{NGC\,6250} ($\alpha\,=\,16^{\mathrm{h}}\,57^{\mathrm{m}}\,58$ and $\delta\,=\,-45^{\circ}\,56\,\fm\,6;\;$ J$2\,000$) has a peculiar distribution in the sky.
The cluster consists of a very compact central region surrounded by  sparse bright stars.
Several arguments suggest that the cluster is associated with a prominent dust cloud and is affected by differential reddening.
\citet{Feinstein2008} identified several dust components along the light of sight and found that most of the stars do not show indications of intrinsic polarization.

Previous cluster parameters were estimated with different techniques (see Table~\ref{Datos-6250}) and they show a low scatter.
The spectral classification of its individual members was performed by \citet{Herbst1977}  and a detailed chemical analysis of $19$ upper main-sequence stars was done by \citet{Martin2017}.
Star No.\,35 is the only known Be star, star No.\,34 was reported as peculiar \citep[in He and in the K-line,][]{Loden1969} and star No.\,3 as a visual binary \citep{Herbst1977}.

\begin{table*}[]
\begin{center}
\caption{NGC 6250: Previous and current determinations of color excess, distance, and age.\label{Datos-6250}}
\begin{tabular}{lcccc}
\hline
\hline
Reference & $E(B-V)$ & $(m_{\rm v} - M_{\rm v})_{0}$ & $d$ & $age$\\
         & [mag] & [mag] & [pc] & [Myr]\\
\hline
\citet{Moffat1975}\tablefootmark{ph}     & $0.38\pm0.02$            &  ~~$9.92$\tablefootmark{*}       & $950$      & $\cdots$\\
\citet{Herbst1977}\tablefootmark{ph}     & $0.37$                   & $10.05$        & $\cdots$   & ~~~~$14$\\
\citet{Battinelli1991}\tablefootmark{c}  & $\cdots$                 & $\cdots$       & $\cdots$   & $\sim 14$\\
\citet{Kharchenko2005}\tablefootmark{h}  & $\cdots$                 & $\cdots$       & $865$      & $\sim 26$\\
\citet{McSwain2005}\tablefootmark{ph}    & ~~~~$0.35$\tablefootmark{**} & $9.69$         & $\cdots$   & $\sim 26$\\
\citet{Piskunov2008}\tablefootmark{h}    & $0.35$                   & $10.77$        & $\cdots$   & $\cdots$\\
\hline
This work                                &  $0.38\pm0.16$            & $10.55\pm0.33$ & $1\,288\pm210$ & $\sim 6$\\
\hline
\end{tabular}
\tablefoot{
\tablefoottext{c}{Results based on catalogs.}
\tablefoottext{h}{Results based on {\sc Hipparcos} data.}
\tablefoottext{ph}{Results based on photometric data.} 
\tablefoottext{*}{The intrinsic distance modulus was calculated in this work, using $(m_{\rm v} - M_{\rm v})$ and $E(B-V)$ values published by \citet{Moffat1975} with $R_{\rm v} = 3.1$.}
\tablefoottext{**}{The color excess $E(B-V)$ was calculated in this work, using the $E(b-y)$ value given by \citet{McSwain2005} and the relation $E(b-y) = 0.74\, E(B-V)$.}
}
\end{center}
\end{table*}

\begin{figure}[]
\centering
\includegraphics[width=9cm,angle=0]{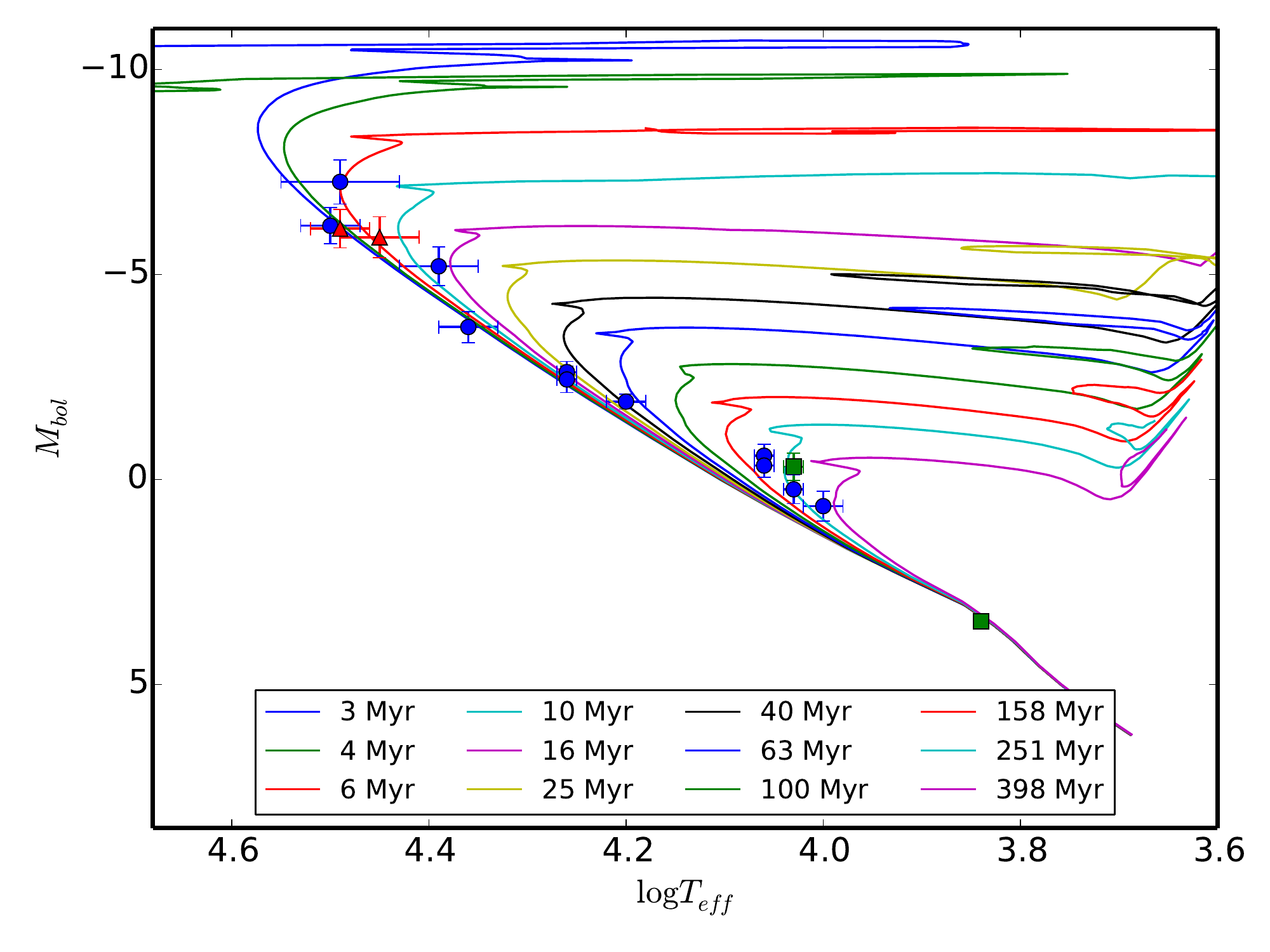}
\caption{\object{NGC\,6250}: HR diagram. The estimated age for this cluster is $\sim 6$~Myr. The isochrone curves are given by \citet{Ekstrom2012}. Members and probable members of the clusters are denoted in $\bullet$ (blue) symbols and probable non-members and non-members in $\blacksquare$ (green). Star members with circumstellar envelopes are indicated in $\blacktriangle$ (red).}\label{HR-6250}
\end{figure}

The results obtained from MIDE3700 code are listed in Tables~\ref{BCD-6250} and \ref{Dist-6250}.
From them we derived the mean cluster parameters: $(m_{\rm v} - M_{\rm v})_{0} = 10.55 \pm 0.33$~mag, $E(B-V)= 0.38 \pm 0.16$~mag and $d= 1\,288\pm 210$~pc.
The obtained color excess is in very good concordance with those determined  by \citet{Moffat1975} and \citet{Herbst1977}. Figure \ref{NGC6250_wise3} is an image obtained with WISE in the band 3 that shows the dust distribution around the open cluster \object{NGC\,6250}. The distance modulus agrees  better with the determination done by \citet{Piskunov2008}.

From our membership criterion we found $8$ members (Nos.\,1, 2, 3, 17, 18, 35, 37 and \object{HD\,329\,211}) based on $(m_{\rm v} - M_{\rm v})_{0} = 10.55$~mag and $\sigma = 0.9$~mag.
Among them, stars Nos\,1, 2 17, 35, 37 and \object{HD\,329\,211} are confirmed members with they $p$ values (see Table~\ref{Dist-6250}).
Stars Nos.\,21, 22, 33, 34, and \object{CD$-$49\,11096} are $pm$, star No.\,32 is a probable non-member ($pnm$). Star No.\,4 is $nm$ (it has a low $p$ value) but  it was reported as $pnm$ by \citet{Moffat1975}.

Our membership criterion agrees with the memberships derived from  polarimetric measurements \citep{Feinstein2008} and from proper motion determinations \citep{Dias2014}, in relation with those stars observed in common (Nos. 1, 2, 3, 17, 18).

An inspection of the angular distances of the stars respect to the cluster center reveals that half of them are located out of the mean angular diameter assumed for the cluster \citep[of $10\arcmin$,][]{Dias2002}.
Since the stars Nos.\,33, 34, and 35 are $m$ or $pm$ according to our membership criterion, one might think that the angular diameter of the cluster is larger than $10\arcmin$.
Extreme cases are \object{CD$-$49\,11096} and \object{HD\,329\,211}. We have not found any information regarding the memberships of these stars in the literature.

On the other hand, according to our HR diagram the cluster age is $\sim 6$~Myr (see Fig.~\ref{HR-6250}) which is lower than the values reported in the literature (see Table~\ref{Datos-6250}).
To justify older ages with our data (by factors 2 or 4) we had to admit that we are overestimating effective temperatures by nearly $60\%$ that correspond to underestimations of the Balmer discontinuity of the order of 0.1 dex which is impossible to admit.
The age discrepancy we find is quite significant and could be due to the early star sample considered by the different authors to identify the isochrone that determines the cluster age. In our work, to derive the age of the cluster  we used stars of spectral type B0, while \citet{Herbst1977} and \citet{Battinelli1991} used stars  of spectral type B1, instead \citet{Kharchenko2005} and \citet{McSwain2005} took a sample of stars of spectral type B2.
These spectral types imply ($T_{\rm eff}, \log L/L_{\odot}$) parameters that range from ($4.48$, $4.63$) to ($4.47$, $4.89$) for stars B0V to B0III, ($4.43$, $4.24$) to ($4.41$, $4.53$) for stars B1V to B1III and ($4.35$, $3.73$) to ($4.32$, $4.1$) for stars B2V to B2III.
These ($T_{\rm eff}, \log L/L_{\odot}$)-intervals identify isochrones in Fig.~\ref{HR-6250} near $6$ Myr, $16$ Myr and $25$ Myr, respectively, which correspond to the estimated ages by the different authors marked in Table~\ref{Datos-6250}.

The BCD spectral classification agrees with that done by \citet{Herbst1977} for eight of the twelve stars we have in common.
Furthermore, we have derived for the first time a spectral classification for \object{CD$-$49\,11096}.
Only the star No.\,1 presents a SBD in its spectrum, which is in emission (see Fig. \ref{Bdd_Espectros2}).
This star has a high effective temperature ($T_{\rm eff}\simeq27\,400$~K).
From simple arguments based on the opacity of circumstellar envelopes of Be stars it is not possible to have conspicuous continuum emissions in the Balmer continuum near $\lambda3600$~\AA\, without some emission in the first members of the Balmer series. The line emission in such classical Be stars is rarely very high. Unfortunately, the BCD spectrophotometric observations of this star did not extend up to the H$\beta$ line. On the other hand, the spectroscopic observation done with REOSC shows H absorption line profiles (see section \S \ref{CE}), although low and high resolution observations are not simultaneous. Therefore, the star was classified as Bdd.

Finally, we confirm that No.\,34 is peculiar and we classified it as a He-strong star. The star shows very intense lines of  \ion{He}{i}. We measured the following equivalent widths: $EW_{4471} = 1.9 \AA$, $EW_{4387} = 1.3 \AA$, $EW_{4026} = 2.1 \AA$, which are about twice greater than the ones expected at the spectral type B0V \citep[see][]{Didelon1982}.

\subsection{\object{NGC\,6383}}


\object{NGC\,6383} ($\alpha\,=\,17^{\mathrm{h}}\,34^{\mathrm{m}}\,48$ and $\delta\,=\,-32^{\circ}\,34\,\fm\,0;\;$ J2\,000) is a rather compact open cluster which could be part of the \object{Sgr\,OB1} association together with \object{NGC\,6530} and \object{NGC\,6531} \citep[cf.][]{Rauw2008}.
The brightest star \object{HD\,159\,176} is a well studied X-ray double-line spectroscopic binary responsible for the ionization of the \ion{H}{ii} region \object{RCW\,132}.

This cluster was widely studied with different techniques (see Table~\ref{Datos-6383}).
In contrast to other studied clusters, the spectral classification of the star members was done by several authors \citep{The1966, Antalova1972, LloydEvans1978, The1985, vandenAncker2000}.
This cluster has many Be and peculiar stars: Nos.\,1 and 100 were reported as double-line spectroscopic binaries \citep[][respectively]{Trumpler1930, Sahade1963} while Nos.\,14 and 83 were reported as possible spectroscopic binaries \citep{LloydEvans1978}.
No.\,3 was reported as an Ap spectroscopic variable \citep{The1985, Landstreet2008}.
Stars Nos.\,1 and 6 were classified as blue stragglers and, in particular,  No.\,1 was also classified as a emission line star.
The star No.\,76  was reported to have H$\alpha$ in emission.

\begin{figure*}[]
\centering
\includegraphics[width=16cm,angle=0]{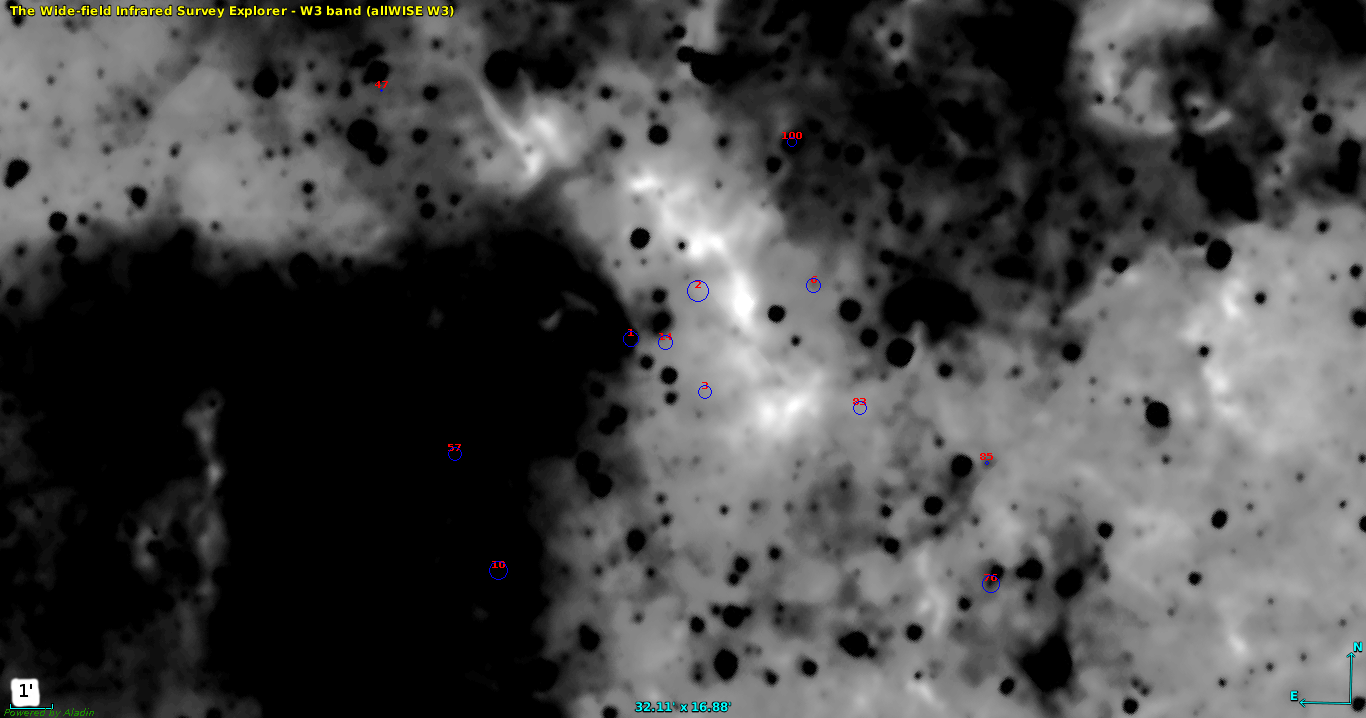}
\caption{Idem Fig. \ref{NGC6087_wise3} but for \object{NGC\,6383}.} \label{NGC6383_wise3}
\end{figure*}

Tables~\ref{BCD-6383} and \ref{Dist-6383} list the derived stellar parameters,  and from them we infer the following cluster parameters: $(m_{\rm v} - M_{\rm v})_{0} = 9.61 \pm 0.38$~mag, $E(B-V)= 0.51 \pm 0.03$~mag and $d= 834 \pm 158$~pc.
The derived color excess is larger than previous published values, therefore our distance estimate is lower.
We attribute this high value to the special characteristics of the selected sample (most of the stars have circumstellar envelopes and are located in a dense dusty region, see Fig. \ref{NGC6383_wise3}.).

Considering the values $(m_{\rm v} - M_{\rm v})_{0} = 9.61$~mag and $\sigma = 1.13$~mag we classified $8$ star members (Nos.\,1, 2, 10, 14, 57, 76, 83, and 100), 2 as $pm$ (Nos.\,3 and 6) and 2 as $pnm$ (Nos.\,47 and 85, both are the coolest stars of this sample).
Stars Nos.\,1, 2, 14, 76, and 83 are also members according to our p-method criterion.
Star No.\,85 was reported as $nm$ by \citet{LloydEvans1978}. The stars classified as $m$ and $pm$ also present $AD$ lower than the cluster angular diameter \citep[of $20\arcmin$,][]{Dias2002}. Moreover, based on proper motion measurements, stars Nos. 1, 2, 3, 6, 10, 14, 47, 57, 83, 100 were considered members by \citet{Dias2014} while star No. 85 has a $1\%$ of probability. Then we have 1 star in discrepancy with \citet[][No. 47]{Dias2014}.
  
\begin{figure}[]
\centering
\includegraphics[width=9cm,angle=0]{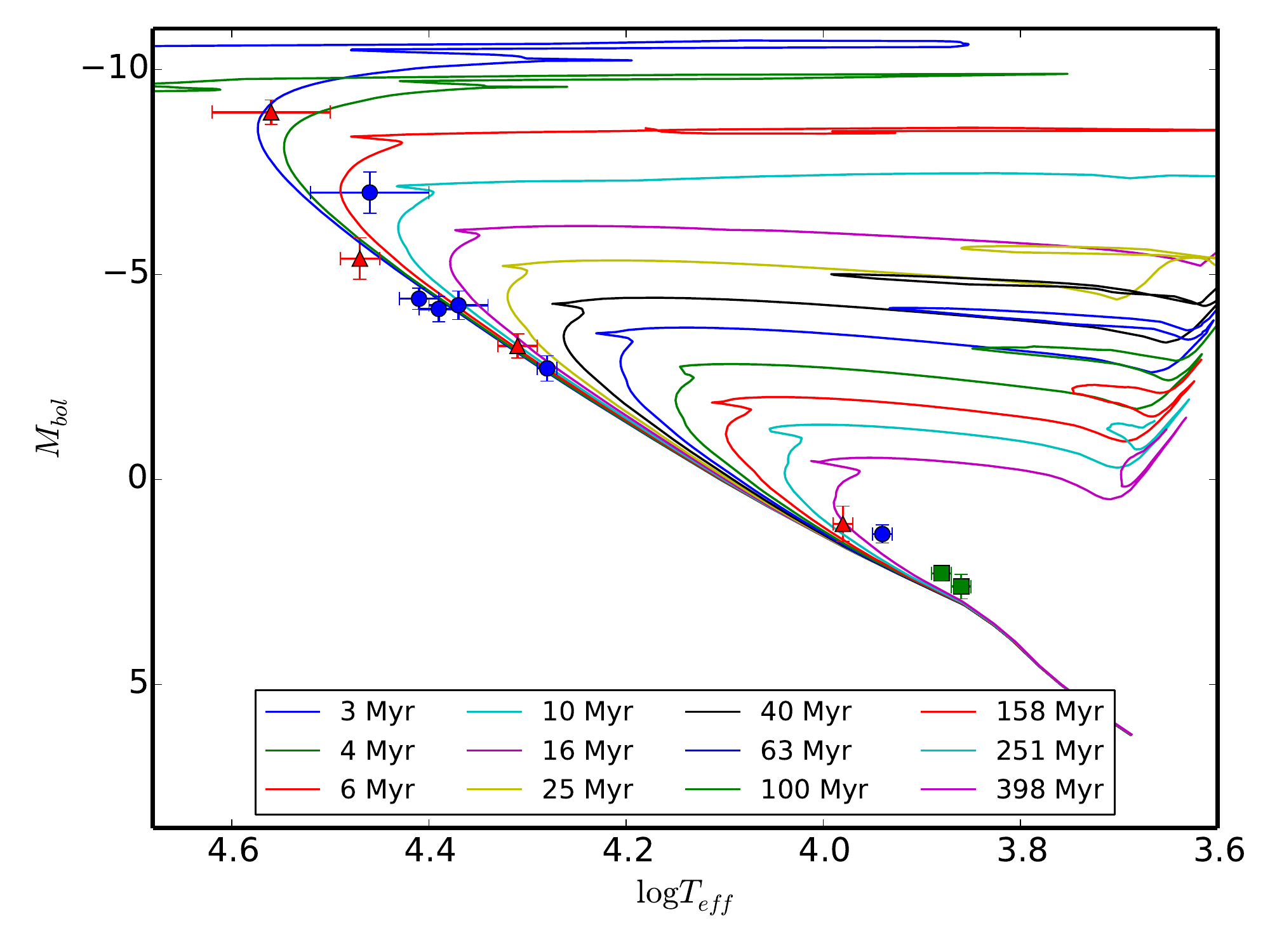}
\caption{\object{NGC\,6383}: HR diagram. The estimated age for this cluster is between $3$ and $10$~Myr. The isochrone curves are given by \citet{Ekstrom2012}. Members and probable members of the clusters are denoted in $\bullet$ (blue) symbols and probable non-members and non-members in $\blacksquare$ (green). Star members with circumstellar envelopes are indicated in $\blacktriangle$ (red).}\label{HR-6383}
\end{figure}

The cluster HR diagram is very well defined and the derived cluster age, $\sim 3$~Myr, agrees with the most of the values found in the literature \citep{Fitzgerald1978, Battinelli1991, Kharchenko2005, Paunzen2007}.
However, if we consider that  the star No.\,6 has started to evolve (see Fig.~\ref{HR-6383}, second more bright star of the sample) and accept that the X-Ray Be binary star No.\,1 (HD 159\,176) is a blue straggler, then the age of the cluster is between $6$~Myr and $10$~Myr.

Our spectral classification agrees with that done in previous works, in particular with \citet{LloydEvans1978}.
We observed a SBD in absorption in the spectrum of the star No.\,57 (see Fig.~\ref{Bdd_Espectros2}).

\begin{table*}[]
\begin{center}
\caption{NGC 6383: Previous and current determinations of color excess, distance, and age.\label{Datos-6383}}
\begin{tabular}{lcccc}
\hline
\hline
Reference & $E(B-V)$ & $(m_{\rm v} - M_{\rm v})_{0}$ & $d$ & $age$\\
         & [mag] & [mag] & [pc] & [Myr]\\
\hline
\citet{Trumpler1930}\tablefootmark{sp}      & $\cdots$                 & $\cdots$       & $2130$       & $\cdots$\\
\citet{Zug1937}\tablefootmark{ph}           & $\cdots$                 & $\cdots$       & $2130$       & $\cdots$\\
\citet{Sanford1949}\tablefootmark{sp}       & $\cdots$                 & $\cdots$       & $760$        & $\cdots$\\
\citet{Eggen1961}\tablefootmark{ph}         & $\cdots$                 & $\cdots$       & $1259$       & $\cdots$\\
\citet{Graham1967}\tablefootmark{ph}        & $\cdots$                 & $10.68\pm0.54$ & $1380$       & $\cdots$\\
\citet{Lindoff1968}\tablefootmark{ph}       & $\cdots$                 & $10.5$         & $1250$       & $\sim 20$\\
\citet{Becker1971}\tablefootmark{c}         & $0.26$                   & $10.92$        & $1065$       & $\cdots$\\
\citet{Fitzgerald1978}\tablefootmark{ph,sp} & $0.33\pm0.02$            & $10.85$        & $1500\pm200$ & $1.6\pm0.4$\\
\citet{LloydEvans1978}\tablefootmark{ph}    & $0.35$                   & $10.6$         & $1350$       & $\cdots$\\
\citet{The1985}\tablefootmark{ph}           & $0.30\pm0.01$            & $\cdots$       & $1400\pm150$ & $\cdots$\\
\citet{Pandey1989}\tablefootmark{ph}        & $0.35$                   & $11.65$        & $\cdots$     & $\sim 4$\\
\citet{Battinelli1991}\tablefootmark{sp}    & $\cdots$                 & $\cdots$       & $1380$       & $\sim 4$\\
\citet{Feinstein1994}\tablefootmark{ph}     & $0.33$                   & $\cdots$       & $1400$       & $\cdots$\\
\citet{Rastorguev1999}\tablefootmark{h}     & $\cdots$                 & $\cdots$       & $1180$       & $\cdots$\\
\citet{Kharchenko2005}\tablefootmark{h}     & $\cdots$                 & $\cdots$       & $985$        & $\sim 5$\\
\citet{Paunzen2007}\tablefootmark{ph}       & $0.28$\tablefootmark{**} & $\cdots$       & $1700$       & $4$\\
\citet{Piskunov2008}\tablefootmark{h}       & $0.30$                   & $10.897$       & $\cdots$     & $\cdots$\\
\citet{Rauw2008}\tablefootmark{ph}          & $0.32\pm0.02$            & $\cdots$       & $1300\pm100$ & $\cdots$\\
\hline
This work                                   & $0.51\pm0.03$            & $9.61\pm0.38$  & $834\pm158$  & $\sim3-10$\\
\hline
\end{tabular}
\tablefoot{
\tablefoottext{c}{Results based on catalog data.}
\tablefoottext{h}{Results based on {\sc Hipparcos} data.}
\tablefoottext{ph}{Results based on photometric data.}
\tablefoottext{sp}{Results based on spectroscopic data.}
\tablefoottext{**}{The color excess $E(B-V)$ was calculated in this work, using the published $E(b-y)$ value and the relation $E(b-y) = 0.74\, E(B-V)$.}
}
\end{center}
\end{table*}

\subsection{\object{NGC\,6530}}


\begin{figure*}[]
\centering
\includegraphics[width=16cm,angle=0]{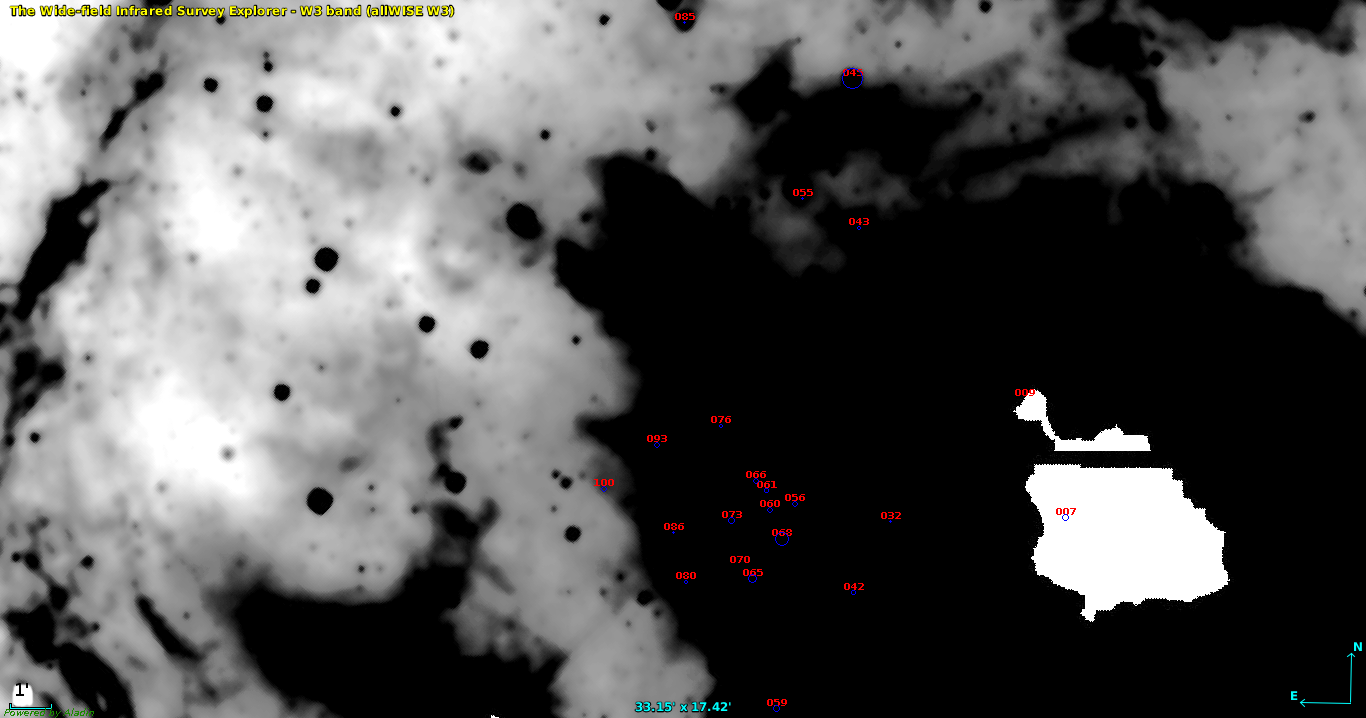}
\caption{Idem Fig. \ref{NGC6087_wise3} but for \object{NGC\,6530}. The image is saturated (white patch).} \label{NGC6530_wise3}
\end{figure*}

\object{NGC\,6530} ($\alpha\,=\,18^{\mathrm{h}}\,04^{\mathrm{m}}\,31$ and $\delta\,=\,-24^{\circ}\,21\,\fm\,5;\;$ J$2\,000$) is a very young cluster located in the \ion{H}{ii} region \object{M8}, also known as the Lagoon Nebulae, near the Sagittarius-Carina spiral Galactic arm.
Several authors studied this cluster and its parameters were estimated with different techniques (see Table~\ref{Datos-6530}).
The color excess $E(B-V)$ and the cluster distance have large uncertainties, with values between $0.30$~mag and $0.541$~mag, and between $730$~pc and $1900$~pc, respectively.
However the cluster age is better defined, being it less than $7$~Myr.

The spectral classification of the individual star members was performed by \citet{Walker1957, Hiltner1965, Chini1981, Torres1987, Boggs1989}, and \citet{Kumar2004}.
The cluster has many variable stars and many emission line stars.
Stars Nos.\,7, 9, 56, 65, 86, and 100 were reported as spectroscopic binaries, and Nos.\,43 and 73 as triple systems.
In addition, stars Nos.\,42, 45, 56, 60, 65, and 100 show H lines in emission, however, it is still debating if these emissions originate in the \ion{H}{ii} region or in a circumstellar envelope.
On the other hand, \citet{Kumar2004} reported the star No.\,65 as a Herbig Ae/Be I or III, and \citet{Niedzielski1988} classified the stars No.\,59 as an unknown peculiar type and No.\,66 as a He-weak.
Finally, stars Nos.\,73 and 93 were reported as blue stragglers.


\begin{figure}[]
\centering
\includegraphics[width=9cm,angle=0]{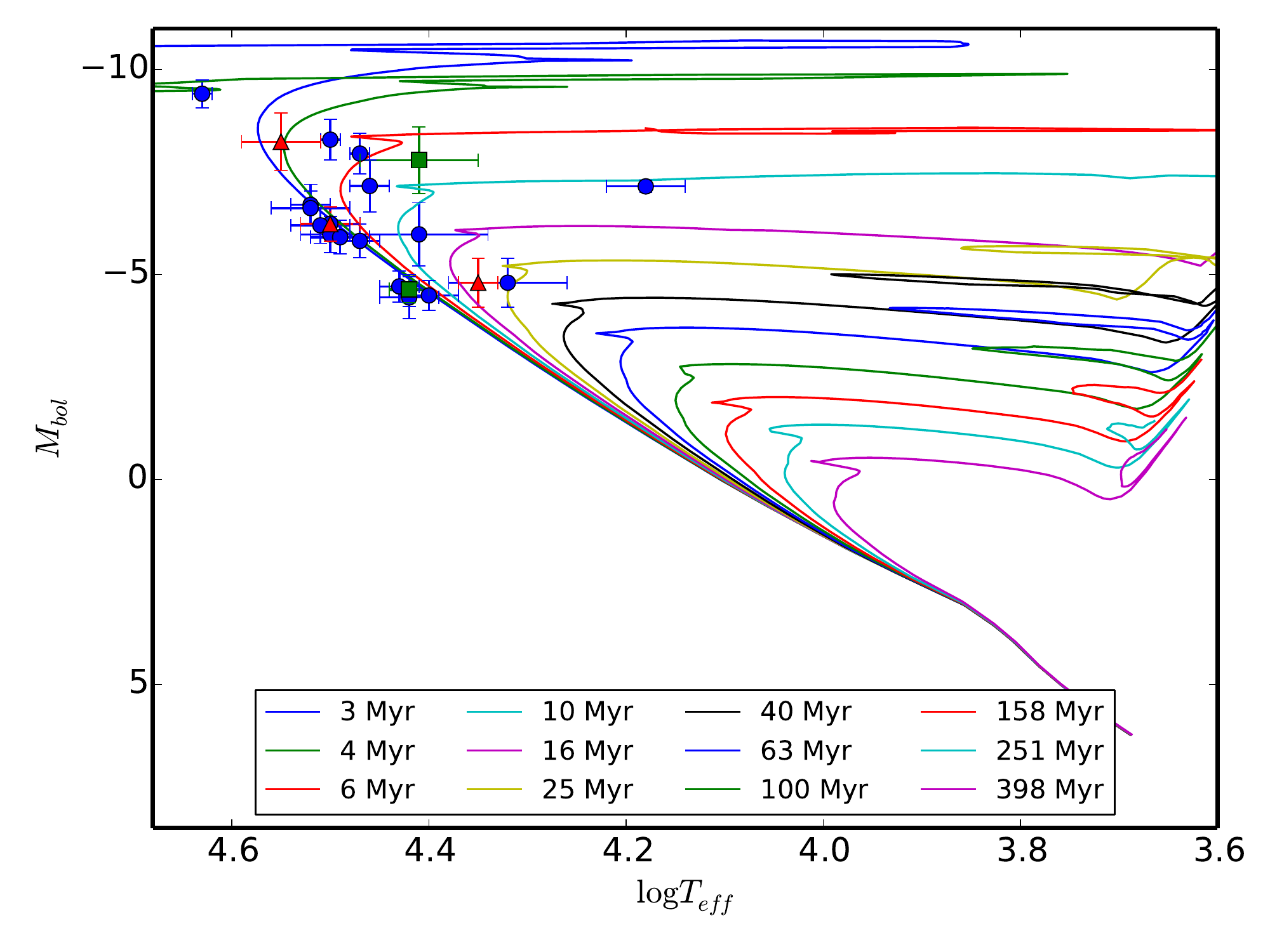}
\caption{\object{NGC\,6530}: HR diagram. The estimated age for this cluster is $4 - 6$~Myr. The isochrone curves are given by \citet{Ekstrom2012}. Members and probable members of the clusters are denoted in $\bullet$ (blue) symbols and probable non-members and non-members in $\blacksquare$ (green). Star members with circumstellar envelopes are indicated in $\blacktriangle$ (red).}\label{HR-6530}
\end{figure}

With the BCD system we derive the stellar fundamental parameters (Tables~\ref{BCD-6530} and \ref{Dist-6530}) and the mean cluster parameters: $(m_{\rm v} - M_{\rm v})_{0} = 11.76 \pm 0.20$~mag, $E(B-V)= 0.26 \pm 0.05$~mag and $d= 2\,245 \pm 215$~pc.
The distance modulus agree very well with the determination done by \citet{Lindoff1968}, but our derived color excess is systematically lower than previous published values.
Figure \ref{NGC6530_wise3} illustrates the WISE band 3 image in the surroundings of \object{NGC\,6530} together with our color excess determinations. The low E(B-V) values we derived indicate that the dusty region is located behind the cluster.

Taking into account our membership criterion (with $(m_{\rm v} - M_{\rm v})_{0} = 11.76$~mag and $\sigma = 0.78$~mag) we found $15$ members (Nos.\,32, 42, 43, 56, 60, 61, 65, 66, 68, 70, 73, 76, 86, 93, and 100), 6 $pm$ (Nos.\,7, 9, 45, 55, 59, and 85) and 2 $nm$ (No.\,80 and \object{LSS\,4627}).
We can confirm as cluster members the stars Nos.\,43, 65, 70, and 86 because they have large $p$ values.
On the other hand, \object{LSS\,4627} is the star farthest from the center of the cluster and we considered it as a $nm$.
The rest of stars have angular distances lower than the cluster angular diameter of $14\arcmin$ \citep{Dias2002}.  We have 17 stars in common with \citet[][Nos.\,32, 42, 43, 55, 56, 59, 60, 61, 65, 66, 68, 70, 73, 76, 80, 86, 93.]{Dias2014} who considered all of them as members. Therefore, only the star No.\,80 is in discrepancy.

Even though, the stars show some scatters in the HR diagram (probably due to the binary nature of many them), the cluster age seems to be about $4-6$~Myr (see Fig.~\ref{HR-6530}).
This value agrees with most of the previous determination.


Our spectral classification confirms that done by \citet{Hiltner1965, Chini1981, Torres1987} and \citet{Boggs1989}.
The stars No.\,68 (B0V) and \object{LSS\,4627} (B1V) have been classified for the first time.
The former exhibits a narrow emission in the H$\beta$ core that could be attained to a nebular contribution.
We find that star No.\,9 has an anomalous color excess for its spectral type.
We observed a SBD in the spectra of the  Nos.\,45 and 65. 
This confirms that the star No.\,65 is a Be star which is shown in Fig. \ref{Bdd_Espectros2}. Instead, we found that star No.\,45 is a supergiant.
According to the location of the stars Nos.\,73 and 93 in the HR diagram, they cannot be classified as blue stragglers.

\begin{table*}[]
\begin{center}
\caption{NGC 6530: Previous and current determinations of color excess, distance, and age.\label{Datos-6530}}
\begin{tabular}{lcccc}
\hline
\hline
Reference & $E(B-V)$ & $(m_{\rm v} - M_{\rm v})_{0}$ & $d$ & $age$\\
         & [mag] & [mag] & [pc] & [Myr]\\
\hline
\citet{Trumpler1930}\tablefootmark{ph}        & $\cdots$                 & $\cdots$        & $1\,090$        & $\cdots$\\
\citet{Sanford1949}\tablefootmark{sp}         & $\cdots$                 & $\cdots$        & $730$         & $\cdots$\\
\citet{Johnson1961}\tablefootmark{ph}         & $0.32$                   & $10.7\pm0.3$    & $1\,400$        & $\cdots$\\
\citet{Penston1964}\tablefootmark{p}          & $\cdots$                 & $\cdots$        & $\cdots$      & $\sim 1$ \\
\citet{Lindoff1968}\tablefootmark{sc}         & $\cdots$                 & $11.91$         & $1\,560$        & $< 7$\\
\citet{VanAltena1972}\tablefootmark{ph}       & $0.35$                   & $11.25$         & $1\,780$        & $2$\\
\citet{Harris1976}\tablefootmark{sp}          & $\cdots$                 & $\cdots$        & $\cdots$      & $6.36\pm0.34$\\
\citet{Kilambi1977}\tablefootmark{ph}         & $0.34$                   & $10.7$          & $\cdots$      & $\cdots$\\
\citet{Sagar1978}\tablefootmark{ph}           & $0.35$                   & $11.3$          & $\cdots$      & $2$\\
\citet{Chini1981}\tablefootmark{ph,sp}        & $0.36\pm0.09$            & $11.4$          & $\cdots$      & $\cdots$\\
\citet{BoehmVitense1984}\tablefootmark{sp}    & $\cdots$                 & $\cdots$        & $\cdots$      & $5\pm2$\\
\citet{Sagar1986}\tablefootmark{ph}           & $\cdots$                 & $\cdots$        & $\cdots$      & $\sim 2$\\
\citet{Battinelli1991}\tablefootmark{p}       & $0.35$                   & $\cdots$        & $1\,600$        & $\sim 2$\\
\citet{Strobel1992}\tablefootmark{p}          & $0.4$                    & $12.7$          & $\cdots$      & $\sim 3$\\
\citet{Feinstein1994}\tablefootmark{ph}       & $0.36$                   & $11.4$          & $1\,900$        & $\cdots$\\
\citet{VandenAncker1997}\tablefootmark{ph,sp} & $0.30$                   & $\cdots$        & $1\,800\pm200$  & $\cdots$\\
\citet{Rastorguev1999}\tablefootmark{h}       & $\cdots$                 & $\cdots$        & $1\,270$        & $\cdots$\\
\citet{Robichon1999}\tablefootmark{h}         & $\cdots$                 & $\cdots$        & $\sim 763.36$ & $\cdots$\\
\citet{Sung2000}\tablefootmark{ph}            & $0.35$                   & $11.25\pm0.1$   & $\cdots$      & $1.5$\\
\citet{Baumgardt2000}\tablefootmark{h}        & $\cdots$                 & $\cdots$        & $\sim 917.43$ & $\cdots$\\
\citet{Loktin2001}\tablefootmark{h}           & $\cdots$                 & $9.005\pm0.255$ & $\cdots$      & $\sim 7.5$\\
\citet{Prisinzano2005}\tablefootmark{ph}      & $\cdots$                 & $\cdots$        & $1\,250$        & $2.3$\\
\citet{Kharchenko2005}\tablefootmark{h}       & $\cdots$                 & $\cdots$        & $1\,322$        & $\sim 4.7$\\
\citet{McSwain2005}\tablefootmark{ph}         & $0.34$\tablefootmark{**} & $10.62$         & $\cdots$      & $\sim 7.4$\\
\citet{Zhao2006}\tablefootmark{ph}            & $\cdots$                 & $\cdots$        & $1\,330$        & $\sim 7.4$\\
\citet{Mayne2008}\tablefootmark{p}            & $0.32$                   & $10.34$         & $\cdots$      & $2$\\
\citet{Kharchenko2009}\tablefootmark{h}       & $0.30$                   & $\cdots$        & $1\,322$        & $\sim 4.7$\\
\citet{Kharchenko2013}\tablefootmark{h}       & $0.541$                  & $\cdots$        & $1\,365$        & $\sim 4.7$\\
\hline
This work                                     & $0.26\pm0.05$            & $11.76\pm0.20$  & $2\,245\pm215$  & $4-6$\\
\hline
\end{tabular}
\tablefoot{
\tablefoottext{h}{Results based on {\sc Hipparcos} data.}
\tablefoottext{p}{Results based on published data.}
\tablefoottext{ph}{Results based on photometric data.}
\tablefoottext{sc}{Results based on synthetic clusters.}
\tablefoottext{sp}{Results based on spectroscopic data.}
\tablefoottext{**}{The color excess $E(B-V)$ was calculated in this work, using the published $E(b-y)$ value and the relation $E(b-y) = 0.74\, E(B-V)$.}
}
\end{center}
\end{table*}


\section{The population of B, Be and Bdd stars}
\label{Dis}

We used the BCD method to derive physical properties of a large number of stars ($\sim230$) in the direction of the open clusters \object{Collinder\,223}, Hogg 16, NGC 2645, NGC 3114, NGC 3766, NGC 4755, NGC 6025, NGC 6087, NGC 6250, NGC 6383 and NGC 6530 (see Paper I, Paper II and this work).
Thus, we gathered a homogeneous set of fundamental parameters that in addition are not affected by the interstellar/circumstellar reddening.
This way we were able to obtain cluster parameters of $11$ open clusters, construct well-defined HR diagrams and perform a better identification of the cluster star members. At the same time, the study of the BD enables the detection of circumstellar envelopes around B-type stars in an independent way of  the typical criterion based on the presence of emission H lines.
This is possible through the observation of a SBD.

In this section, we discuss the derived BCD stellar fundamental parameters and describe the main characteristics of the population of B stars. Particularly, we present global properties of  stars with circumstellar envelopes.

\subsection{The BCD stellar fundamental parameters}

The discussion of the fundamental parameters is based on a reduced sample of $198$ early-type stars, classified as members ($m$) and probable members ($pm$), of the $11$ open clusters studied. We discarded $pnm$ or $nm$ stars from this analysis.

\begin{figure*}[]
\centering
\subfigure[Comparison of effective temperatures.]{\includegraphics[width=9cm,angle=0]{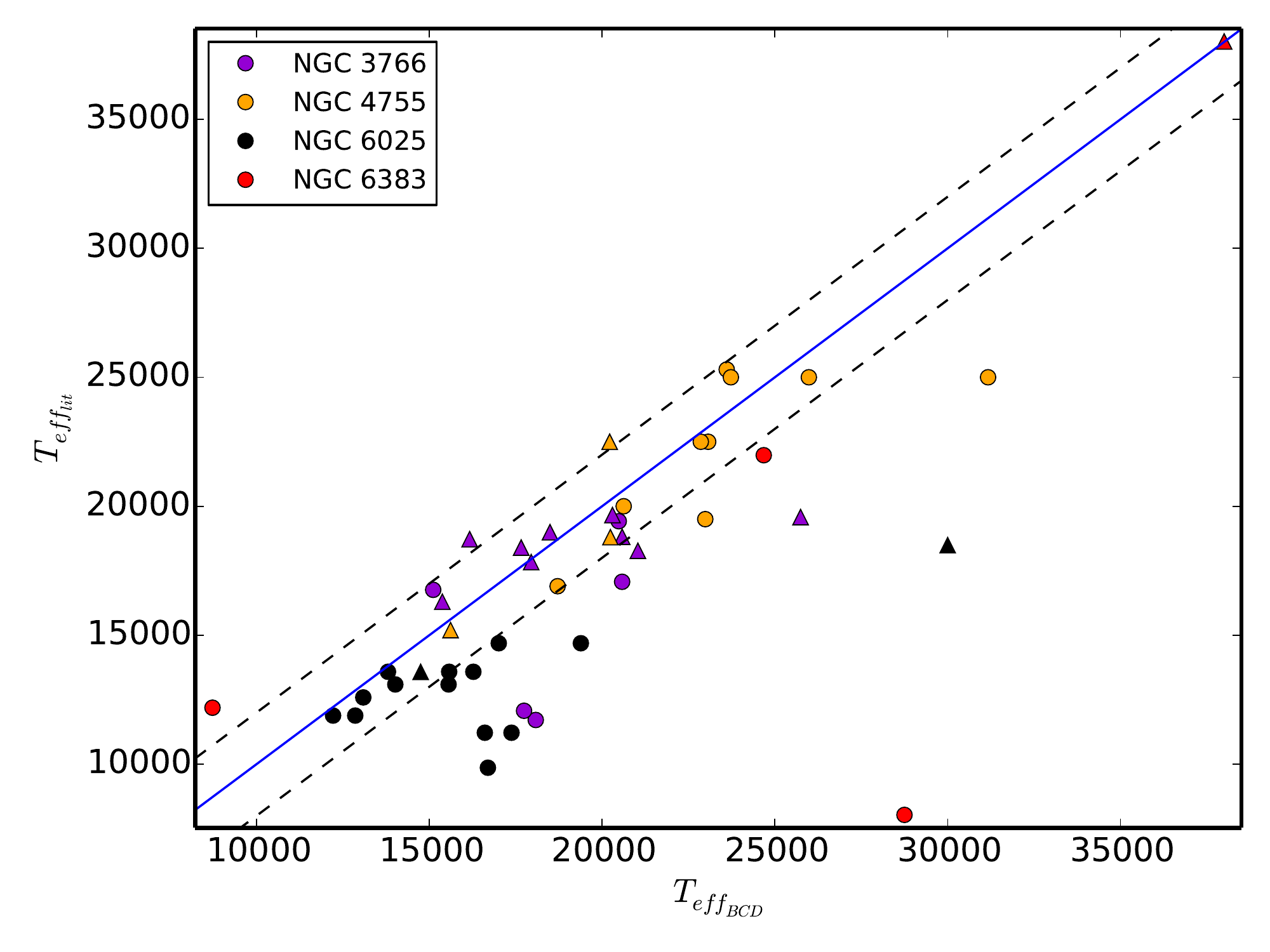}}
\subfigure[Comparison of absolute visual magnitudes.]{\includegraphics[width=9cm,angle=0]{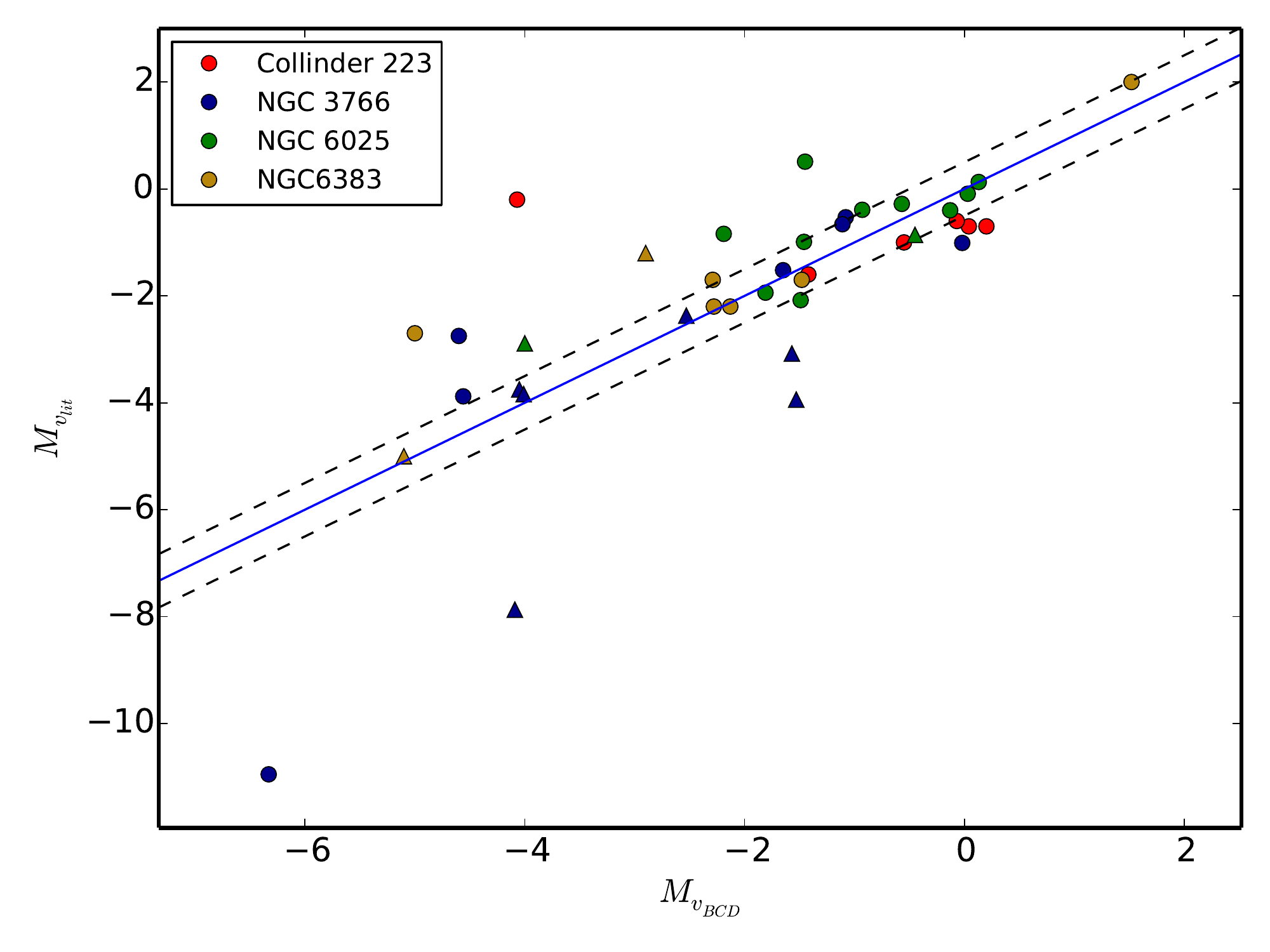}}
\caption{Comparison of BCD $T_{\rm eff}$ and $M_{\rm v}$ values determined for member stars with those determined in previous works. The straight line is the identity function and the dotted lines deviate from it in $2000$~K and $0.5$~mag, respectively. Circle symbols identify normal B-type stars and  triangles  are used for stars with circumstellar envelopes.}\label{Comp_Mv}
\end{figure*}

Firstly, we compare the BCD fundamental parameters of all the stars, $T_{\rm eff}$ and $M_{\rm v}$, with the scarce values reported in the literature.
This comparison is shown in Fig.~\ref{Comp_Mv} where we distinguish normal B-type stars (identified with circles) from those that have circumstellar envelopes (identified with triangles).
The effective temperatures were taken from \citet[][\object{NGC\,3766}]{McSwain2008, McSwain2009}, \citet[][\object{NGC\,4755}]{Dufton2006}, \citet[][\object{NGC\,6025}]{Piskunov1980}, and \citet[][\object{NGC\,6383}]{vandenAncker2000}.
Whilst the absolute visual magnitudes were gathered from \citet[][\object{Collinder\,223}]{Claria1991}, \citet[][\object{Hogg\,16}]{Fitzgerald1979b}, \citet[][\object{NGC\,2645}]{Fitzgerald1979a}, \citet[][\object{NGC\,3766}]{Shobbrook1985}, \citet[][\object{NGC\,6025}]{Kilambi1975} and \citet[][\object{NGC\,6383}]{LloydEvans1978}\footnote{We only consider works that explicitly published $M_{\rm v}$ values for individual stars.}.

In relation to the BCD $T_{\rm eff}$ values (see Fig.~\ref{Comp_Mv}a), we note that in general they are systematically greater than those derived by other techniques.
It has been shown that the effective temperatures derived with the BCD method have no significant differences with the values derived from multicolor photometric systems, line-blanketed models and the infrared flux method \citep{Cidale2007}.
Therefore, we observe that our $T_{\rm eff}$ determinations show discrepancies of about $\Delta\, T_{\rm eff} \lesssim 2000$~K, with those obtained by \citet[][H$\alpha$ and Str\"{o}mgren photometry]{McSwain2008, McSwain2009} and \citet[][non-LTE atmospheric models]{Dufton2006}.
Instead, significant discrepancies are observed in some of the stars  of \object{NGC\,6025}.
On the other hand, we have three stars in common with \citet{vandenAncker2000} and only the star NGC\,6383\,6 shows a huge discrepancy.
Perhaps, this could be due to the too late spectral classification assigned to this star by these authors.

\begin{figure*}[]
\centering
\subfigure[Comparison of absolute bolometric magnitudes.]{\includegraphics[width=9cm,angle=0]{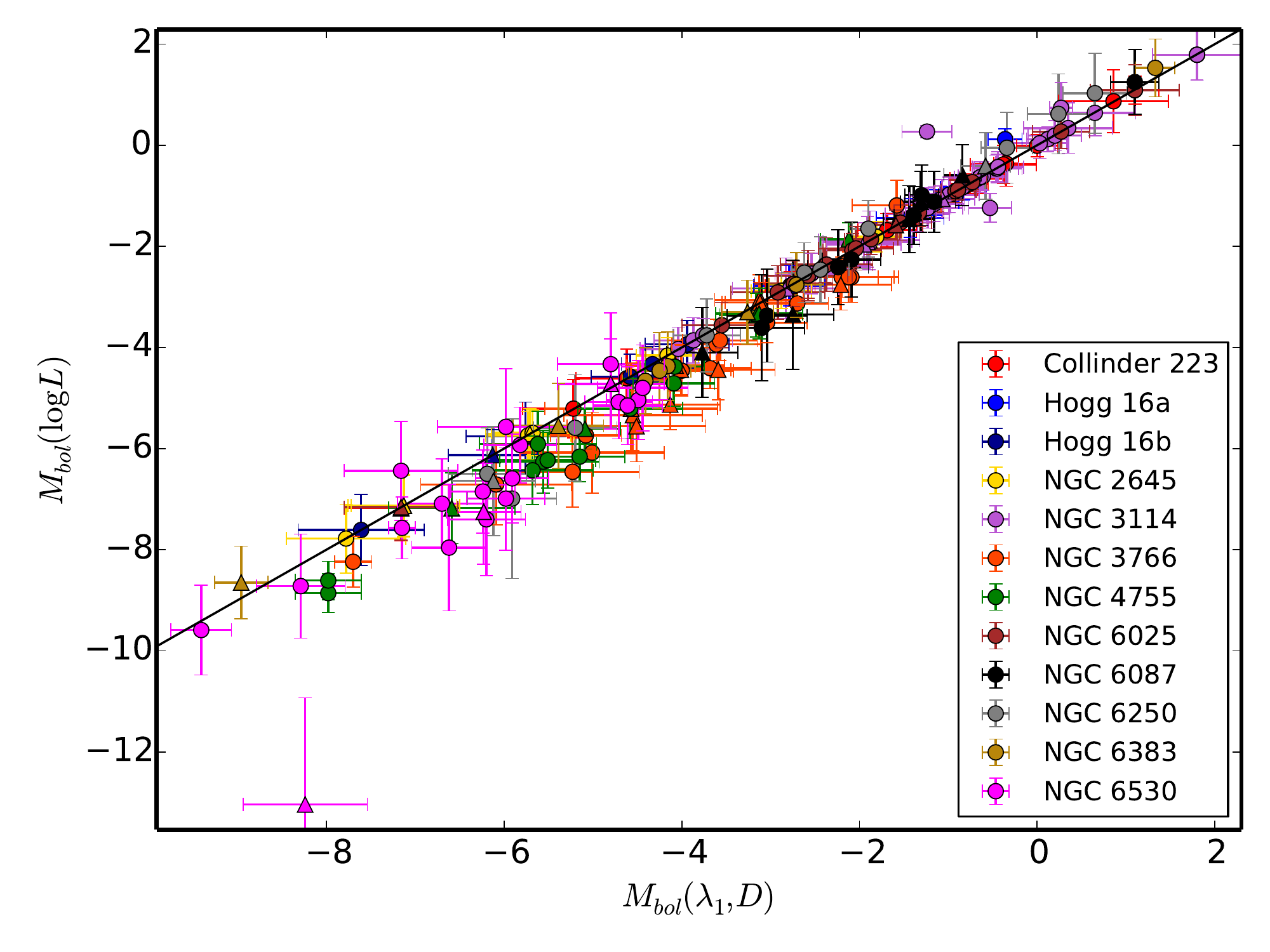}}
\subfigure[Comparison of surface gravities.]{\includegraphics[width=9cm,angle=0]{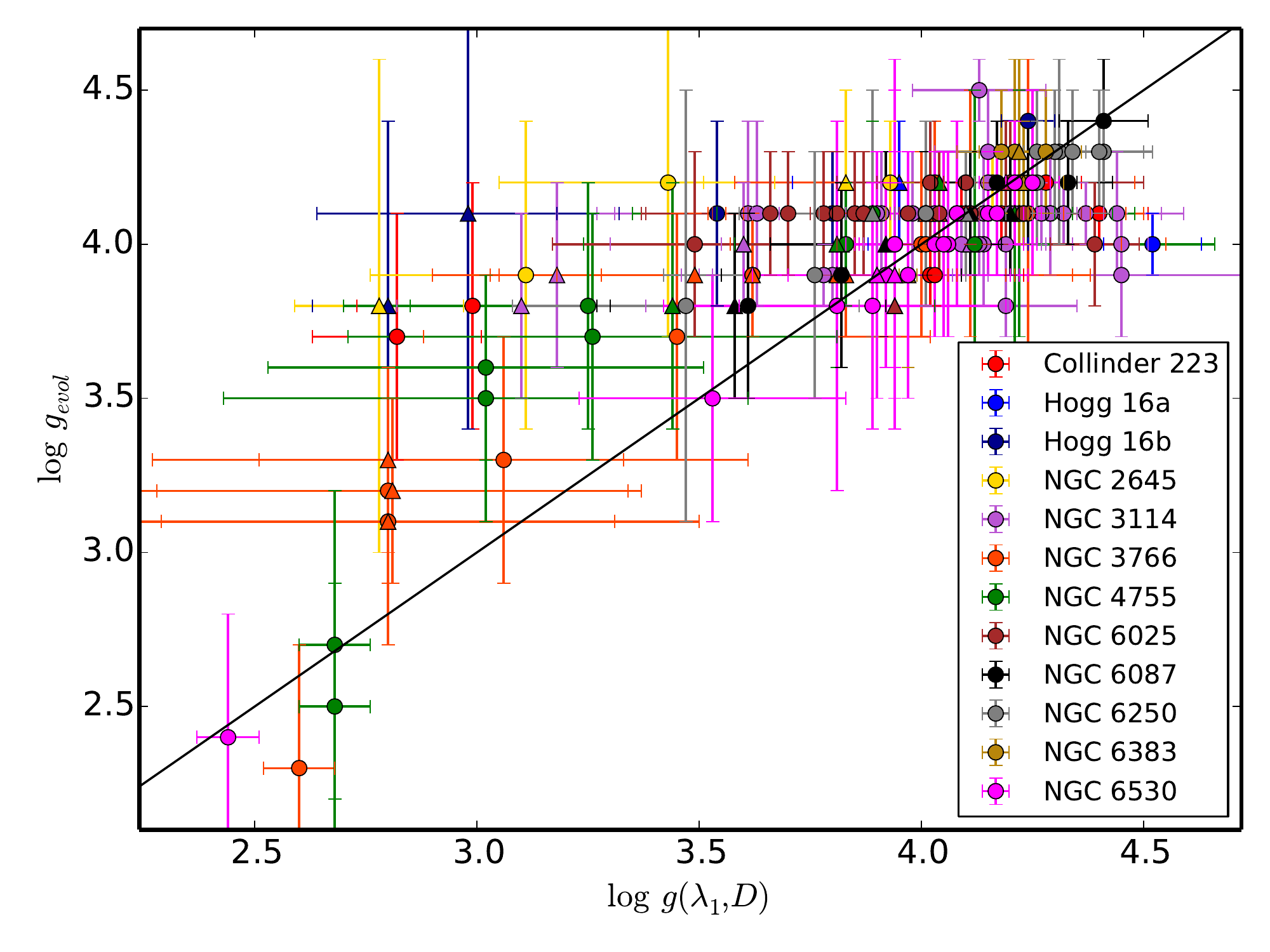}}
\caption{Comparison of $M_{{\rm bol}_{\rm evol}}$ and $\log g_{\rm evol}$ parameters, derived  from stellar evolution models, against BCD $M_{\rm bol} (\lambda_{1}, D)$ and $\log g(\lambda_{1}, D)$ estimates. Circles identify normal B-type stars and triangles  stars with circumstellar envelopes. The straight line corresponds to the identity function.}\label{Comp_Logg}
\end{figure*}

Regarding the $M_{\rm v}$ values, we have to consider that in most of the cases these were calculated using the distance modulus of the cluster.
Therefore, we can only compare values obtained for stars belonging to  clusters that have similar $(m_{\rm v} - M_{\rm v})_{0}$ as ours.
Such clusters are \object{Collinder\,223}, \object{NGC\,3766}, \object{NGC\,6025} and \object{NGC\,6383}.
This comparison is shown in Fig.~\ref{Comp_Mv}b.
The greatest discrepancies, this means  values with $| M_{\rm v_{\rm BCD}} - M_{\rm v_{\rm lit}} | > 2$~mag, are found among stars reported as Em, Be, supergiant or variable\footnote{These stars are: Cr\,223\,35 (reported as blue straggler), Hogg\,16\,2 (reported as Em), Hogg\,16\,3 (reported as Em), Hogg\,16\,9 (reported as Em), NGC\,2645\,1 (classified as supergiant), NGC\,2645\,2 (classified as supergiant), NGC\,3766\,15 (reported as Be), NGC\,3766\,232 (reported as variable), NGC\,3766\,291 (reported as Be), NGC\,6383\,6 (reported with an IR excess), and NGC\,6383\,47 (classified as a bright giant)}.
It worth mentioning that most of the cited works consider average values of $E(B-V)$ and these could lead to the discrepancies observed in stars with circumstellar envelopes.

In relation with comparisons of absolute bolometric magnitudes and surface gravities  (see Fig.~\ref{Comp_Logg})  derived from stellar evolution models with the BCD parameters, we can behold that both show the same trends found in Paper I.
In that opportunity only the results for \object{NGC\,3766} and \object{NGC\,4755} were considered.
We highlight the excellent agreement between BCD and evolutive $M_{\rm bol}$ values as is depicted in Fig.~\ref{Comp_Logg}a.
Whilst the discrepant behavior in $\log\,g$ (see Fig.~\ref{Comp_Logg}b) was already mentioned in Paper I and it is related to the technique used to derive the surface gravity.
The $\log\,g$ values derived from the BCD method ($\log\,g(\lambda_1,D)$) are also called atmospheric gravities, since the BCD calibration of this quantity is based on parameters $\log\,g$ determined by fitting observed H$\gamma$ and H$\beta$ line profiles with synthetic ones, obtained with classical model atmospheres. On the contrary, $\log\,g_{\rm evol}$ are  calculated using bolometric magnitudes $M_{\rm bol}(\lambda_1,D)$ and effective $T_{\rm eff}(\lambda_1,D)$ read on the BCD calibrations, which produce estimates of the stellar radius $R/R_{\odot}$, while they lead to masses $M/M_{\odot}$ read on the stellar evolutionary tracks. For stars with $M \lesssim 12\,M_{\odot}$ and radii $R \lesssim 12\,R_{\odot}$, the value of $\log\,g_{\rm evol}$ is often estimated in the deep stellar layers while $\log\,g_{\rm atm}$ is more related to the outermost atmospheric layers.
A reverse situation would occur for evolved massive stars with $M \gtrsim 20\,M_{\odot}$ and $R \gtrsim40\,R_{\odot}$.

The observed deviation between the $\log\,g_{\rm evol}$ and $\log\,g(\lambda_1,D)$, mostly at $\log\,g\lesssim3.7$ has already been noted by \citet{Gerbaldi1993} for late B- and early A-type stars.  This deviation is due to an overestimate of the stellar mass possibly carried by difficulties  on the stellar structure calculations at the helium ionization region (Maeder, private communication).

It is worth noting that rotation, whose effects are neglected in the present analysis, would act with systematic differences in the $\log g$ determinations that range in the opposite way as noted above for stars with $M \lesssim 12\,M_{\odot}$. The $\log\,g$ corrected for rapid rotation effect, called pnrc (parent non-rotating counterpart) surface gravities \citep{Fremat2005}, are generally larger by about $0.14\pm0.14$ dex than the observed or apparent ones. However, differences are largely aspect-angle dependent \citep{Zorec2016}.

Therefore, we should find higher apparent bolometric luminosities and lower effective temperatures than those expected for non-rotating stellar counterparts \citep{Fremat2005,Zorec2005}. As a consequence we will have lower $\log\,g_{\rm evol}$ determinations than the values for non-rotating cases.

\subsection{Comments on the Balmer discontinuity of Be stars}
\label{Dis_SBD}

\begin{figure}[]
\centering
\includegraphics[width=8cm,angle=0]{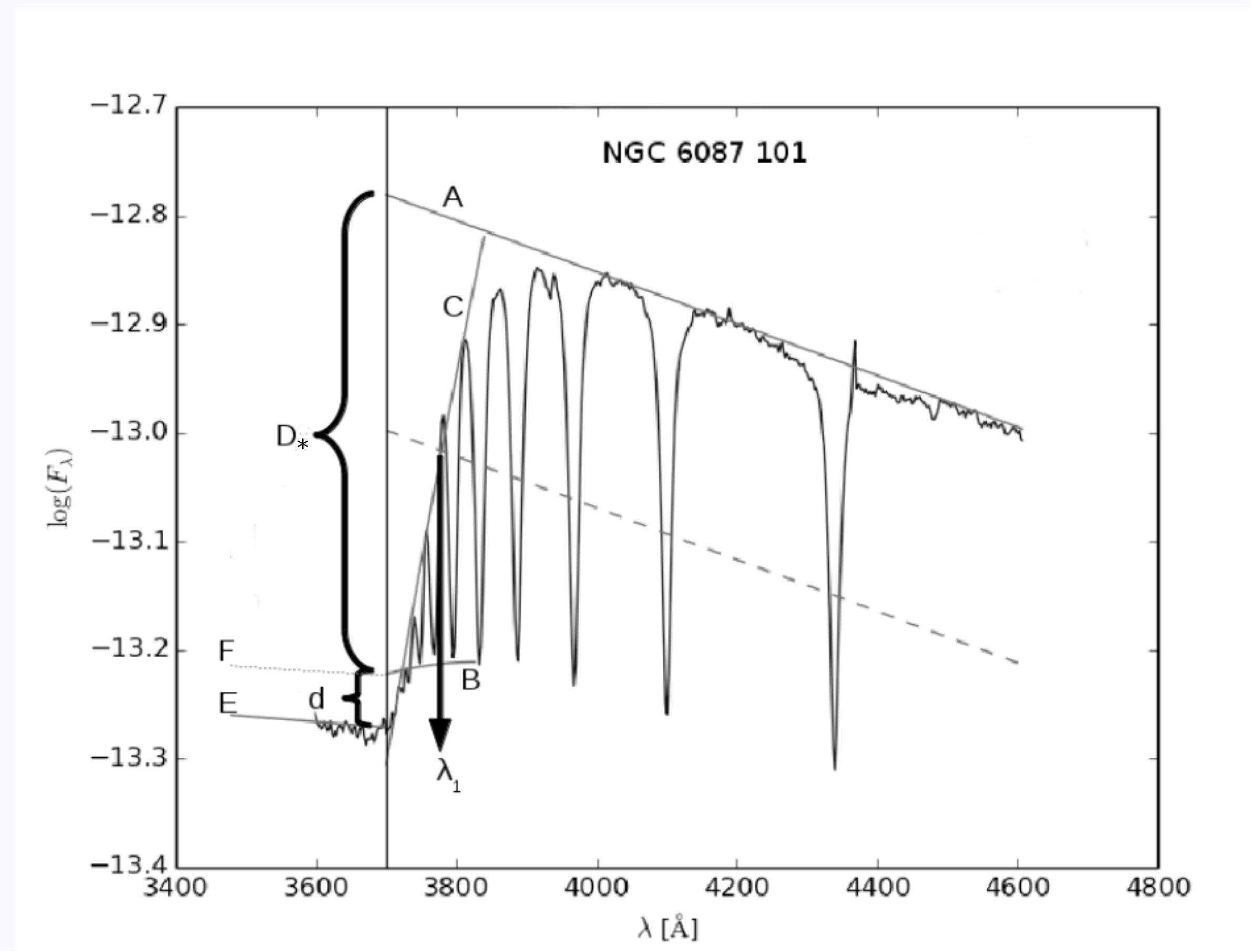}
\includegraphics[width=8cm,angle=0]{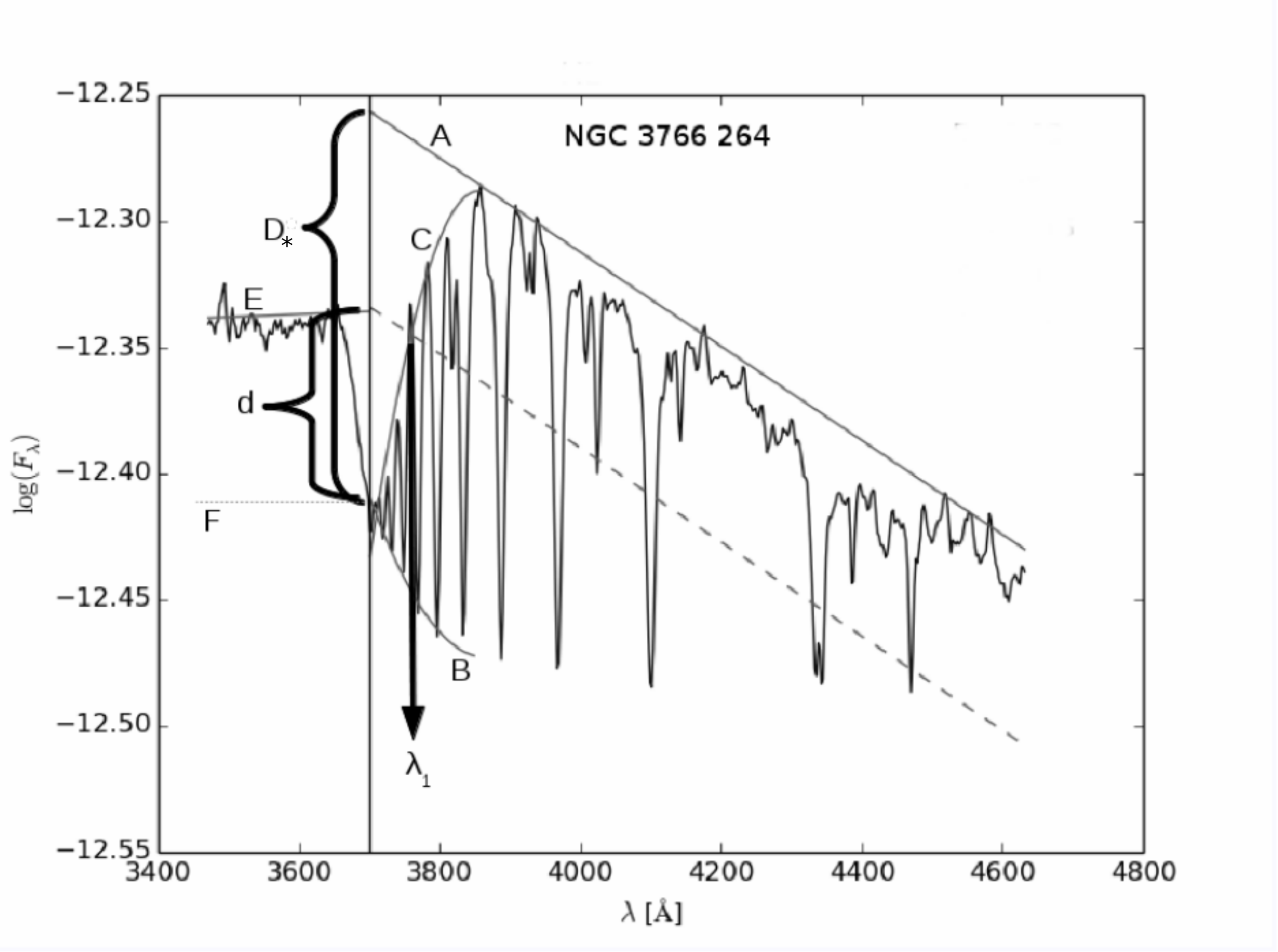}
\caption{Stars with a second Balmer discontinuity. $D_*$ is the height of the photospheric component of the Balmer discontinuity and $d$ is the height of the second component. In the upper panel $d$ is in absorption (Be-shell phase) and in the lower panel is in emission (Be phase).\label{fig-SBD}}
\end{figure}

The photospheric component of the BD is determined by the flux ratio $D_* = \log\,(F^+_{3700}/F^-_{3700})$. $F^+_{3700}$ is given by the interception of the extrapolated straight line ``A'' that fits the Paschen continuum, shown in Fig. \ref{fig-SBD}, and a vertical line at $\lambda$ = 3\,700 \AA\, and $F^-_{3700}$ is given by the interception of same vertical line  and the envelope curve  ``B'' that passes over the absorption cores of the higher members of the Balmer lines until the point they merged.

Let us recall some characteristics of these fluxes:

1) $F^+_{3700}$ represents the flux density that would exist if the wings of the higher members of the Balmer lines series (H$_n$; $n > 8$) do not overlap. This line overlapping produces a pseudo-continuum (represented by the last Balmer lines merged at redward the upper curve ``C'' surrounding the Balmer lines in Fig. \ref{fig-SBD}). Due to this overlapping and the low resolution of the spectra, the last Balmer lines merged redwards to the theoretical value,  $\lambda \sim 3\,648$~\AA.

2) This pseudo-continuum intercepts also the Balmer continuum shortly after the BD in O-, B-, A-, F-type stars without circumstellar emission or absorption.

3) $F^-_{3700}$ also belongs  to the above mentioned pseudo-continuum, and as both fluxes $F^\pm_{3700}$ are measured at the same wavelength they are affected by nearly the same amount of emission or absorption, $\Delta F_{3700}$, originated by a circumstellar envelope. For this reason, Papers I and II of this series explain our care in determining where the last lines of the Balmer series overlap. The circumstellar emission or absorption have very little if any incidence on the determination of the photospheric BD, $D_*$. Therefore,  the BD of some stars with circumstellar envelopes can be  defined as the quantity $D = D_*+ d$, where $D_*$ is the height of the photospheric (or first) component of the BD and $d$ represents the height of the SBD. The first component of the BD ($D_*$) has always been observed constant in Be stars, within the limits of uncertainties of the BCD system  \citep[$0.015$ dex in $D$ and $1-2$~\AA\, in $\lambda_1$,][]{DeLoore1979,Divan1983,Divan1982d,Zorec1986,Zorec1989,Zorec1991, Vinicius2006,Cochetti2013,Cochetti2015}. Apart from the rare case of \object{$\gamma$~Cas} (O9Ve, \object{HD\,5394}) in 1935 and 1938 \citep{Barbier1939}, the constancy of $D_*$ has even been noted in stars with Be$\rightleftarrows$B$\rightleftarrows$Be-shell-phase changes such as: \object{$\gamma$~Cas}, \object{Pleione}, \object{88~Her} and \object{59~Cyg}.

However, as $D_*$ is determined empirically it does not avoid two effects caused by the presence of a circumstellar envelope, which nonetheless are very small:

a) The genuine photospheric fluxes $F^\pm_{3700}$ transform into $F^\pm_{3700} + \Delta F_{3700}$, so that:

\begin{equation}
D^{\rm obs}_* \simeq D^o_*\,+\, \left(\frac{\Delta F_{3700}}{F^{+}_{3700}}\right)\,(1-10^{D^o_*}),
\label{eq1}
\end{equation}

\noindent where $D^o_*$ represents the actual photospheric BD, which in the hottest stars is 
\begin{equation}
 \Delta F_{3700}/F^{+}_{3700}\,\lesssim  0.05
\label{eq1.1}
\end{equation}

b) On the other hand, since $F^+_{3700}$ is obtained by extrapolation, it is affected by the color effects in the Paschen continuum carried by the $\lambda^3$ dependency of the circumstellar envelope-opacity. \citet{Moujtahid1998, Moujtahid1999} and \citet{Gkouvelis2016} estimated these effects and obtained that

\begin{eqnarray}
\begin{array}{rcl}
  \displaystyle D_* & = & D_*^{\rm obs}+ \epsilon_{D}\\
  \\
  \noindent        {\rm with}\\
  \\
  \displaystyle \epsilon_D & \simeq & 0.02\times[\Phi^{\rm obs}-\Phi(\lambda_1,D^o_*)],
\end{array}
\label{eq2}
\end{eqnarray}

Note that as the compilation made by \citet{Moujtahid1998} actually corresponds to spotted measurements concerning otherwise long-term Be-star "bumper" activity \citep{Cook1995,Hubert1998,Hubert2000,Keller2002,Mennickent2002,deWit2006} the color changes are of the order of  $\Phi^{\rm obs}-\Phi(\lambda_1,D^o_*)  \lesssim0.5$~$\mu$m and the error committed by this effect on the estimate of the photospheric BD is  $\epsilon_{D}\lesssim0.010$ dex. This error carries uncertainties on the estimate of the visual absolute magnitude which is smaller than the deviations of the absolute magnitude within a spectral type-luminosity class curvilinear box of the BCD system \citep{Chalonge1973,Zorec2009}.

The SBD is caused by the flux excess  produced in the circumstellar envelope due to bound-free and free-free transitions, mostly of hydrogen and helium atoms and electron scattering. Therefore, as spectrophotometric variations of Be stars are currently different according to the Be phases, emission phases, i.e. $d < 0$ are generally accompanied by a brightening and a reddening of the Paschen continuum, while shell phases ($d >0$) generally show a little decreases of the stellar brightening, but without changing the color of the Paschen continuum. An extensive compilation of these spectrophotometric behaviors was published by \citet{Moujtahid1998}.

The height of the SBD, $d$, is then measured as the flux difference (in a logarithmic scale) between the observed Balmer continuum fitted by a straight-line ``E'' and a parallel line ``F'' to Balmer continuum through the point defined by the merged of the Balmer lines provides a measurement of $d$ (see Fig. \ref{fig-SBD}).

\subsection{Stars with circumstellar envelopes}
\label{CE}

\begin{figure*}[]
\centering
\includegraphics[width=9cm,,height=15cm,angle=0]{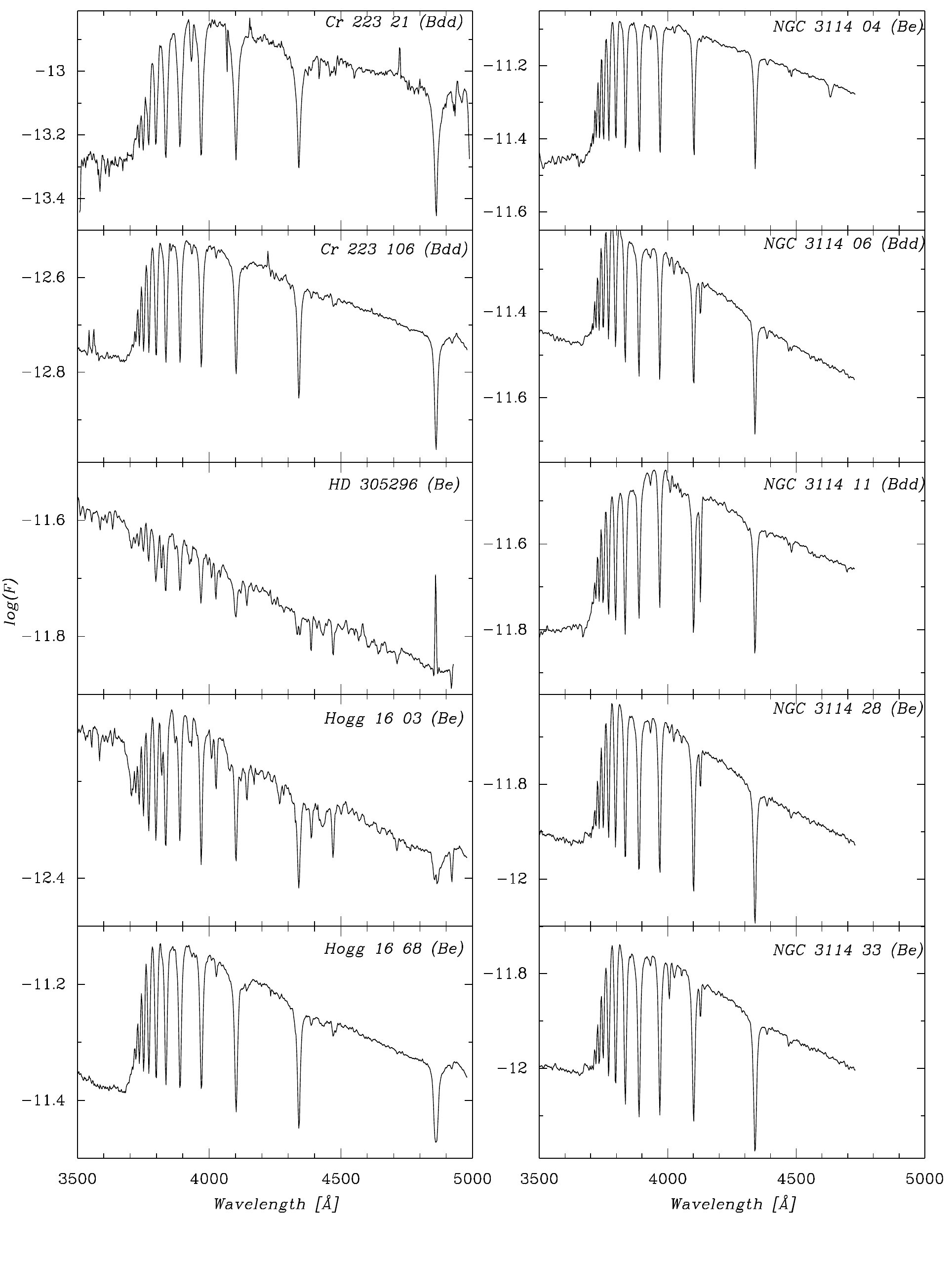}
\includegraphics[width=9cm,height=15cm,angle=0]{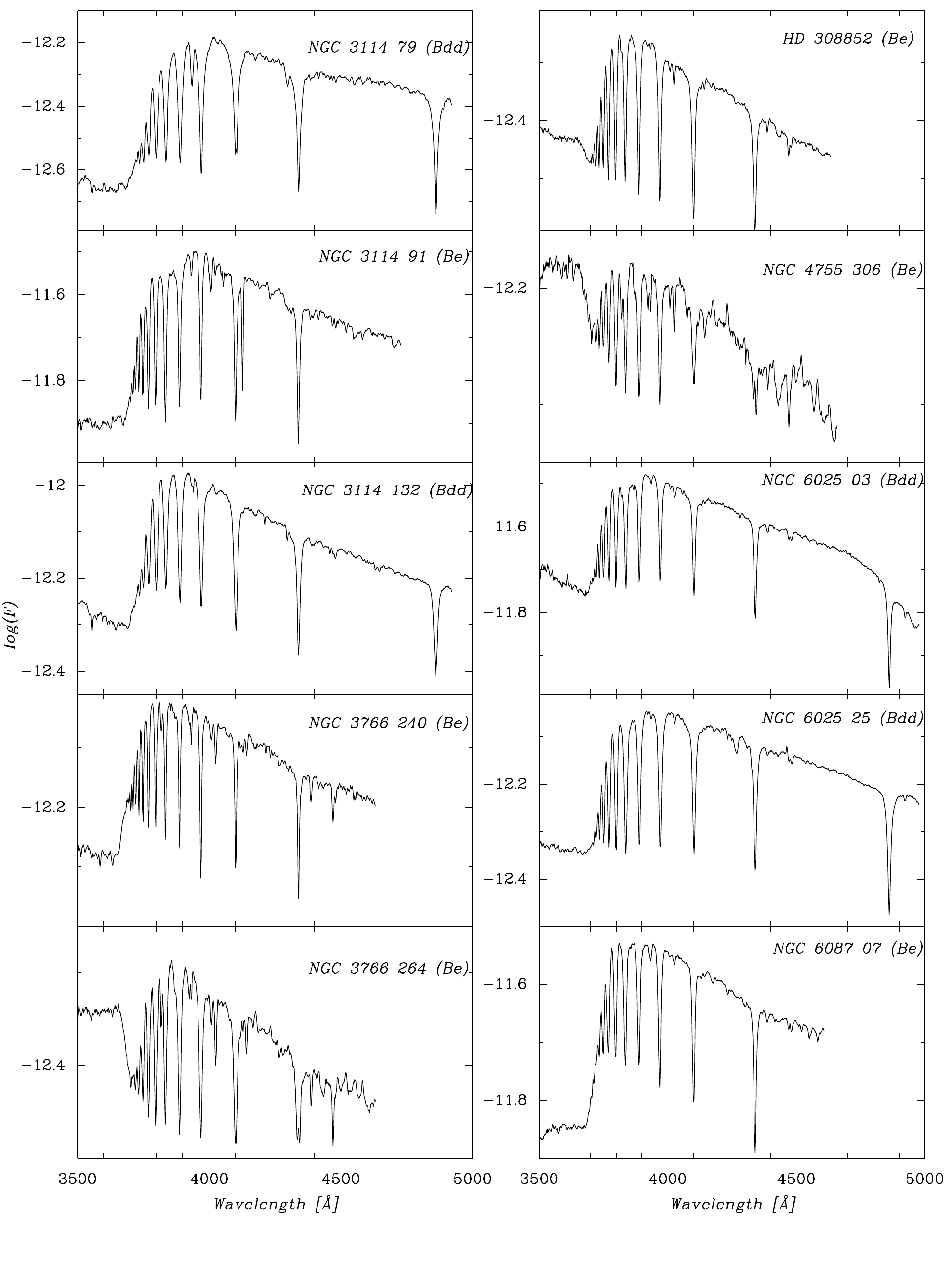}
\caption{Low resolution spectra of stars with a SBD.}\label{Bdd_Espectros}
\end{figure*}

During the process of seeking emission line stars in the direction of the $11$ studied open clusters we found a total of $46$ Be stars ($6$ new Be stars: \object{Cr\,223\,2}, \object{NGC\,2645\,5}, \object{NGC\,3766\,301},  \object{HD\,308\,852}, \object{NGC\,6087\,7}, and \object{NGC\,6383\,76})\footnote{We observe H$\alpha$ in emission in the spectrum of these stars.} when considering the whole sample of 230 stars. Among these 46 Be stars we detected 19 objects with a SBD.

In addition, we found $15$ B-type stars with a SBD that show neither line emission features nor have been previously reported as Be stars, and $3$ B supergiants (BSG) with a SBD (\object{HD\,74\,180}, \object{NGC\,4755\,1}, \object{NGC\,6530\,45}). The $15$ B-type stars that show a SBD could be Be candidates, these stars will be called hereafter Bdd (B stars with a double Balmer discontinuity).
In this way,  the total number of stars with a SBD (Bdd+Be+BSG stars) is 37, which were identified in the corresponding tables of Papers I, II and this work, with a supra index $dd$ (stars \object{NGC\,3766\,55} and \object{NGC\,3114\,126} were discarded since they have a doubtful SBD).
Therefore, our whole sample of objects with a SBD in absorption consists of $29$ stars ($3$ in \object{Cr\,223}, $1$ in \object{Hogg16a}, $2$ in \object{NGC\,2645}, $10$ in \object{NGC\,3114}, $1$ in \object{NGC\,3776}, $1$ in \object{NGC\,4755}, $2$ in \object{NGC\,6025}, $7$ in \object{NGC\,6087}, $1$ in \object{NGC\,6383}, and $1$ in \object{NGC\,6530}) and $8$ in emission ($2$ in \object{Cr\,223}, $1$ in \object{Hogg16b}, $2$ in \object{NGC\,3776}, $1$ in \object{NGC\,4755}, $1$ in \object{NGC\,6250}, $1$ in \object{NGC\,6530}).

Henceforth we will restrict ourselves to study the sample of Be and Bdd stars belonging to the $11$ studied galactic clusters. This sample includes $42$ Be ($17$ with a SBD) and $13$ Bdd stars, mainly of B-type stars, although a few late O and early A stars were also observed. Among the $42$ Be stars we found $30$ dwarfs (e.g. luminosity classes IV, V and VI) and $12$ slightly more evolved stars (e.g. luminosity classes III and II).
Thus, considering only the sample actually observed in the BCD system, which encompass $N_{\rm tot} = 198$ B+Be+Bdd stars (luminosity classes from ZAMS to bright giants) of which $N_{\rm e} = 42$ are Be stars, we obtain $f=N_{\rm e}/N_{\rm tot}=21 \%$.
This percentage is higher than that reported by \citet{Zorec1997} for the field of galactic Be stars ($\sim 17\%$).
While among the  $13$ ($\sim 7\%$), $11$ are dwarfs and $2$ are giants. Table~\ref{Be_y_Bdd} lists the Be and Bdd stars used for this analysis.

We notice that according to the weather conditions, several observing restrictions and other stellar sample selections, the above fraction $f=21\%$ of B emission-line stars can be biased, mainly because part of the  sample of Be stars were not selected randomly. In order to evaluate a possible overestimate of the actual fraction of Be stars that we have calculated for  open clusters, we first compare this fraction with the fraction that can be derived using the available Be  and B stars in the WEBDA database for the same 11 clusters. Thus, in the WEBDA database we counted $57$ Be stars among $236$ B stars. This implies a fraction of $24\%$ which is larger than the one we have estimated in this work. However, we must keep in mind that not all stars in the WEBDA database are necessarily cluster members.

A second test is to compute a Monte Carlo simulation of the distribution of an "actual" fraction, $f_{\rm actual} = 21\%$ B emission-line stars for $N_{\rm tot} = 198$ B+Be+Bdd stars in $11$ open clusters, over 100 independent samplings.
We obtained thus an average $\langle f_{\rm simulated}\rangle=22\%$ B emission-line stars with dispersion $\sigma_{\rm f} = 9\%$. 
If we  had observed $N_{\rm tot} = 2000$ B+Be+Bdd stars, the simulation predicts that we should have $\langle f_{\rm simulated}\rangle=22\%$ B emission-line stars with $\sigma_{\rm f}=4\%$.
In this simulation we have assumed that all clusters have the same age, so that each can be able to bear the same fraction of B emission-line stars. Considering differences in cluster ages, the bias $\sigma_{\rm f}$ could perhaps be larger.

Hence, both results suggest that the fraction of Be stars in open clusters is probably larger than that of  field Be stars.

\begin{table}[]
\begin{center}
\tabcolsep 2pt
\caption{\label{Be_y_Bdd} Be and Bdd stars belonging to open clusters.}
\begin{tabular}{llll}
\hline
\hline
ID & ST & Group & Paper\\
   &    &       & Reference\\

\hline
\hline
Cr\,223\,021\tablefootmark{dd} & A1V   & Bdd & II\\
Cr\,223\,106\tablefootmark{dd} & B5IV  & Bdd & II\\
HD\,305\,296\tablefootmark{dd} & B0Ve  & Be  & II\\

\hline
Hogg\,16a\,68\tablefootmark{dd} & B4Ve & Be & II\\

Hogg\,16b\,02 & B2II   & Be & II\\
Hogg\,16b\,03\tablefootmark{dd} & B1IIIe & Be & II\\

\hline
NGC\,2645\,03 & B0II  & Be & II\\
NGC\,2645\,04 & B0IIe & Be & II\\
NGC\,2645\,05 & B1IVe & Be & II\\

\hline
NGC\,3114\,003\tablefootmark{dd} & B6IIIe & Be  & II\\
NGC\,3114\,004\tablefootmark{dd} & B8IIIe & Be  & II\\
NGC\,3114\,006\tablefootmark{dd} & B8IV   & Bdd & II\\
NGC\,3114\,011\tablefootmark{dd} & B8IV   & Bdd & II\\
NGC\,3114\,028\tablefootmark{dd} & B6IVe  & Be  & II\\
NGC\,3114\,033\tablefootmark{dd} & B5IVe  & Be  & II\\
NGC\,3114\,079\tablefootmark{dd} & A1V    & Bdd & II\\
NGC\,3114\,091\tablefootmark{dd} & B8IVe  & Be  & II\\
NGC\,3114\,129                   & B8Ve   & Be  & II\\
NGC\,3114\,132\tablefootmark{dd} & B7:VI: & Bdd & II\\

\hline
NGC\,3766\,001                   & B5III & Be  & I\\
NGC\,3766\,015                   & B4III & Be  & I\\
NGC\,3766\,026                   & B2IVe & Be  & I\\
NGC\,3766\,027                   & B3III & Be  & I\\
NGC\,3766\,151                   & B3Ve  & Be  & I\\
NGC\,3766\,239                   & B3IV  & Be  & I\\
NGC\,3766\,240\tablefootmark{dd} & B5IIe & Be  & I\\
NGC\,3766\,264\tablefootmark{dd} & B2Ve  & Be  & I\\
NGC\,3766\,291                   & B3Ve  & Be  & I\\
NGC\,3766\,301                   & B3Ve  & Be  & I\\
HD\,308\,852\tablefootmark{dd}   & B5IVe & Be  & I\\

\hline
NGC\,4755\,008                   & B3V    & Be & I\\
NGC\,4755\,011                   & B6V    & Be & I\\
NGC\,4755\,117                   & B3V    & Be & I\\
NGC\,4755\,202                   & B2IV   & Be & I\\
NGC\,4755\,306\tablefootmark{dd} & B0IIIe & Be & I\\

\hline
NGC\,6025\,01                   & B0Ve: & Be  & II\\
NGC\,6025\,03\tablefootmark{dd} & B5III & Bdd & II\\
NGC\,6025\,06                   & B6V   & Be  & II\\
NGC\,6025\,25\tablefootmark{dd} & B7V   & Bdd & II\\

\hline
NGC\,6087\,007\tablefootmark{dd} & B6Ve   & Be  & III\\
NGC\,6087\,009\tablefootmark{dd} & B6Ve   & Be  & III\\
NGC\,6087\,010\tablefootmark{dd} & B5IIIe & Be  & III\\
NGC\,6087\,011\tablefootmark{dd} & B4V    & Bdd & III\\
NGC\,6087\,014\tablefootmark{dd} & B8V    & Be  & III\\
NGC\,6087\,022                   & B2IVe  & Be  & III\\
NGC\,6087\,101\tablefootmark{dd} & A1:VI: & Bdd & III\\
NGC\,6087\,156\tablefootmark{dd} & B6:V:  & Bdd & III\\

\hline
NGC\,6250\,01\tablefootmark{dd} & B0III & Bdd & III\\
NGC\,6250\,35                   & B0IVe & Be  & III\\

\hline
NGC\,6383\,001                   & O7V  & Be  & III\\
NGC\,6383\,057\tablefootmark{dd} & A2V  & Bdd & III\\
NGC\,6383\,076                   & B3Ve & Be  & III\\

\hline
NGC\,6530\,042                   & B0IV & Be & III\\
NGC\,6530\,065\tablefootmark{dd} & O    & Be & III\\
NGC\,6530\,100                   & B2VI & Be & III\\
\hline
\hline
\end{tabular}
\tablefoot{
\tablefoottext{dd}{Star with a SBD.}
}
\end{center}
\end{table}

\begin{figure}[]
\centering
\includegraphics[width=9cm,height=15cm,angle=0]{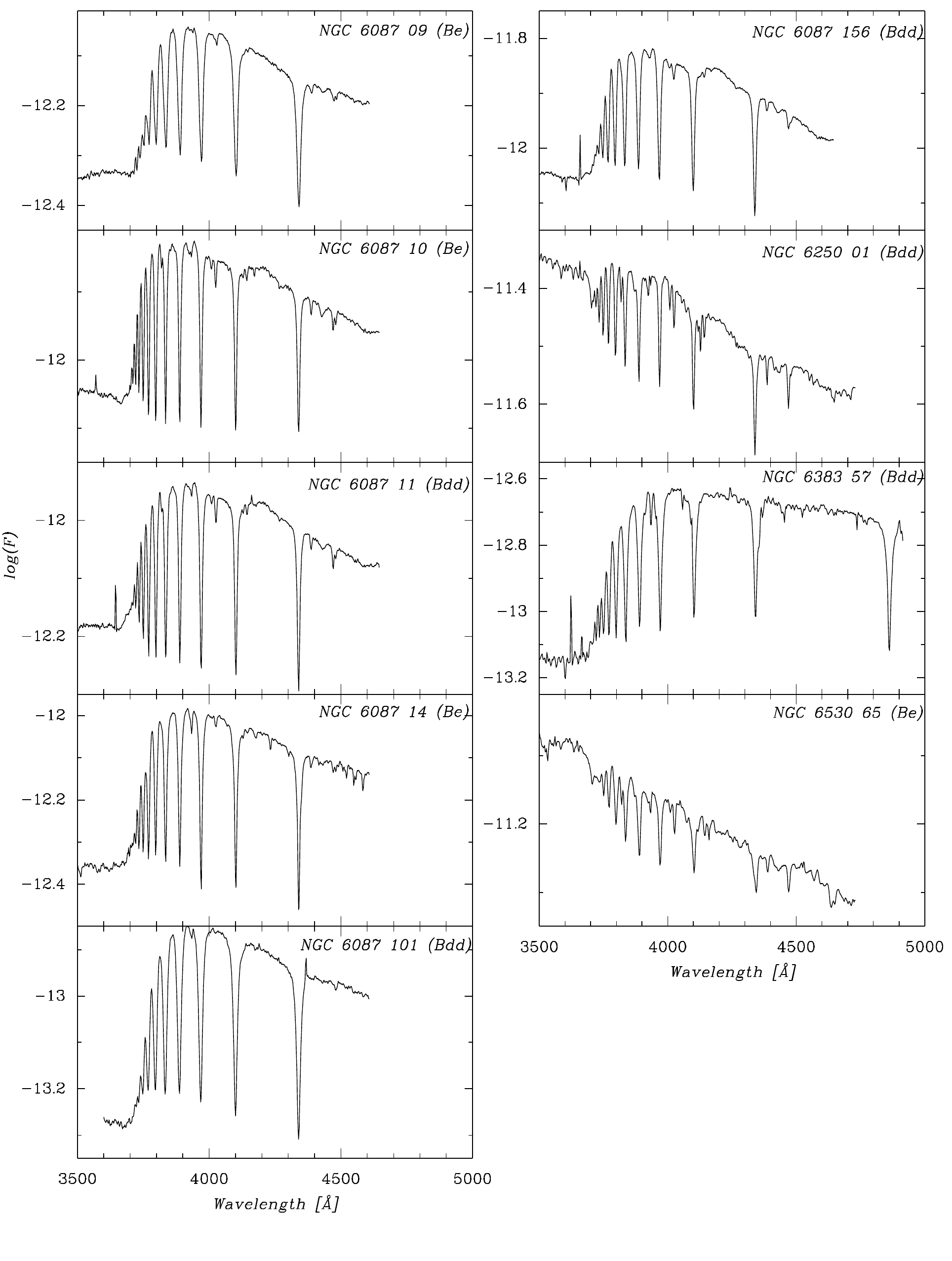}
\caption{Idem Fig. \ref{Bdd_Espectros}.\label{Bdd_Espectros2}}
\end{figure}

\begin{figure}[]
\centering
\includegraphics[width=9cm,height=9cm,angle=-90]{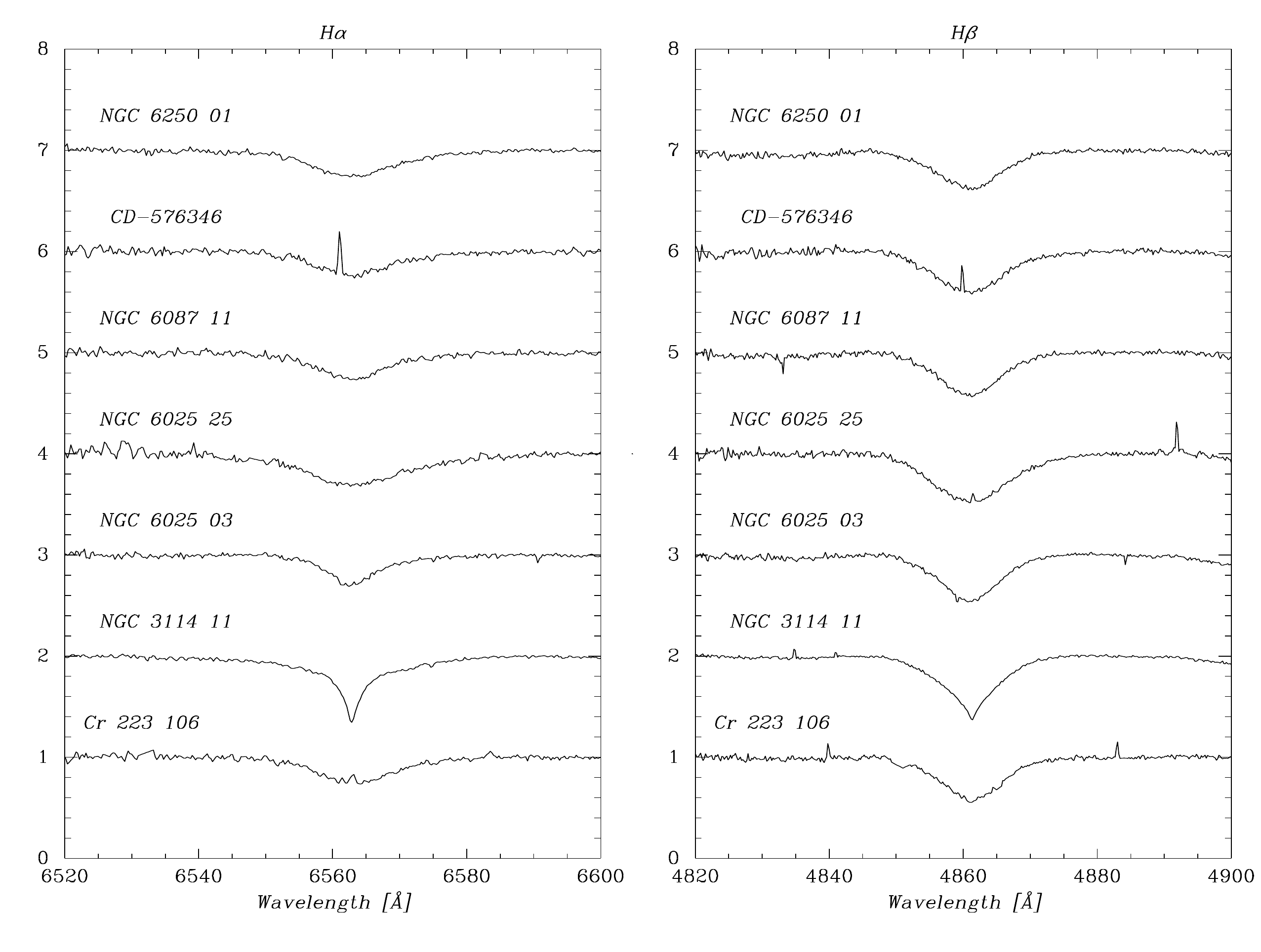}
\caption{Bdd H$\alpha$ (first column) and H$\beta$ (second column) profile lines.}\label{Bdd_Halpha}
\end{figure}

The spectra of these $13$ Bdd stars are shown in Figs.~\ref{Bdd_Espectros} and \ref{Bdd_Espectros2} together with the spectra of the $17$ Be stars with a  SBD. Moreover, we have high resolution spectra of only $7$ Bdd stars in the region of H$\alpha$.
With the exception of \object{NGC\,3114\,11} that show a typical shell profile, the rest of the stars display an absorption H$\alpha$ line (see Fig.~\ref{Bdd_Halpha}) that might be partially filled in by emission.
However, as these spectra were not taken simultaneously with the spectrophotometric BCD observations either emission or shell features, if they existed, could have disappeared in the meantime.

\subsection{The $T_{\rm eff}$- $\log\,g$ diagram}

\begin{figure}[]
\centering
\includegraphics[width=9cm,angle=0]{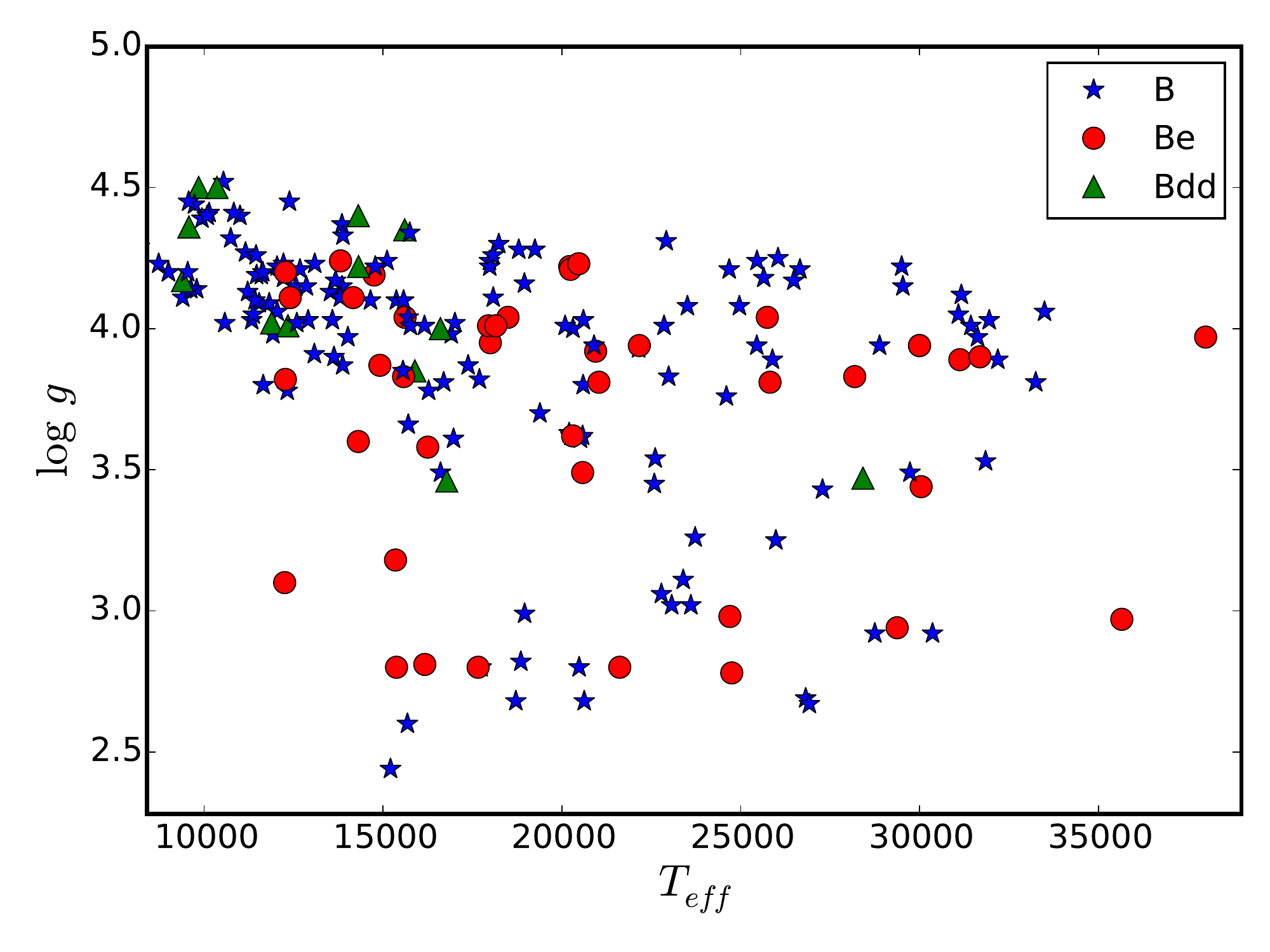}
\caption{Cluster star members in $T_{\rm eff}$ against $\log g$ plot. Be stars are distributed in the whole temperature interval.}\label{LoggVSTeff}
\end{figure}

In Fig.~\ref{LoggVSTeff} we present a $T_{\rm eff}$- $\log\,g$ diagram derived with the BCD parameters, where we can observe that most of the B and Be stars are dwarfs ($3.5\,\lesssim\,\log\,g\,\lesssim\,4.5$) and are distributed in the whole temperature interval.
A smaller number of B and  Be giants seem also to be uniform distributed in the same $T_{\rm eff}$ range.
This supports the results found by \citet{Mermilliod1982a, Slettebak1985, Zorec2005} who state that Be stars occupy the whole main sequence band and the Be phenomenon occurs in very different evolutionary states.
On the other hand, the sample of Bdd stars is mainly constituted by cool B dwarfs with circumstellar envelopes, as shown in next section.

\subsection{Distribution of stars per spectral subtype and age}

It has to be emphasized that our sampling is far from being complete due to the observing conditions (short running campaigns or full moon nights) that introduced strong selection effects, since sometimes we only could observe the brightest members of each clusters.
However, we took care of observing  almost all  previously reported Be stars.
Notwithstanding of the small sample size and risk of bias in the included trials, it is still possible to discuss age-dependence trends related to the appearance of the Be phenomenon, as was previously suggested by \citet{Mathew2008, Zorec2005, Fabregat2000}, and compare these trends with the presence of the new group of Bdd stars.
To this purpose we classified our sample in three age groups.
The first group comprises open clusters with ages between $3$~Myr and $10$~Myr, the second one is related with intermediate-age clusters ($10-40$~Myr) and the last group encompasses  clusters older than $40$~Myr.
      
\begin{figure*}[]
\centering
\includegraphics[width=6cm,height=4.5cm,angle=0]{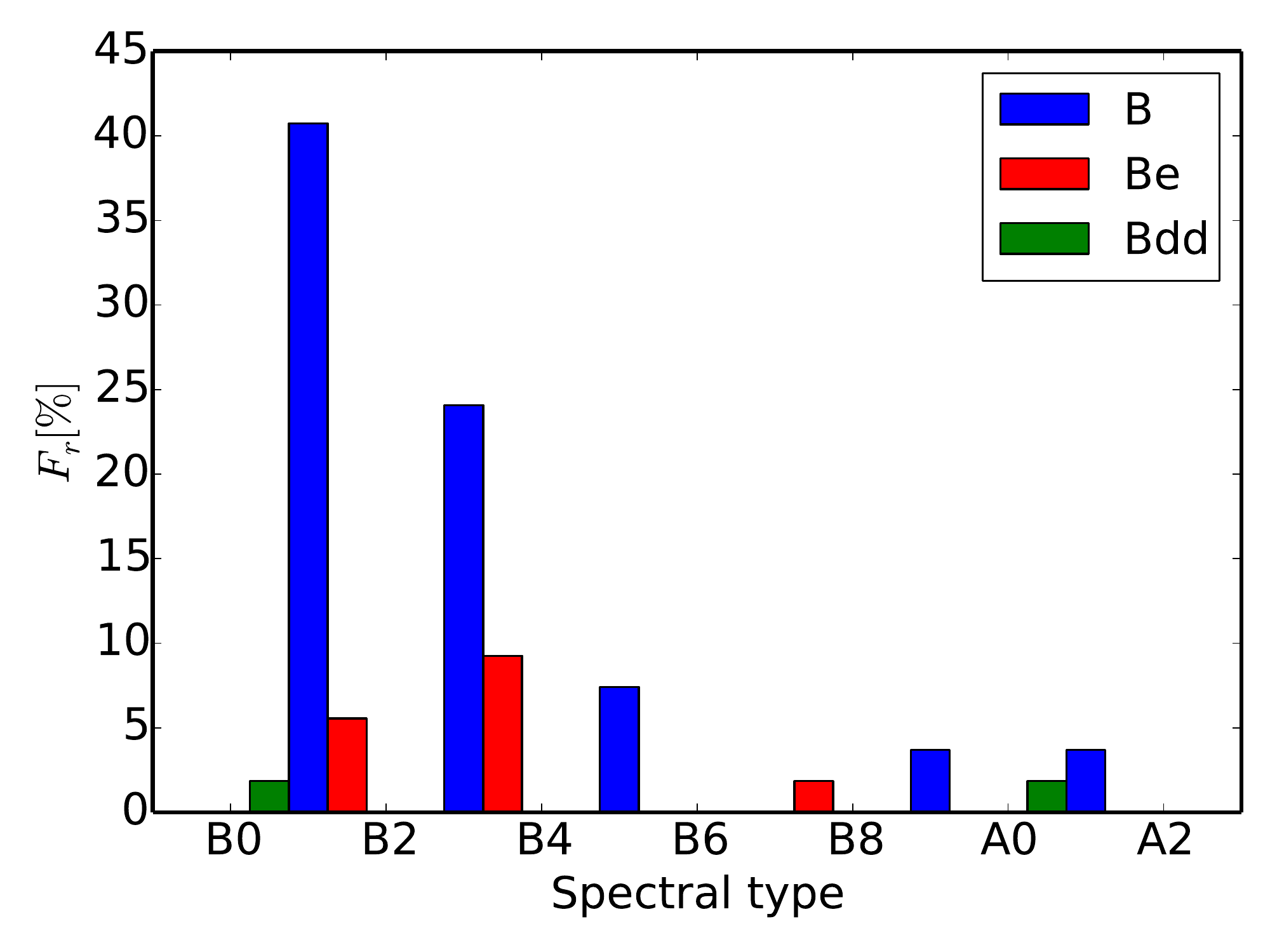}
\includegraphics[width=6cm,height=4.5cm,angle=0]{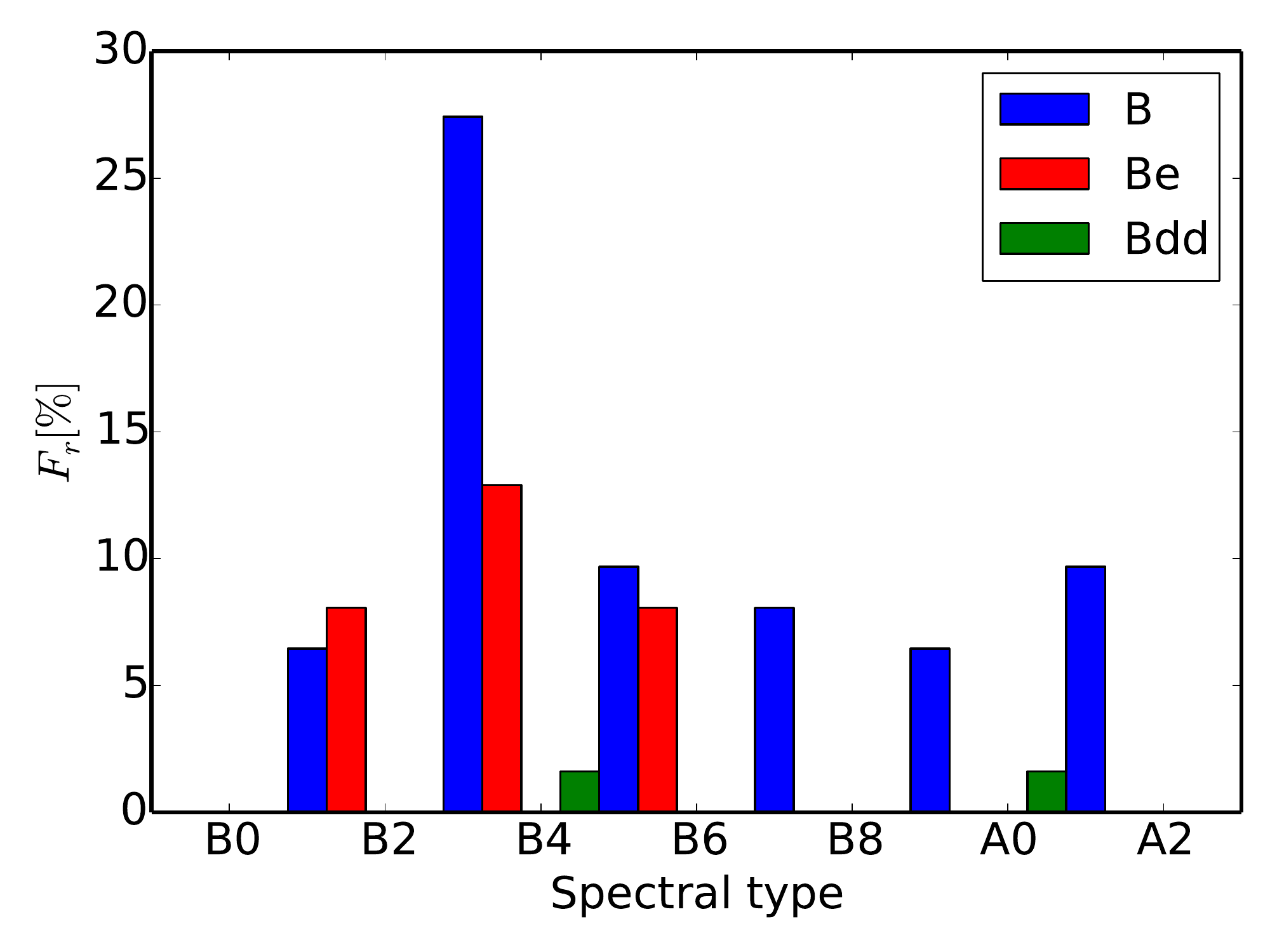}
\includegraphics[width=6cm,height=4.5cm,angle=0]{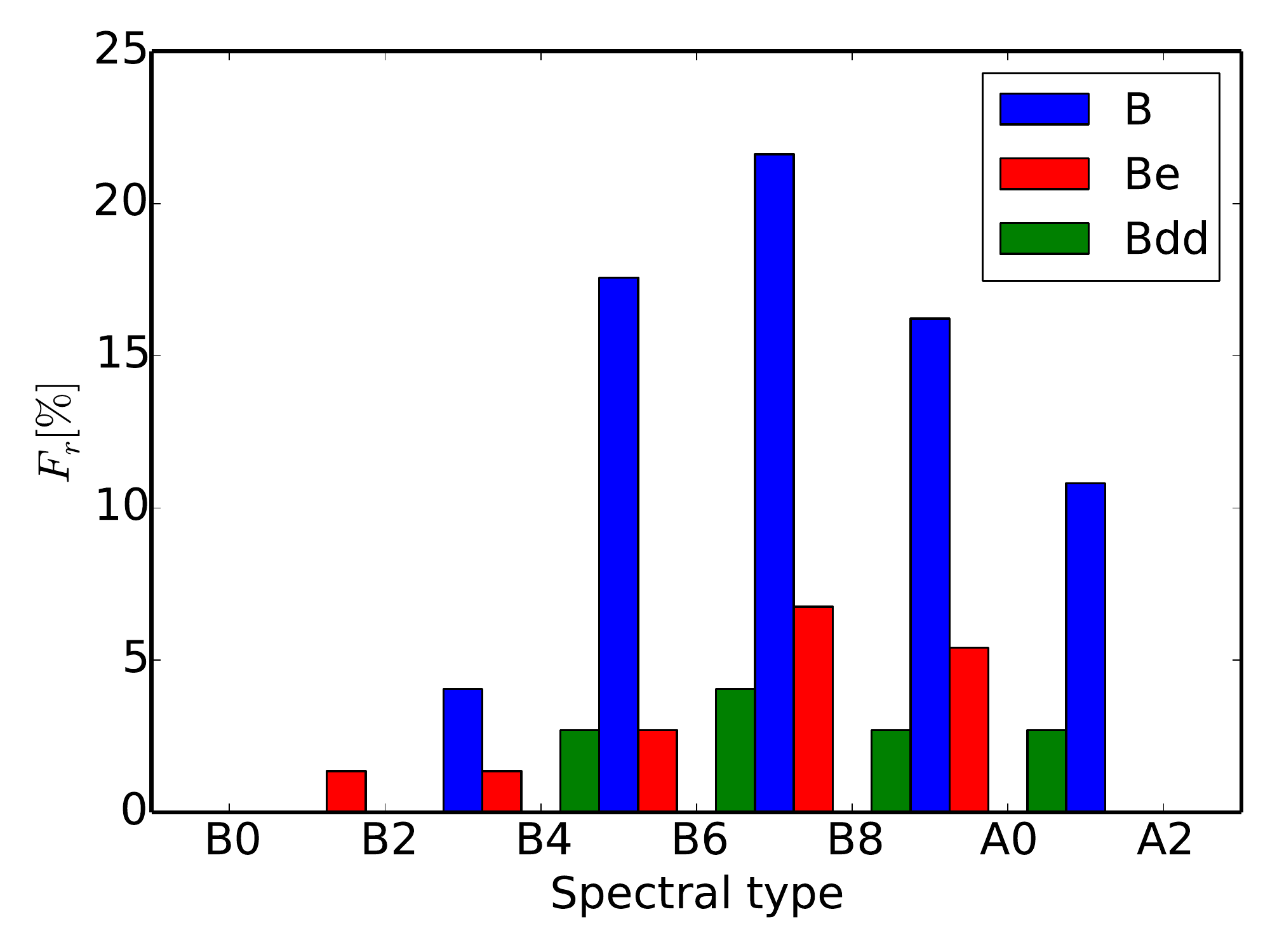}
(a) \hspace{6cm} (b) \hspace{6cm} (c)
\caption{Number and frequency of stars with and without circumstellar envelopes per spectral subtype in open clusters with different ages: (a) between 3 Myr and 10 Myr, (b)  between 10 Myr and 40 Myr, and (c)  older  than 40 Myr. The plots show a clear trend of the appearance of the Be phenomenon with age.} \label{Hist_TEyCL}
\end{figure*}

For each age group we calculated the relative frequency of different kinds of stars: B, Be and Bdd, per interval of spectral subtype.
To avoid skewing the results we searched for  optimal  bin number and bin  size per spectral subtype.
Hence, we adopted a bin-width optimization method  that minimized the integrated squared error \citep{Shimazaki2007}.
We obtained a total of $6$ bins that correspond to the spectral intervals: B0-B2, B2-B4, B4-B6, B6-B8, B8-A0, and A0-A2.
The corresponding bar-graph histograms per spectral subtype and per age group are illustrated in Fig.~\ref{Hist_TEyCL}, where the blue, red and green bars represent, respectively, the relative frequency of normal B, Be and Bdd stars within each bin.

In the plots we observe that young and intermediate-age clusters are composed of Be stars of mostly  B2-B4 spectral types  and only a few Bdd stars were detected (Fig.~\ref{Hist_TEyCL}a and b).
We have a fraction of about $19 \%$ of Be stars in young open clusters ($< 10$~Myr).
This number then enhances to about a $28 \%$ for the intermediate-age group (with ages between $10-40$~Myr, see Fig.~\ref{Hist_TEyCL}b) where we have the maximum Be star fraction, supporting thus previous results found by  \citet{Malchenko2008}  and  \citet{Fabregat2000}.
In the intermediate-age group we found also a great number of evolved Be stars with luminosity III-II.
Furthermore, in the same plot, we can observe a small fraction of Be stars at the B4-B6 types.
Then, the percentage of Be stars seems to decrease in the old open clusters ($\sim 17\%$, Fig.~\ref{Hist_TEyCL}c), as was stressed by \citet{Mermilliod1982a} and \citet{Grebel1997}.
In these old clusters, the Be phenomenon is dominated by stars with spectral types later than B4 and the maximum frequency of Be stars is around the B6-B8 types.
This result indicates a clear trend of the appearance of the Be phenomenon with age.

On the other hand, if we focus on the relative frequency of Bdd stars  (stars with circumstellar envelopes that are out off an emission phase) per spectral subtype (Fig.~\ref{Hist_TEyCL}c), a significant number ($\sim12\%$) of Bdd stars later than B5 is present and its distribution seems to follow the frequency distribution of B and ``active'' Be stars.
Therefore, the distribution of Bdd stars seems to complete, in number, the lack of late ``active'' Be stars.
We interpret the result in the same sense as \citet{Zorec2003}'s hypothesis which states that the strong deficiency of the field Be stars cooler than B7 perhaps could be taken on by Bn stars (i.e., B stars with broad absorption lines).
\citet{Zorec2003} also proposed that the absence of late Be stars and the increment of Bn stars is mainly due to the fact that their low effective temperatures are not able to keep their circumstellar envelopes completely ionized.
Thus, Bn stars could be latent Be and Be-shell stars \citep{Zorec2003, Ghosh1999}.
Therefore, it would be necessary to study in more detail the properties of the Bn star population and a possible link with the Bdd star group, if they are not the same population.

In summary, the Be phenomenon is observed in both dwarf and giant stars along the whole main sequence.
In addition we confirm the trend of an age evolution of the Be phenomenon, since  among the  massive stars the Be phenomenon seems to appear on average at an earlier age than in the less massive stars.
This result confirms previous age-dependence trends found by \citet{Mathew2008, Zorec2005, Fabregat2000}.
In addition, we found that the number of stars with circumstellar envelopes (active and passive B emission stars)  in the old age cluster group amounts to $29\%$, suggesting that Bdd stars might be the coldest counterpart of stars with the Be phenomenon.

From a theoretical point of view, the age evolution of the Be phenomenon was supported by  \citet{Ekstrom2008}.
These authors showed that the age-dependence of the Be phenomenon could be related to the  increase of  the ratio $V\,\sin\,i/V_{\rm crit}$  due to a lowering of $V_{\rm crit}$ (critical rotation velocity) of stars with age, via the transport of angular momentum by meridional circulation to the star's surface and stellar wind.
On the other hand, the rotation modifies stellar pulsations, since centrifugal force distorts the resonant cavity of the pulsations and the Coriolis force modifies the modes' dynamic \citep{Ouazzani2012}.
As well, the evolutionary changes affect the pulsational properties of the stars due to changes in the gravity, as well as the isentropic sound velocity and density \citep{Bruggen1987}.
Therefore, both rotation and pulsation modes might compete to release material from the star to fill a disk.
Therefore, our knowledge about the evolution effects of rotation and  pulsational modes, and their interplay, is crucial to understand the mechanism of ejection of mass and disk formation.

\section{Conclusions}
\label{Con}

In this work we used the BCD spectrophotometric system, based on measurable quantities of the stellar continuum spectrum around the Balmer discontinuity, to directly determine the fundamental parameters (spectral type, luminosity class, $T_{\rm eff}$, $\log g$, $M_{\rm v}$, $M_{\rm bol}$ and $\Phi_{\rm b}^{0}$) of $68$ B and Be stars in four open clusters.
In addition, we were able to derive individual values of the stellar distance modulus, color excess $E(B-V)$, luminosity, mass and age for the whole sample.
From these measurements we obtained the cluster parameters.

Thus, this work complete the study of the determination of cluster parameters for  eleven open clusters (\object{Collinder\,223}, \object{Hogg\,16}, \object{NGC\,2645}, \object{NGC\,3114}, \object{NGC\,3766}, \object{NGC\,4755}, \object{NGC\,6025}, \object{NGC\,6087}, \object{NGC\,6250}, \object{NGC\,6383} and \object{NGC\,6530}), which are summarized in Table~\ref{Resultados}. Particularly, the distance determinations obtained through the BCD method for \object{NGC\,2645}, \object{NGC\,6087}, \object{NGC\,6250}, and \object{NGC\,6383} agree with distance estimates obtained with classical photometric methods.

\begin{table}[h!]
\small
\begin{center}
\caption{Open cluster parameters derived with the BCD method.}\label{Resultados}
\begin{tabular}{lcrr}
\hline
\hline
Open           & $E(B-V)$      & $(m_{\rm v} - M_{\rm v})_{0}$ & $age$~~~~\\
cluster        & [mag]         & [mag]~~~~      & [Myr]~~\\
\hline
\object{Collinder\,223} & $0.25\pm0.03$ & $11.21\pm0.25$ & $\sim 25$\\
\object{Hogg\,16a}      & $0.26\pm0.03$ & $ 8.91\pm0.26$ & $< 40$\\
\object{Hogg\,16b}      & $0.65\pm0.09$ & $12.78\pm0.32$ & $10-16$\\
\object{NGC\,2645}      & $0.54\pm0.07$ & $12.39\pm0.30$ & $6-14$\\
\object{NGC\,3114}      & $0.05\pm0.01$ & $ 9.21\pm0.15$ & $\sim 100$\\
\object{NGC\,3766}      & $0.25\pm0.02$ & $11.50\pm0.15$ & $6-28$\\
\object{NGC\,4755}      & $0.30\pm0.04$ & $12.10\pm0.22$ & $\sim 8$\\
\object{NGC\,6025}      & $0.34\pm0.02$ & $ 9.25\pm0.17$ & $40-68$\\
\object{NGC\,6087}      & $0.35\pm0.03$ & $ 9.00\pm0.19$ & $\sim 55$\\
\object{NGC\,6250}      & $0.38\pm0.16$ & $10.55\pm0.33$ & $6$\\
\object{NGC\,6383}      & $0.51\pm0.03$ & $ 9.61\pm0.38$ & $3-10$\\
\object{NGC\,6530}      & $0.26\pm0.05$ & $11.76\pm0.20$ & $4-6$\\
\hline
\end{tabular}
\end{center}
\end{table}

In addition, from our whole spectroscopic study, we find six new Be stars and fifteen late B stars with circumstellar envelopes out off an emission phase (named Bdd stars).
The Be star population represents the $21\%$ of the sample of cluster stars, which is larger than the current population of  field Be stars ($17\%$), while the  Bdd star group represents $\sim 1/3$ of the Be star population.

We also identify four blue straggler candidates:  \object{HD\,305\,296} (in Collinder\,223), \object{NGC\,6025\,12}, \object{NGC\,6087\,12}, and \object{NGC\,6530\,7}.

The maximum distribution of Be stars per spectral subtype presents a maximum at the B2-B4 spectral type in young and intermediate-age open clusters whereas this maximum is at the  B6-B8 type in the old clusters.
In addition, we find that $\sim 30\%$ of the cluster B stars have circumstellar envelopes (Be + Bdd stars).

Finally, our result supports  the statement that  the Be phenomenon occurs in very different evolutionary states and occupy the whole main sequence band.
On the other hand, we find clearly indication of an enhancement of the Be phenomenon with age which is confirmed by the increase of the fraction number of Be stars at the spectral types B2-B4 in the age interval $10-40$~Myr. We also find the appearance of an important number of low massive (later than B4) stars  with circumstellar envelopes and out of emission phase  (named Bdd stars) in the old open clusters. This last group was detected because of the presence of a second Balmer discontinuity.


\begin{acknowledgements}
We would like to thank our anonymous Referee for his/her careful reading and valuable comments and suggestions that certainly help to improve this manuscript.
YJA: Visiting Astronomer, Complejo Astron\'omico El Leoncito operated under agreement between the Consejo Nacional de Investigaciones Cient\'ificas y T\'ecnicas de la Rep\'ublica Argentina and the National Universities of La Plata, C\'ordoba and San Juan.
This research has made use of the SIMBAD database, operated at CDS, Strasbourg, France, and WEBDA database, operated at the Department of Theoretical Physics and Astrophysics of the Masaryk University.
This work was granted by the National University of La Plata (j\'ovenes investigadores 2016).
LC thanks financial support from  CONICET (PIP 0177) and the Universidad Nacional de La Plata (Programa de Incentivos G11/137), Argentina.
\end{acknowledgements}



\onltab{
\begin{table*}
\begin{center}
\tabcolsep 3pt
\caption{\object{NGC\,6087}: Stellar fundamental parameters from the BCD system. Nomenclature according to \citet{Fernie1961} and \citet{Breger1966}.\label{BCD-6087}}
\begin{tabular}{clccclccccc}
\hline
\hline
ID & ~~~~~Other & $D$  & $\lambda_{1}$ & $\Phi_{\rm b}$ & S.T & $T_{\rm eff}$ & $\log g$ & $M_{\rm v}$ & $M_{\rm bol}$ & $\Phi_{b}^{0}$\\
Fernie/ & ~~~designation & [dex] & [\AA] & [$\mu$] & & [K] & [dex] & [mag] & [mag] & [$\mu$]\\
Breger &&&&&&&&&&\\
\hline
001 & \object{HD\,146\,271}                         & $0.23$ & $45$ & $1.37$ & B4IV                     & $16\,971\pm   744$                  & $3.61\pm0.31$                  & $-1.81\pm0.52$                  & $-3.10\pm0.48$                  & $0.75\pm0.01$\\
007 & \object{HD\,146\,483}\tablefootmark{dd}       & $0.30$ & $55$ & $1.18$ & B6Ve                     & $14\,162\pm   521$                  & $4.11\pm0.17$                  & $-0.41\pm0.22$                  & $-1.44\pm0.29$                  & $0.79\pm0.01$\\
008 & \object{HD\,146\,448}                         & $0.26$ & $53$ & $1.14$ & B5V                      & $15\,765\pm   600$                  & $4.01\pm0.21$                  & $-0.94\pm0.35$                  & $-2.09\pm0.33$                  & $0.77\pm0.01$\\
009 & \object{HD\,146\,484}\tablefootmark{dd}       & $0.30$ & $61$ & $1.13$ & B6Ve\tablefootmark{Be}   & $13\,810\pm   511$                  & $4.24\pm0.10$                  & $-0.31\pm0.26$                  & $-1.31\pm0.23$                  & $0.80\pm0.02$\\
010 & \object{HD\,146\,324}\tablefootmark{dd}       & $0.26$ & $44$ & $1.39$ & B5IIIe\tablefootmark{Be} & $16\,250\pm   631$                  & $3.58\pm0.31$                  & $-1.69\pm0.51$                  & $-2.75\pm0.46$                  & $0.76\pm0.01$\\
011 & \object{HD\,146\,294}\tablefootmark{dd}       & $0.24$ & $53$ & $1.25$ & B4V                      & $16\,604\pm   782$                  & $4.00\pm0.22$                  & $-1.29\pm0.37$                  & $-2.50\pm0.37$                  & $0.75\pm0.01$\\
013 & \object{HD\,146\,261}                         & $0.26$ & $53$ & $1.37$ & B5V                      & $16\,159\pm   506$                  & $4.01\pm0.21$                  & $-1.07\pm0.36$                  & $-2.24\pm0.34$                  & $0.76\pm0.01$\\
014 & \object{CPD$-$57\,7791}\tablefootmark{dd}     & $0.35$ & $58$ & $1.10$ & B8V\tablefootmark{Be}    & $12\,273\pm   447$                  & $4.20\pm0.11$                  & $ 0.03\pm0.25$                  & $-0.84\pm0.21$                  & $0.83\pm0.02$\\
015 & \object{HD\,146\,204}                         & $0.30$ & $59$ & $1.23$ & B6V                      & $14\,783\pm   663$                  & $4.22\pm0.13$                  & $-0.35\pm0.26$                  & $-1.39\pm0.22$                  & $0.79\pm0.01$\\
022 & \object{HD\,146\,531}                         & $0.18$ & $53$ & $1.35$ & B2IVe\tablefootmark{Be}  & $20\,945\pm1\,236$                  & $3.92\pm0.26$                  & $-2.16\pm0.41$                  & $-3.77\pm0.40$                  & $0.72\pm0.01$\\
025 & \object{CPD$-$57\,7817}                       & $0.30$ & $64$ & $1.26$ & B7V\tablefootmark{CP2}   & $13\,879\pm   485$                  & $4.33\pm0.10$:                 & $-0.41\pm0.19$                  & $-1.30\pm0.19$                  & $0.80\pm0.02$\\
033 & \object{GEN\#\,+2.60870033}                   & $0.52$ & $71$ & $1.10$ & A2:V:                    & $ 9\,063\pm   348$:                 & $4.34\pm0.10$:                 & $ 1.15\pm0.50$:                 & $ 1.10\pm0.27$\tablefootmark{a} & $1.14\pm0.06$\\
035 & \object{HD\,146\,428}                         & $0.37$ & $77$ & $1.79$ & A8:VI:                   & $ 7\,652\pm   176$\tablefootmark{a} & $4.32\pm0.01$\tablefootmark{a} & $ 2.40\pm0.15$\tablefootmark{a} & $ 2.29\pm0.15$\tablefootmark{a} & $1.54\pm0.03$\\
036 & \object{GEN\#\,+2.60870036}                   & $0.32$ & $57$ & $1.30$ & B6V                      & $13\,673\pm   592$                  & $4.17\pm0.14$                  & $-0.27\pm0.23$                  & $-1.16\pm0.24$                  & $0.80\pm0.01$\\
101 & \object{GEN\#\,+2.60870101}\tablefootmark{dd} & $0.44$ & $73$ & $1.10$ & A1:VI:                   & $10\,355\pm   336$:                 & $4.50\pm0.10$:                 & $ 0.50\pm0.43$:                 & $ 0.73\pm0.32$\tablefootmark{a} & $1.00\pm0.04$\\
128 & \object{CD$-$57\,6341}                        & $0.23$ & $49$ & $1.41$ & B4IV                     & $17\,694\pm   779$                  & $3.82\pm0.27$                  & $-1.60\pm0.47$                  & $-3.04\pm0.45$                  & $0.75\pm0.01$\\
129 & \object{GEN\#\,+2.60870129}                   & $0.49$ & $72$ & $1.60$ & A2:V:                    & $10\,137\pm   625$:                 & $4.41\pm0.10$:                 & $ 1.15\pm0.50$:                 & $ 1.10\pm0.27$\tablefootmark{a} & $1.10\pm0.07$\\
156 & \object{CD$-$57\,6346}\tablefootmark{dd}      & $0.26$ & $71$ & $1.06$ & B6:V:                    & $15\,605\pm   761$:                 & $4.35\pm0.05$:                 & $-0.95\pm0.16$:                 & $-1.89\pm0.20$:                 & $0.78\pm0.02$\\
\hline
\end{tabular}
\tablefoot{
The symbol : is used to indicate extrapolated values.
$g$ is the stellar surface gravity given in cm\,s$^{-2}$.
\tablefoottext{dd}{Star with a SBD.}
\tablefoottext{Be}{Known Be star.}
\tablefoottext{CP2}{Chemically peculiar star.}
\tablefoottext{a}{Values interpolated from \citet{Cox2000}.}
}
\end{center}
\end{table*}
}

\onltab{
\begin{table*}
\begin{center}
\tabcolsep 3pt
\caption{\object{NGC\,6087}: Physical parameters and distance to the stars.\label{Dist-6087}}
\begin{tabular}{clcccccccc}
\hline
\hline
ID & ~~~Other & $m_{\rm v}$ & $E(B-V)$ & $(m_{\rm v} - M_{\rm v})_{0}$ & $AD$ & $p$ & $\log {\cal L}_{\star}$ & $M_{\star}$ & $age$\\
Fernie/ & ~~~~designation & [mag] & [mag] & [mag] & [arcmin] & [\%] & [${\cal L}_{\odot}$] & [$M_{\odot}$] & [Myr] \\
Breger &&&&&&&&\\
\hline                                      
001 &  \object{HD\,146\,271}                         & $ 8.35$ & $ 0.46\pm0.01$ & $ 8.7\pm0.5$                    & $14.43$ & $  2.1$ & $3.3\pm0.4$ & $6\pm1$   & $ 45\pm  9$\\
007 &  \object{HD\,146\,483}\tablefootmark{dd}       & $ 8.29$ & $ 0.26\pm0.01$ & $ 7.9\pm0.2$\tablefootmark{pm}  & $ 6.66$ & $  0.0$ & $2.5\pm0.3$ & $4\pm0.2$ & $ 74\pm 27$\\
008 &  \object{HD\,146\,448}                         & $ 9.02$ & $ 0.28\pm0.01$ & $ 9.1\pm0.4$                    & $ 5.53$ & $ 40.8$ & $2.8\pm0.3$ & $5\pm0.4$ & $ 55\pm 15$\\
009 &  \object{HD\,146\,484}\tablefootmark{dd}       & $ 9.48$ & $ 0.23\pm0.01$ & $ 9.1\pm0.3$                    & $ 6.19$ & $ 40.8$ & $2.3\pm0.2$ & $4\pm0.2$ & $ 64\pm 37$\\
010 &  \object{HD\,146\,324}\tablefootmark{dd}       & $ 7.92$ & $ 0.43\pm0.01$ & $ 8.3\pm0.5$                    & $ 0.27$ & $  0.0$ & $3.2\pm0.4$ & $6\pm1$   & $ 51\pm 10$\\
011 &  \object{HD\,146\,294}\tablefootmark{dd}       & $ 9.43$ & $ 0.38\pm0.01$ & $ 9.6\pm0.4$                    & $ 1.62$ & $  0.0$ & $2.9\pm0.3$ & $5\pm0.4$ & $ 46\pm 13$\\
013 &  \object{HD\,146\,261}                         & $ 9.39$ & $ 0.46\pm0.01$ & $ 9.0\pm0.4$                    & $ 7.35$ & $100.0$ & $2.9\pm0.3$ & $5\pm0.4$ & $ 49\pm 13$\\
014 &  \object{CPD$-$57\,7791}\tablefootmark{dd}     & $ 9.70$ & $ 0.20\pm0.02$ & $ 9.0\pm0.3$                    & $ 5.87$ & $100.0$ & $2.1\pm0.2$ & $3\pm0.2$ & $121\pm 53$\\
015 &  \object{HD\,146\,204}                         & $10.19$ & $ 0.33\pm0.01$ & $ 9.5\pm0.3$                    & $ 7.71$ & $  0.1$ & $2.5\pm0.2$ & $4\pm0.2$ & $ 41\pm 27$\\
022 &  \object{HD\,146\,531}                         & $ 9.69$ & $ 0.48\pm0.01$ & $10.4\pm0.4$\tablefootmark{pm}  & $16.40$ & $  0.0$ & $3.5\pm0.4$ & $8\pm1$   & $ 19\pm  7$\\
025 &  \object{CPD$-$57\,7817}                       & $ 9.82$ & $ 0.34\pm0.01$ & $ 9.2\pm0.2$                    & $ 1.60$ & $ 10.8$ & $2.3\pm0.2$ & $4\pm0.2$ & $ 48\pm 42$\\
033 &  \object{GEN\#\,+2.60870033}                   & $11.97$ & $\cdots$       & $ 6.5\pm0.2$\tablefootmark{nm}  & $ 7.57$ & $  0.0$ & $1.3\pm0.3$ & $2\pm0.1$ & $264\pm240$\\
035 &  \object{HD\,146\,428}                         & $ 9.95$ & $ 0.14\pm0.02$ & $ 7.1\pm0.2$\tablefootmark{nm}  & $ 4.49$ & $  0.0$ & $0.9\pm0.2$ & $2\pm0.1$ & $458\pm459$\\
036 &  \object{GEN\#\,+2.60870036}                   & $10.36$ & $ 0.38\pm0.01$ & $ 9.5\pm0.2$                    & $ 5.51$ & $  0.1$ & $2.3\pm0.2$ & $4\pm0.2$ & $ 75\pm 35$\\
101 &  \object{GEN\#\,+2.60870101}\tablefootmark{dd} & $11.25$ & $ 0.05\pm0.01$ & $10.6\pm0.4$\tablefootmark{pm}  & $ 8.62$ & $  0.0$ & $1.4\pm0.3$ & $2\pm0.2$ & $103\pm121$\\
128 &  \object{CD$-$57\,6341}                        & $ 8.71$ & $ 0.50\pm0.01$ & $ 8.8\pm0.5$                    & $ 0.57$ & $ 10.8$ & $3.2\pm0.4$ & $6\pm0.6$ & $ 39\pm  7$\\
129 &  \object{GEN\#\,+2.60870129}                   & $ 9.75$ & $ 0.27\pm0.04$ & $ 7.8\pm0.5$\tablefootmark{pm}  & $ 0.76$ & $  0.0$ & $1.4\pm0.3$ & $2\pm0.1$ & $ 78\pm121$\\
156 &  \object{CD$-$57\,6346}\tablefootmark{dd}      & $ 9.14$ & $ 0.20\pm0.01$ & $ 9.5\pm0.2$                    & $ 4.51$ & $  0.1$ & $2.5\pm0.2$ & $4\pm0.3$ & $ 26\pm 28$\\
\hline
\end{tabular}
\tablefoot{
$m_{\rm v}$ values obtained from SIMBAD database.
The distances of the stars with negative color gradients were determined using the distance modulus assigned to the cluster.
\tablefoottext{dd}{Stars with a SBD.}
\tablefoottext{pm}{Probable member stars.}
\tablefoottext{nm}{Non-member stars.}
}
\end{center}
\end{table*}
}


\onltab{
\begin{table*}
\begin{center}
\tabcolsep 3pt
\caption{\object{NGC\,6250}: Stellar fundamental parameters from the BCD system. Nomenclature according to \citet{Moffat1975} and \citet{Herbst1977}.\label{BCD-6250}}
\begin{tabular}{clccclccccc}
\hline
\hline
ID & ~~~~~Other & $D$  & $\lambda_{1}$ & $\Phi_{\rm b}$ & S.T & $T_{\rm eff}$ & $\log g$ & $M_{\rm v}$ & $M_{\rm bol}$ & $\Phi_{b}^{0}$\\
Moffat/ & ~~~designation & [dex] & [\AA] & [$\mu$] & & [K] & [dex] & [mag] & [mag] & [$\mu$]\\
Herbst &&&&&&&&&&\\
\hline
01       & \object{HD\,152\,853}\tablefootmark{dd} & $0.09$ & $52$ & $1.03$ & B0III                   & $28\,418\pm2\,781$                  & $3.47\pm0.39$                  & $-3.75\pm0.60$                  & $-5.91\pm0.50$                  & $0.69\pm0.00$\\
02       & \object{HD\,152\,799}                   & $0.16$ & $76$ & $0.73$ & B2VI:                   & $22\,919\pm1\,577$                  & $4.31\pm0.04$:                 & $-1.86\pm0.15$:                 & $-3.72\pm0.38$:                 & $0.71\pm0.01$\\
03       & \object{HD\,152\,822}                   & $0.22$ & $68$ & $0.93$ & B4VI:\tablefootmark{VB} & $18\,236\pm   625$                  & $4.30\pm0.07$                  & $-1.32\pm0.11$                  & $-2.62\pm0.26$                  & $0.75\pm0.01$\\
04       & \object{HD\,152\,917}                   & $0.33$ & $70$ & $1.63$ & F1V                     & $ 6\,930\pm    70$\tablefootmark{a} & $4.34\pm0.00$\tablefootmark{a} & $ 3.58\pm0.02$\tablefootmark{a} & $ 3.46\pm0.02$\tablefootmark{a} & $1.71\pm0.07$\\
17       & \object{HD\,329\,271}                   & $0.22$ & $64$ & $1.36$ & B4V                     & $18\,074\pm   623$                  & $4.26\pm0.09$                  & $-1.27\pm0.19$                  & $-2.44\pm0.32$                  & $0.75\pm0.01$\\
18       & \object{CD$-$45\,11088}                 & $0.42$ & $70$ & $0.78$ & A0VI:                   & $10\,828\pm   367$                  & $4.41\pm0.11$:                 & $ 0.19\pm0.38$                  & $ 0.24\pm0.35$:                 & $0.94\pm0.04$\\
21       & \object{HD\,152\,743}                   & $0.12$ & $52$ & $0.75$ & B1IV                    & $24\,600\pm2\,371$                  & $3.76\pm0.34$                  & $-3.12\pm0.52$                  & $-5.20\pm0.47$                  & $0.70\pm0.01$\\
22       & \object{HD\,152\,706}                   & $0.46$ & $71$ & $0.79$ & A1V                     & $10\,084\pm   376$                  & $4.40\pm0.10$:                 & $ 0.46\pm0.25$:                 & $ 0.65\pm0.36$:                 & $1.01\pm0.04$\\
32       & \object{HD\,152\,687}                   & $0.44$ & $58$ & $1.03$ & B9Ve                    & $10\,722\pm   275$                  & $4.10\pm0.16$                  & $ 0.59\pm0.22$:                 & $-0.31\pm0.33$:                 & $0.92\pm0.03$\\
33       & \object{HD\,152\,561}                   & $0.41$ & $56$ & $1.04$ & B9Ve                    & $11\,420\pm   284$                  & $4.10\pm0.16$                  & $ 0.27\pm0.25$                  & $-0.58\pm0.28$                  & $0.88\pm0.02$\\
34       & \object{HD\,153\,073}                   & $0.08$ & $61$ & $0.90$ & B0V\tablefootmark{p}    & $31\,435\pm1\,884$                  & $4.01\pm0.24$                  & $-3.46\pm0.41$                  & $-6.19\pm0.44$                  & $0.67\pm0.01$\\
35       & \object{HD\,152\,979}                   & $0.09$ & $57$ & $0.99$ & B0IVe\tablefootmark{Be} & $31\,127\pm2\,571$                  & $3.89\pm0.30$                  & $-3.59\pm0.56$                  & $-6.12\pm0.47$                  & $0.68\pm0.01$\\
37       & \object{HD\,329\,379}                   & $0.06$ & $48$ & $2.50$ & B0II                    & $30\,363\pm4\,229$                  & $2.92\pm0.34$                  & $-5.11\pm0.88$                  & $-7.26\pm0.54$                  & $0.67\pm0.01$\\
$\cdots$ & \object{CD$-$49\,11096}                 & $0.40$ & $62$ & $1.59$ & B9V                     & $11\,450\pm   308$                  & $4.26\pm0.14$                  & $ 0.46\pm0.39$                  & $-0.34\pm0.29$:                 & $0.89\pm0.03$\\
$\cdots$ & \object{HD\,329\,211}                   & $0.26$ & $69$ & $0.51$ & B5VI:                   & $15\,747\pm   671$                  & $4.34\pm0.07$:                 & $-0.95\pm0.13$                  & $-1.90\pm0.18$                  & $0.77\pm0.02$\\
\hline
\end{tabular}
\tablefoot{
The symbol : is used to indicate extrapolated values.
$g$ is the stellar surface gravity given in cm\,s$^{-2}$.
\tablefoottext{dd}{Stars with a SBD.}
\tablefoottext{Be}{Known Be star.}
\tablefoottext{VB}{Reported as visual binary star.}
\tablefoottext{p}{Chemically peculiar star.}
\tablefoottext{a}{Values interpolated from \citet{Cox2000}.}
}
\end{center}
\end{table*}
}

\onltab{
\begin{table*}
\begin{center}
\tabcolsep 3pt
\caption{\object{NGC\,6250}: {Physical parameters and distance to the stars.}\label{Dist-6250}}
\begin{tabular}{clcccccccc}
\hline
\hline
ID & ~~~~~Other & $m_{\rm v}$ & $E(B-V)$ & $(m_{\rm v} - M_{\rm v})_{0}$ & $AD$ & $p$ & $\log {\cal L}_{\star}$ & $M_{\star}$ & $age$\\
Moffat & ~~~designation & [mag] & [mag] & [mag] & [arcmin] & [\%] & [${\cal L}_{\odot}$] & [$M_{\odot}$] & [Myr] \\
Herbst &&&&&&&&\\
\hline                                      
01       & \object{HD\,152\,853}\tablefootmark{dd} & $ 7.94$ & $ 0.25\pm0.01$ & $10.9\pm0.6$                    & $  2.91$ & $ 15.4$ & $4.7\pm0.6$ & $20\pm12$   & $  6\pm  3$\\
02       & \object{HD\,152\,799}                   & $ 8.74$ & $ 0.01\pm0.01$ & $10.6\pm0.1$                    & $  2.85$ & $ 83.3$ & $3.4\pm0.3$ & $ 8\pm 1$   & $  2\pm  4$\\
03       & \object{HD\,152\,822}                   & $ 9.07$ & $ 0.14\pm0.01$ & $ 9.0\pm0.1$                    & $  0.58$ & $  3.3$ & $2.9\pm0.2$ & $ 6\pm 0.3$ & $ 11\pm 11$\\
04       & \object{HD\,152\,917}                   & $ 7.62$ & $\cdots$       & $ 2.9\pm0.5$\tablefootmark{nm}  & $  6.24$ & $  0.0$ & $0.5\pm0.1$ & $ 1\pm 0.1$ & $  6\pm  6$\\
17       & \object{HD\,329\,271}                   & $10.65$ & $ 0.46\pm0.01$ & $10.5\pm0.2$                    & $  3.67$ & $ 83.3$ & $2.9\pm0.3$ & $ 5\pm 0.3$ & $ 12\pm 11$\\
18       & \object{CD$-$45\,11088}                 & $11.10$ & $\cdots$       & $ 9.7\pm0.6$                    & $  4.65$ & $  0.3$ & $1.6\pm0.3$ & $ 3\pm 0.2$ & $110\pm112$\\
21       & \object{HD\,152\,743}                   & $ 9.06$ & $ 0.04\pm0.01$ & $12.1\pm0.5$\tablefootmark{pm}  & $  8.01$ & $  0.0$ & $4.1\pm0.4$ & $12\pm 2$   & $ 12\pm  5$\\
22       & \object{HD\,152\,706}                   & $10.09$ & $\cdots$       & $ 8.5\pm0.6$\tablefootmark{pm}  & $  8.93$ & $  0.0$ & $1.5\pm0.3$ & $ 2\pm 0.2$ & $148\pm152$\\
32       & \object{HD\,152\,687}                   & $ 8.69$ & $ 0.08\pm0.02$ & $ 7.9\pm0.2$\tablefootmark{pnm} & $ 15.20$ & $  0.0$ & $1.9\pm0.3$ & $ 3\pm 0.2$ & $236\pm 59$\\
33       & \object{HD\,152\,561}                   & $ 9.11$ & $ 0.12\pm0.02$ & $ 8.5\pm0.3$\tablefootmark{pm}  & $ 18.90$ & $  0.0$ & $2.1\pm0.3$ & $ 3\pm 0.2$ & $184\pm 46$\\
34       & \object{HD\,153\,073}                   & $ 9.10$ & $ 0.17\pm0.01$ & $12.0\pm0.4$\tablefootmark{pm}  & $ 15.82$ & $  0.0$ & $4.5\pm0.4$ & $17\pm 3$   & $  3\pm  2$\\
35       & \object{HD\,152\,979}                   & $ 8.16$ & $ 0.23\pm0.01$ & $11.0\pm0.6$                    & $ 15.11$ & $  7.3$ & $4.6\pm0.4$ & $17\pm 4$   & $  4\pm  3$\\
37       & \object{HD\,329\,379}                   & $ 9.70$ & $ 1.37\pm0.01$ & $10.6\pm0.9$                    & $ 35.81$ & $ 83.3$ & $ \cdots  $ & $  \cdots $ & $ \cdots  $\\
$\cdots$ & \object{CD$-$49\,11096}                 & $10.91$ & $ 0.53\pm0.02$ & $ 8.8\pm0.4$\tablefootmark{pm}  & $229.43$ & $  0.0$ & $1.9\pm0.3$ & $ 3\pm 0.2$ & $135\pm 80$\\
$\cdots$ & \object{HD\,329\,211}                   & $10.91$ & $\cdots$       & $10.7\pm0.5$                    & $ 66.09$ & $ 52.9$ & $2.6\pm0.2$ & $ 4\pm 0.3$ & $ 24\pm 25$\\
\hline
\end{tabular}
\tablefoot{
$m_{\rm v}$ values obtained from SIMBAD database.
The distances of the stars with negative color gradients were determined using the distance modulus assigned to the cluster.
\tablefoottext{dd}{Stars with a SBD.}
\tablefoottext{pm}{Probable member star.}
\tablefoottext{nm}{Non-member star.}
}
\end{center}
\end{table*}
}


\onltab{
\begin{table*}
\begin{center}
\tabcolsep 1.3pt
\caption{\object{NGC\,6383}: Stellar fundamental parameters from the BCD system. Nomenclature according to \citet{Eggen1961, The1965}, and \citet{LloydEvans1978}.\label{BCD-6383}}
\begin{tabular}{clccclccccc}
\hline
\hline
ID & ~~~~~Other & $D$  & $\lambda_{1}$ & $\Phi_{\rm b}$ & S.T & $T_{\rm eff}$ & $\log g$ & $M_{\rm v}$ & $M_{\rm bol}$ & $\Phi_{b}^{0}$\\
Eggen/The/ & ~~~designation & [dex] & [\AA] & [$\mu$] & & [K] & [dex] & [mag] & [mag] & [$\mu$]\\
Lloyd Evans &&&&&&&&&&\\
\hline
001 & \object{CD$-$32\,12935}                   & $0.01$ & $38$ & $1.37$ & O7V\tablefootmark{BE2,BEc,Em,mk} & $38\,000\pm2\,000$\tablefootmark{a} & $3.97\pm0.02$\tablefootmark{a} & $-5.10\pm0.30$\tablefootmark{a} & $-8.96\pm0.30$\tablefootmark{a} & $0.62\pm0.02$\tablefootmark{m}\\
002 & \object{CD$-$32\,12931}                   & $0.14$ & $67$ & $1.58$ & B2V                              & $24\,678\pm1\,125$                  & $4.21\pm0.08$                  & $-2.13\pm0.27$                  & $-4.16\pm0.31$                  & $0.70\pm0.01$ \\
003 & \object{CD$-$32\,12927}                   & $0.39$ & $96$ & $1.84$ & A3V\tablefootmark{VB,BE,MA,mk}   & $ 8\,727\pm   273$\tablefootmark{a} & $4.23\pm0.03$\tablefootmark{a} & $ 1.52\pm0.22$\tablefootmark{a} & $ 1.33\pm0.22$\tablefootmark{a} & $1.12\pm0.02$:\\
006 & \object{CD$-$32\,12921}                   & $0.07$ & $47$ & $1.43$ & B0II\tablefootmark{V?,IR,BS}     & $28\,750\pm3\,906$                  & $2.92\pm0.34$                  & $-5.00\pm0.92$                  & $-7.00\pm0.50$                  & $0.68\pm0.01$ \\
010 & \object{CD$-$32\,12943}                   & $0.15$ & $59$ & $1.51$ & B2V                              & $23\,510\pm1\,427$                  & $4.08\pm0.16$                  & $-2.29\pm0.34$                  & $-4.25\pm0.35$                  & $0.71\pm0.01$ \\
014 & \object{CD$-$32\,12929}                   & $0.20$ & $69$ & $1.40$ & B3VI:\tablefootmark{BE?}         & $19\,247\pm   670$                  & $4.28\pm0.08$                  & $-1.48\pm0.12$                  & $-2.71\pm0.31$                  & $0.73\pm0.01$ \\
047 & \object{CD$-$32\,12954}                   & $0.39$ & $71$ & $1.84$ & A8V                              & $ 7\,652\pm   176$\tablefootmark{a} & $4.32\pm0.01$\tablefootmark{a} & $ 2.40\pm0.15$\tablefootmark{a} & $ 2.29\pm0.15$\tablefootmark{a} & $1.54\pm0.07$ \\
057 & \object{CD$-$32\,12946}\tablefootmark{dd} & $0.49$ & $70$ & $1.73$ & A2V                              & $ 9\,573\pm   322$                  & $4.36\pm0.10$:                 & $ 0.77\pm0.23$:                 & $ 1.08\pm0.43$:                 & $1.07\pm0.05$ \\
076 & \object{CD$-$32\,12908}                   & $0.18$ & $65$ & $1.49$ & B3Ve\tablefootmark{Em}           & $20\,473\pm   822$                  & $4.23\pm0.09$                  & $-1.62\pm0.20$                  & $-3.26\pm0.29$                  & $0.72\pm0.01$ \\
083 & \object{CD$-$32\,12919}                   & $0.13$ & $66$ & $1.35$ & B1V\tablefootmark{BE?}           & $25\,644\pm1\,062$                  & $4.18\pm0.10$                  & $-2.28\pm0.28$                  & $-4.41\pm0.26$                  & $0.70\pm0.01$ \\
085 & \object{CD$-$32\,12910}                   & $0.26$ & $91$ & $2.01$ & F0V\tablefootmark{BE,mk}         & $ 7\,300\pm   163$\tablefootmark{a} & $4.34\pm0.01$\tablefootmark{a} & $ 2.70\pm0.30$\tablefootmark{a} & $ 2.61\pm0.30$\tablefootmark{a} & $1.62\pm0.02$:\\
100 & \object{CD$-$32\,12924}                   & $0.10$ & $74$ & $1.27$ & B0V\tablefootmark{BEc,BE2}       & $29\,500\pm1\,590$                  & $4.22\pm0.05$:                 & $-2.90\pm0.37$:                 & $-5.39\pm0.51$:                 & $0.68\pm0.01$ \\
\hline
\end{tabular}
\tablefoot{
The symbol : is used to indicate extrapolated values.
$g$ is the stellar surface gravity given in cm\,s$^{-2}$.
\tablefoottext{dd}{Stars with a SBD.}
\tablefoottext{mk}{Spectral types derived using the MK-system.}
\tablefoottext{BE}{Spectroscopic Binary.}
\tablefoottext{BE2}{Double line spectroscopic binary.}
\tablefoottext{BEc}{Eclipsing binary.}
\tablefoottext{BS}{Blue straggler star..}
\tablefoottext{Em}{Emission line star.}
\tablefoottext{IR}{Star with infrared excess.}
\tablefoottext{MA}{Magnetic Ap star.}
\tablefoottext{V}{Variable star.}
\tablefoottext{VB}{Visual binary.}
\tablefoottext{a}{Values interpolated from \citet{Cox2000}.}
\tablefoottext{m}{Value calculated in this work, using a relation between $\Phi^{0}_{\rm b}$ and $(B-V)_{0}$ given by \citet{Moujtahid1998}.}
}
\end{center}
\end{table*}
}

\onltab{
\begin{table*}
\begin{center}
\tabcolsep 3pt
\caption{\object{NGC\,6383}: Physical parameters and distance to the stars.\label{Dist-6383}}
\begin{tabular}{clcccccccc}
\hline
\hline
ID & ~~~~~Other & $m_{\rm v}$ & $E(B-V)$ & $(m_{\rm v} - M_{\rm v})_{0}$ & $AD$ & $p$ & $\log {\cal L}_{\star}$ & $M_{\star}$ & $age$\\
Eggen/The/ & ~~~designation & [mag] & [mag] & [mag] & [arcmin] & [\%] & [${\cal L}_{\odot}$] & [$M_{\odot}$] & [Myr] \\
Lloyd Evans &&&&&&&&\\
\hline                                      
001 & \object{CD$-$32\,12935}                   & $ 5.68$ & $0.53\pm0.01$ & $ 9.4\pm0.3$                    & $ 1.47$ & $ 76.5$ & $5.4\pm0.3$ & $35\pm5$     & $  3\pm0.4$\\
002 & \object{CD$-$32\,12931}                   & $10.43$ & $0.66\pm0.01$ & $10.5\pm0.3$                    & $ 2.76$ & $  1.9$ & $3.6\pm0.3$ & $ 9\pm1$     & $  3\pm  3$\\
003 & \object{CD$-$32\,12927}                   & $10.30$ & $0.49\pm0.02$ & $ 7.3\pm0.2$\tablefootmark{pm}  & $ 3.61$ & $  0.0$ & $1.3\pm0.2$ & $ 2\pm0.1$   & $409\pm206$\\
006 & \object{CD$-$32\,12921}                   & $ 9.07$ & $0.51\pm0.01$ & $12.5\pm0.9$\tablefootmark{pm}  & $ 5.49$ & $  0.0$ & $ \cdots  $ & $\cdots $    & $ \cdots  $\\
010 & \object{CD$-$32\,12943}                   & $10.00$ & $0.60\pm0.01$ & $10.4\pm0.3$                    & $ 6.64$ & $  3.2$ & $3.7\pm0.3$ & $ 9\pm1$     & $  9\pm  6$\\
014 & \object{CD$-$32\,12929}                   & $ 9.85$ & $0.50\pm0.01$ & $ 9.8\pm0.1$                    & $ 2.22$ & $ 55.2$ & $3.0\pm0.3$ & $ 6\pm0.5$   & $  6\pm  7$\\
047 & \object{CD$-$32\,12954}                   & $ 9.99$ & $0.23\pm0.05$ & $ 6.9\pm0.2$\tablefootmark{pnm} & $ 6.81$ & $  0.0$ & $0.9\pm0.2$ & $ 2\pm0.1$   & $458\pm459$\\
057 & \object{CD$-$32\,12946}\tablefootmark{dd} & $10.64$ & $0.49\pm0.04$ & $ 8.3\pm0.2$                    & $ 4.68$ & $  0.2$ & $1.4\pm0.3$ & $ 2\pm0.1$   & $164\pm173$\\
076 & \object{CD$-$32\,12908}                   & $ 9.83$ & $0.58\pm0.01$ & $ 9.7\pm0.2$                    & $11.71$ & $ 76.5$ & $3.2\pm0.3$ & $ 7\pm1$     & $  9\pm  7$\\
083 & \object{CD$-$32\,12919}                   & $ 9.53$ & $0.49\pm0.01$ & $10.3\pm0.3$                    & $ 7.03$ & $  5.5$ & $3.8\pm0.3$ & $10\pm1$     & $  3\pm  3$\\
085 & \object{CD$-$32\,12910}                   & $ 9.53$ & $0.27\pm0.02$ & $ 5.0\pm0.3$\tablefootmark{pnm} & $10.28$ & $  0.0$ & $0.8\pm0.3$ & $ 1.5\pm0.1$ & $467\pm536$\\
100 & \object{CD$-$32\,12924}                   & $ 8.97$ & $0.40\pm0.01$ & $10.6\pm0.4$                    & $ 6.20$ & $  1.1$ & $4.1\pm0.3$ & $13\pm2$     & $  2\pm  2$\\
\hline
\end{tabular}
\tablefoot{
$m_{\rm v}$ values obtained from SIMBAD database.
\tablefoottext{dd}{Stars with a SBD.}
\tablefoottext{pm}{Probable member star.}
\tablefoottext{pnm}{Probable non-member star.}
}
\end{center}
\end{table*}
}


\onltab{
\begin{table*}
\begin{center}
\tabcolsep 1pt
\caption{\object{NGC\,6530}: Stellar fundamental parameters from the BCD system. Nomenclature according to \citet{Walker1957}, and \citet{Kilambi1977}.\label{BCD-6530}}
\begin{tabular}{clccclccccc}
\hline
\hline
ID & ~~~~~~Other & $D$  & $\lambda_{1}$ & $\Phi_{\rm b}$ & S.T & $T_{\rm eff}$ & $\log g$ & $M_{\rm v}$ & $M_{\rm bol}$ & $\Phi_{b}^{0}$\\
Walker/ & ~~~~designation & [dex] & [\AA] & [$\mu$] & & [K] & [dex] & [mag] & [mag] & [$\mu$]\\
Kilambi &&&&&&&&&&\\
\hline
007      & \object{HD\,164\,794}                     & $0.01$ & $ 9$     & $1.10$ & O4V((f))\tablefootmark{so,V,BE,Em}   & $42\,857\pm1\,000$\tablefootmark{n}  & $3.92\pm0.30$\tablefootmark{n}  & $-5.50\pm0.50$\tablefootmark{n}  & $-9.41\pm0.34$\tablefootmark{n}  & $0.61\pm0.03$\tablefootmark{m}\\
009      & \object{HD\,164\,816}                     & $0.05$ & $\cdots$ & $0.66$ & O9III\tablefootmark{mk,V,BE2,p,Em}   & $31\,846\pm1\,000$\tablefootmark{n}  & $3.53\pm0.30$\tablefootmark{n}  & $-5.25\pm0.50$\tablefootmark{n}  & $-8.29\pm0.50$\tablefootmark{n}  & $0.93\pm0.03$\tablefootmark{m}\\
032      & \object{GEN\#\,+2.65300032}               & $0.14$ & $60$     & $0.81$ & B2V\tablefootmark{V,Em}              & $24\,964\pm1\,539$                   & $4.08\pm0.15$                   & $-2.33\pm0.33$                   & $-4.49\pm0.36$                   & $0.70\pm0.00$\\
042      & \object{HD\,315\,032}                     & $0.08$ & $58$     & $0.95$ & B0IV\tablefootmark{V,H$\alpha$,Be}   & $31\,688\pm2\,368$                   & $3.90\pm0.30$                   & $-3.74\pm0.54$                   & $-6.23\pm0.42$                   & $0.68\pm0.00$\\
043      & \object{HD\,315\,026}                     & $0.09$ & $66$     & $0.86$ & B0V\tablefootmark{V,Vv,ST}           & $29\,534\pm1\,271$                   & $4.15\pm0.11$                   & $-2.92\pm0.40$                   & $-5.82\pm0.41$                   & $0.68\pm0.00$\\
045      & \object{HD\,164\,865}\tablefootmark{dd}   & $0.14$ & $21$     & $2.56$ & B4Ia\tablefootmark{V,H$\alpha$}      & $15\,208\pm1\,636$                   & $2.44\pm0.07$\tablefootmark{a}  & $-6.20\pm0.03$\tablefootmark{a}  & $-7.15\pm0.15$\tablefootmark{a}  & $0.76\pm0.03$\\
055      & \object{HD\,315\,023}                     & $0.10$ & $78$     & $0.80$ & B0VI:\tablefootmark{V,Em}            & $28\,882\pm1\,511$                   & $3.94\pm0.00$\tablefootmark{a}  & $-4.00\pm0.64$\tablefootmark{a}  & $-7.16\pm0.64$\tablefootmark{a}  & $0.69\pm0.00$\\
056      & \object{CD$-$24\,13829}                   & $0.07$ & $64$     & $1.05$ & B0V\tablefootmark{V,Vv,BE2,Em}       & $33\,494\pm1\,836$                   & $4.06\pm0.20$                   & $-3.86\pm0.36$                   & $-6.70\pm0.50$                   & $0.67\pm0.01$\\
059      & \object{HD\,315\,033}                     & $0.13$ & $70$     & $1.19$ & B1V\tablefootmark{V,p}               & $26\,656\pm1\,163$                   & $4.21\pm0.08$                   & $-2.41\pm0.29$                   & $-4.71\pm0.38$                   & $0.70\pm0.01$\\
060      & \object{LSS\,4615}                        & $0.08$ & $62$     & $1.04$ & B0V\tablefootmark{V,Vr,Em}           & $31\,948\pm1\,922$                   & $4.03\pm0.20$                   & $-3.54\pm0.43$                   & $-6.24\pm0.36$                   & $0.67\pm0.01$\\
061      & \object{ALS\,16976}                       & $0.13$ & $66$     & $0.92$ & B1V\tablefootmark{V,Em}              & $26\,483\pm1\,034$                   & $4.17\pm0.10$                   & $-2.41\pm0.29$                   & $-4.61\pm0.39$                   & $0.70\pm0.00$\\ 
065      & \object{HD\,164\,906}\tablefootmark{dd}   & $0.03$ & $51$     & $1.27$ & O\tablefootmark{V,Vr,BE,p,Be}        & $35\,655\pm3\,828$:                  & $2.97\pm0.36$:                  & $-5.70\pm0.88$                   & $-8.24\pm0.70$:                  & $0.66\pm0.01$:\\
066      & \object{CD$-$24\,13831}                   & $0.10$ & $\cdots$ & $0.95$ & B1V\tablefootmark{V,He,Em}           & $25\,450\pm4\,550$\tablefootmark{a}  & $3.94\pm0.00$\tablefootmark{a}  & $-3.23\pm0.77$\tablefootmark{a}  & $-5.98\pm0.77$\tablefootmark{a}  & $0.62\pm0.02$\tablefootmark{m}\\
068      & \object{CD$-$24\,13830}                   & $0.08$ & $60$     & $1.70$ & B0Ve\tablefootmark{V}                & $31\,622\pm2\,282$                   & $3.97\pm0.27$                   & $-3.47\pm0.60$                   & $-5.98\pm0.44$                   & $0.68\pm0.01$\\
070      & \object{ALS\,18783}                       & $0.13$ & $73$     & $0.67$ & B2V\tablefootmark{Em}                & $26\,044\pm1\,632$                   & $4.25\pm0.06$:                  & $-2.17\pm0.32$:                  & $-4.44\pm0.51$:                  & $0.70\pm0.01$\\
073      & \object{HD\,315\,031}                     & $0.07$ & $56$     & $1.08$ & B0IV\tablefootmark{Vv,ST}            & $33\,249\pm2\,996$                   & $3.81\pm0.39$                   & $-3.92\pm0.61$                   & $-6.62\pm0.42$                   & $0.68\pm0.01$\\
076      & \object{HD\,315\,024}                     & $0.08$ & $58$     & $0.83$ & B0IV\tablefootmark{Em}               & $32\,189\pm2\,391$                   & $3.89\pm0.30$                   & $-3.76\pm0.48$                   & $-6.20\pm0.44$                   & $0.68\pm0.01$\\
080      & \object{CD$-$24\,13837}                   & $0.05$ & $42$     & $0.82$ & O9Ib\tablefootmark{Vr,Em}            & $25\,977\pm4\,032$                   & $2.71\pm0.14$:                  & $-6.22\pm0.89$:                  & $-7.79\pm0.81$                   & $0.68\pm0.01$\\
085      & \object{HD\,164\,933}                     & $0.07$ & $\cdots$ & $1.16$ & B0III\tablefootmark{Vv,mk}           & $29\,732\pm1\,000$:\tablefootmark{n} & $3.49\pm0.30$:\tablefootmark{n} & $-5.11\pm0.50$:\tablefootmark{n} & $-7.95\pm0.50$:\tablefootmark{n} & $1.09\pm0.03$\tablefootmark{m}\\
086      & \object{CD$-$24\,13840}                   & $0.17$ & $88$     & $0.70$ & B2V\tablefootmark{BE2,BEc,Em}        & $20\,900\pm3\,225$\tablefootmark{a}  & $3.94\pm0.00$\tablefootmark{a}  & $-2.45\pm0.60$\tablefootmark{a}  & $-4.80\pm0.60$\tablefootmark{a}  & $0.62\pm0.02$\tablefootmark{m}\\
093      & \object{HD\,315\,021}                     & $0.08$ & $62$     & $1.06$ & B0V\tablefootmark{Vv}                & $31\,086\pm1\,967$                   & $4.05\pm0.20$                   & $-3.43\pm0.53$                   & $-5.91\pm0.41$                   & $0.68\pm0.01$\\
100      & \object{HD\,164\,947}                     & $0.16$ & $77$     & $1.04$ & B2VI\tablefootmark{VB,V,BE,Be}       & $22\,166\pm1\,268$                   & $3.94\pm0.00$\tablefootmark{a}  & $-2.45\pm0.60$\tablefootmark{a}  & $-4.80\pm0.60$\tablefootmark{a}  & $0.72\pm0.01$\\
$\cdots$ & \object{LSS\,4627}                        & $0.13$ & $64$     & $0.58$ & B1V                                  & $26\,483\pm1\,241$                   & $4.14\pm0.11$                   & $-2.57\pm0.31$                   & $-4.64\pm0.33$                   & $0.70\pm0.00$\\
\hline
\end{tabular}
\tablefoot{
The symbol : is used to indicate extrapolated values.
$g$ is the stellar surface gravity given in cm\,s$^{-2}$.
\tablefoottext{dd}{Stars with a SBD.}
\tablefoottext{so}{Spectral types derived using ``The Galactic O-Star Spectroscopic Survey''  \citep{Sota2011}.}
\tablefoottext{mk}{Spectral types derived using the MK-system.}
\tablefoottext{BE}{Spectroscopic binary.}
\tablefoottext{BE2}{Double line spectroscopic binary.}
\tablefoottext{BEc}{Eclipsing binary.}
\tablefoottext{Be}{Known Be star.}
\tablefoottext{Em}{Emission line star.}
\tablefoottext{H$\alpha$}{Star with emission in H$\alpha$.}
\tablefoottext{He}{Helium star.}
\tablefoottext{p}{Chemically peculiar star.}
\tablefoottext{ST}{Triple system.}
\tablefoottext{V}{Variable star.}
\tablefoottext{VB}{Visual binary.}
\tablefoottext{Vr}{Radial velocity variable.}
\tablefoottext{Vv}{Probable variable like Vega star.}
\tablefoottext{a}{Values interpolated from \citet{Cox2000}.}
\tablefoottext{m}{Value calculated in this work, using a relation between $\Phi^{0}_{\rm b}$ and $(B-V)_{0}$ given by \citet{Moujtahid1998}.}
\tablefoottext{n}{Stellar parameter given by \citet{Martins2005}.}
}
\end{center}
\end{table*}
}

\onltab{
\begin{table*}
\begin{center}
\tabcolsep 3pt
\caption{\object{NGC\,6530}: Physical parameters and distance to the stars.\label{Dist-6530}}
\begin{tabular}{clcccccccc}
\hline
\hline
ID & ~~~~~Other & $m_{\rm v}$ & $E(B-V)$ & $(m_{\rm v} - M_{\rm v})_{0}$ & $AD$ & $p$ & $\log {\cal L}_{\star}$ & $M_{\star}$ & $age$\\
Walker/ & ~~~designation & [mag] & [mag] & [mag] & [arcmin] & [\%] & [${\cal L}_{\odot}$] & [$M_{\odot}$] & [Myr] \\
Kilambi &&&&&&&&\\
\hline                                      
007      & \object{HD\,164\,794}                     & $ 5.93$ &  $ 0.33\pm0.02$ & $10.4\pm0.5$\tablefootmark{pm}  & $  8.78$ & $   0.0$ & $5.7\pm0.4$ & $54\pm13$ & $1\pm0.6$\\
009      & \object{HD\,164\,816}                     & $ 7.09$ &  $ \cdots     $ & $11.5\pm0.5$\tablefootmark{pm}  & $  8.25$ & $  12.4$ & $5.3\pm0.4$ & $31\pm10$ & $5\pm0.8$\\
032      & \object{GEN\#\,+2.65300032}               & $10.37$ &  $ 0.08\pm0.01$ & $12.5\pm0.3$                    & $  4.53$ & $   0.0$ & $3.8\pm0.3$ & $10\pm 1$ & $  7\pm5$\\
042      & \object{HD\,315\,032}                     & $ 9.18$ &  $ 0.20\pm0.01$ & $12.3\pm0.5$                    & $  4.14$ & $   0.3$ & $4.6\pm0.4$ &$186\pm 4$ & $  4\pm2$\\
043      & \object{HD\,315\,026}                     & $ 9.01$ &  $ 0.13\pm0.01$ & $11.5\pm0.4$                    & $  7.84$ & $  12.4$ & $4.2\pm0.3$ & $14\pm 1$ & $  3\pm2$\\
045      & \object{HD\,164\,865}\tablefootmark{dd}   & $ 7.74$ &  $ 1.23\pm0.02$ & $10.1\pm0.1$\tablefootmark{pm}  & $ 11.10$ & $   0.0$ & $4.8\pm0.2$ & $17\pm 3$ & $ 10\pm2$\\
055      & \object{HD\,315\,023}                     & $10.07$ &  $ 0.09\pm0.01$ & $13.8\pm0.6$\tablefootmark{pm}  & $  7.95$ & $   0.0$ & $4.6\pm0.4$ & $17\pm 2$ & $  7\pm1$\\
056      & \object{CD$-$24\,13829}                   & $ 9.08$ &  $ 0.29\pm0.01$ & $12.1\pm0.4$                    & $  2.22$ & $   4.8$ & $4.7\pm0.4$ & $19\pm 3$ & $  3\pm2$\\
059      & \object{HD\,315\,033}                     & $ 8.90$ &  $ 0.34\pm0.01$ & $10.3\pm0.3$\tablefootmark{pm}  & $  5.10$ & $   0.0$ & $3.9\pm0.3$ & $10\pm 1$ & $  2\pm2$\\
060      & \object{LSS\,4615}                        & $ 9.65$ &  $ 0.27\pm0.01$ & $12.4\pm0.4$                    & $  1.60$ & $   0.1$ & $4.5\pm0.3$ & $17\pm 2$ & $  3\pm2$\\
061      & \object{ALS\,16976}                       & $10.28$ &  $ 0.17\pm0.01$ & $12.2\pm0.3$                    & $  1.61$ & $   1.3$ & $3.8\pm0.3$ & $10\pm 1$ & $  3\pm3$\\
065      & \object{HD\,164\,906}\tablefootmark{dd}   & $ 7.45$ &  $ 0.46\pm0.01$ & $11.7\pm0.9$                    & $  2.02$ & $  71.6$ & $ \cdots  $ & $\cdots $ & $\cdots $\\
066      & \object{CD$-$24\,13831}                   & $10.17$ &  $ 0.25\pm0.02$ & $12.6\pm0.8$                    & $  1.45$ & $   0.0$ & $4.2\pm0.5$ & $12\pm 2$ & $ 12\pm6$\\
068      & \object{CD$-$24\,13830}                   & $ 9.75$ &  $ 0.76\pm0.01$ & $10.9\pm0.6$                    & $  2.00$ & $   0.0$ & $4.5\pm0.4$ & $17\pm 4$ & $  3\pm2$\\
070      & \object{ALS\,18783}                       & $10.48$ &  $\cdots$       & $11.8\pm0.4$                    & $  1.58$ & $  80.8$ & $3.7\pm0.3$ & $10\pm 1$ & $  2\pm3$\\
073      & \object{HD\,315\,031}                     & $ 8.31$ &  $ 0.30\pm0.01$ & $11.3\pm0.6$                    & $  0.71$ & $   1.0$ & $4.8\pm0.5$ & $22\pm10$ & $  3\pm2$\\
076      & \object{HD\,315\,024}                     & $ 9.55$ &  $ 0.11\pm0.01$ & $12.0\pm0.5$                    & $  2.11$ & $   0.0$ & $4.6\pm0.4$ & $19\pm 5$ & $  3\pm2$\\
080      & \object{CD$-$24\,13837}                   & $ 9.39$ &  $ 0.11\pm0.01$ & $15.3\pm0.9$\tablefootmark{nm}  & $  1.77$ & $   0.0$ & $ \cdots  $ & $\cdots $ & $\cdots $\\
085      & \object{HD\,164\,933}                     & $ 8.70$ &  $ 0.04\pm0.02$ & $13.7\pm0.5$\tablefootmark{pm}  & $ 11.87$ & $   0.0$ & $ \cdots  $ & $\cdots $ & $\cdots $\\
086      & \object{CD$-$24\,13840}                   & $ 9.60$ &  $ 0.06\pm0.02$ & $11.9\pm0.6$                    & $  0.89$ & $  39.9$ & $3.7\pm0.4$ & $ 9\pm 1$ & $ 24\pm1$\\
093      & \object{HD\,315\,021}                     & $ 8.59$ &  $ 0.28\pm0.01$ & $11.2\pm0.5$                    & $  1.99$ & $   0.2$ & $4.4\pm0.4$ & $15\pm 2$ & $  3\pm2$\\
100      & \object{HD\,164\,947}                     & $ 8.87$ &  $ 0.24\pm0.01$ & $10.6\pm0.6$                    & $  2.49$ & $   0.0$ & $3.8\pm0.4$ & $ 9\pm 1$ & $ 19\pm4$\\
$\cdots$ & \object{LSS\,4627}                        & $11.37$ & $\cdots$        & $13.1\pm0.4$\tablefootmark{nm}  & $196.64$ & $   0.0$ & $3.9\pm0.3$ & $10\pm 1$ & $  3\pm3$\\
\hline
\end{tabular}
\tablefoot{
$m_{\rm v}$ values obtained from SIMBAD database.
The distances of the stars with negative color gradients were determined using the distance modulus assigned to the cluster.
\tablefoottext{dd}{Stars with a SBD.}
\tablefoottext{pm}{Probable member stars.}
\tablefoottext{nm}{Non-member stars.}
}
\end{center}
\end{table*}
}


\bibliographystyle{bibtex/aa}
\bibliography{referencias}

\Online

\end{document}